\documentclass[aps,twocolumn,showpacs,preprintnumbers,nofootinbib,prd,superscriptaddress,groupedaddress,10pt]{revtex4-2}

\makeatletter
\def\l@subsubsection#1#2{}
\def\l@subsubsubsection#1#2{}
\makeatother

\usepackage{graphicx,amssymb,amsmath,amsthm,amsfonts,epsfig,epsf}
\usepackage[linktocpage]{hyperref}
\usepackage[usenames]{color}
\usepackage{epstopdf}

\usepackage{bm}
\usepackage{dcolumn}
\usepackage[latin1]{inputenc}
\usepackage{latexsym}
\usepackage{rotating}
\usepackage{hyperref}
\usepackage{color}
\usepackage{longtable}

\usepackage{enumerate}
\usepackage{tensor,multirow}
\usepackage{url}

\usepackage{slashed}

\newcommand{\comment}[1]{#1}

\newcommand{\nn}{\nonumber}

\def\be{\begin{equation}}
\def\ee{\end{equation}}
\def\bea{\begin{eqnarray}}
\def\eea{\end{eqnarray}}
\newcommand{\ft}[2]{{\textstyle\frac{#1}{#2}}}
\begin{document}
{\hfill }

\title{
Black-hole microstate spectroscopy: ringdown, quasinormal modes, and echoes
}

\author{
Taishi Ikeda$^{1}$,
Massimo Bianchi$^{2}$,
Dario Consoli$^{2}$,
Alfredo Grillo$^{2}$,
Jos\`e Francisco Morales$^{2}$,
Paolo Pani$^{1}$,
Guilherme Raposo$^{1}$
}

\affiliation{$^{1}$ Dipartimento di Fisica, ``Sapienza" Universit\`a di Roma \& Sezione INFN Roma1, Piazzale Aldo Moro 
5, 00185, Roma, Italy}
\affiliation{$^{2}$ Dipartimento di Fisica,  Universit\`a di Roma ``Tor Vergata"  \& Sezione INFN Roma2, Via della ricerca scientifica 1, 
00133, Roma, Italy}

\begin{abstract}
Deep conceptual problems associated with classical black holes can be addressed in string theory by the ``fuzzball'' 
paradigm, which provides a microscopic description of a black hole in terms of a thermodynamically large number of 
regular, horizonless, geometries with much less symmetry than the corresponding black hole. Motivated by the 
tantalizing possibility to observe quantum gravity signatures near astrophysical compact objects in this scenario, we 
perform the first $3+1$ numerical 
simulations of a scalar field propagating on a large class of multicenter geometries with no spatial isometries arising 
from ${\cal N}=2$ four-dimensional supergravity. We identify the prompt response to the perturbation and the 
ringdown modes associated with the photon sphere, which are similar to the black-hole 
case, and the appearance of echoes at later time, which is a smoking gun of \comment{some structure at the horizon scale} and of the regular 
interior of these solutions. The response is in agreement with an analytical model based on geodesic motion in these 
complicated geometries. Our results provide the first numerical evidence for the dynamical linear stability of 
fuzzballs, and pave the way for an accurate discrimination between fuzzballs and black holes using 
gravitational-wave spectroscopy.
\end{abstract}

\maketitle

\section{Introduction}
Within Einstein's theory of General Relativity, black holes~(BHs) are the simplest macroscopic objects one can conceive. 
In stationary configurations, they are fully described only by their mass, spin, and possibly electric 
charge~\cite{Carter71,Robinson:1975bv,Heusler:1998ua,Chrusciel:2012jk}, being in this respect more akin to 
elementary particles than to astrophysical objects~\cite{Holzhey:1991bx}. This simplicity is also associated with a high 
degree of symmetry: stationary BHs must be axisymmetric~\cite{Hawking:1973uf}, and become spherical in the static 
(i.e., non-spinning) limit. 

Owing to this and other remarkable properties, the equations governing the linear response 
of a BH to external perturbations and its quick relaxation towards stationarity after being formed (e.g., in a merger or 
in a stellar collapse) are separable in terms of a simple set of ordinary differential 
equations~\cite{Brill:1972xj,Teukolsky:1972my}, which enormously simplifies the analysis of BH linear perturbations. The latter are 
crucial, for instance, to describe the so-called ``ringdown'' during the post-merger phase of a binary 
coalescence~\cite{TheLIGOScientific:2016pea}. The BH ringdown is governed by a discrete set of complex frequencies 
--~the so-called quasinormal modes~(QNMs)~-- which are uniquely determined by the BH parameters. 
BH spectroscopy~\cite{Vishveshwara:1970cc,ChandraBook,1980ApJ...239..292D,Dreyer:2003bv,Berti:2005ys,Kokkotas:1999bd, 
Berti:2009kk,Isi:2019aib,Giesler:2019uxc} performed by measuring the ringdown with current and future 
gravitational-wave detectors~\cite{TheLIGOScientific:2016src,LIGOScientific:2019fpa,Berti:2016lat} is at present 
the most robust way to study the strong-field regime of General Relativity and the nature of a merger 
remnant~\cite{Berti:2015itd,TheLIGOScientific:2016src,LIGOScientific:2019fpa,Abbott:2020jks}.
Indeed, BHs can be considered as the ``hydrogen atom'' of gravity, and their gravitational-wave spectrum is a unique footprint of possible deviations from General Relativity in the strong-field 
regime, similarly to the energy levels of the hydrogen whose measurement had a paramount impact in shaping the development of quantum electrodynamics~\cite{LambShift}.

This state of affairs is enormously more involved when the spacetime fails to be as simple and as symmetric as a BH.
This happens arguably in any quantum gravity proposal aiming at resolving some outstanding issues with classical 
BHs~\cite{Mathur:2009hf}, namely the curvature singularities that are conjectured to be always covered by event 
horizons~\cite{Penrose:1969pc,Wald:1997wa,Penrose_CCC}, the conundrum of the huge BH 
entropy~\cite{Bekenstein,Hawking:1976de}, and the unitarity-loss problem associated with Hawking evaporation at the 
semiclassical level~\cite{Hawking:1974sw}.

In the string-theory ``fuzzball'' proposal~\cite{Lunin:2001jy,Lunin:2002qf,Mathur:2005zp,Mathur:2008nj}, a 
classical BH is described by an ensemble of smooth and horizonless geometries which represent the microstates of the BH with the same mass and 
asymptotic charges. The classical properties of a BH are expected to emerge 
either through an averaging procedure over a large number of microstates or as a `collective behavior' of 
fuzzballs.
Although finding a statistically significant number of microstate geometries in order to account for the whole BH 
entropy is challenging,
large families of microstates have been discovered in the last 
few years~\cite{Bena:2015bea, Bena:2016agb, Bena:2016ypk, Bena:2017xbt, Bianchi:2017bxl, Bena:2017upb}. 
A microscopic description of the whole entropy is provided by D-brane counting for BPS black holes in four and five dimensions~\cite{Strominger:1996sh, Horowitz:1996ay,Maldacena:1997de}.  Should the 
fuzzball program be successful, it would be a natural solution to the singularity, entropy, and unitarity problems that 
plague the classical BH interpretation.

Fuzzball microstates\footnote{With some abuse of language we interchangeably use the terms ``fuzzballs'' and  ``microstates'' (or ``microstate geometries''), although as we said above the former are ensembles of the latter.} are much less symmetric than a BH. Besides being stationary solutions to consistent 
low-energy truncations of string theory, fuzzballs do not generically possess any spatial isometry. This lack of 
symmetry and the complexity of the microstate geometries have so far hampered the possibility to study their ringdown
and multipole structure~\cite{Bianchi:2017sds, Bianchi:2018kzy, Bena:2018mpb, Bena:2019azk, Bena:2020uup, Bianchi:2020des, Bena:2020see, Bianchi:2020miz, Bianchi:2020yzr}
and to compare it with the one of a BH, which is a task of utmost importance to devise phenomenological tests of the 
fuzzball paradigm --~and hence of quantum gravity~-- with gravitational-wave data~\cite{Mayerson:2020tpn}.

In order to overcome this problem, in this work we perform for the first time $3+1$ numerical simulations of small 
fields propagating on a large class of a microstate geometries. We unveil the entire ringdown 
phenomenology~\cite{Cardoso:2019rvt,Maggio:2020jml} predicted in somehow less motivated\footnote{\comment{To the best of our knowledge, BH microstates are the only model of exotic compact object that can be \emph{arbitrarily} close to a BH and yet arises from a fully consistent theory. Other models often studied in the literature are either~\cite{Cardoso:2019rvt}: i) coming from a consistent theory (e.g. boson stars), but do not approach the BH limit continuously; or ii) have a proper BH limit but are ``ad hoc'', i.e. either phenomenological or obtained by prescribing a metric with very peculiar properties (e.g. energy-condition violations, thin shells, etc), as in the case of gravastars and certain wormhole solutions.}} models of exotic compact objects.

This includes the universal prompt ringdown similar to the BH case, which is nonetheless followed by a modulated series of repeated 
``echoes''~\cite{Cardoso:2016rao,Cardoso:2016oxy} associated with long-lived modes almost trapped within the fuzzball 
gravitational potential, providing a \comment{smoking gun of some extra structure at the horizon scale~\cite{Barausse:2014tra,Holdom:2016nek,Conklin:2017lwb,Oshita:2018fqu,Burgess:2018pmm,Wang:2019rcf,Cardoso:2019apo,Coates:2019bun,Buoninfante:2019teo,Delhom:2019btt,Dey:2020lhq,Buoninfante:2020tfb,Maggio:2020jml,Liu:2021aqh}.}
These ringdown features are currently searched for in 
gravitational-wave data~\cite{Abedi:2016hgu,Ashton:2016xff,Conklin:2017lwb,Westerweck:2017hus,Abedi:2018pst,
Conklin:2019fcs,Tsang:2019zra,Uchikata:2019frs,Abbott:2020jks} (see \cite{Cardoso:2019rvt,Abedi:2020ujo} for some reviews).
Finally, we provide a simple physical interpretation of these effects in terms of the geodesics and multipolar structure of these complicated microstate geometries.
We use natural units throughout.

\section{Fuzzball geometries}
Our framework is that of ${\cal N}=2$ four-dimensional supergravity, wherein gravity is (non-)minimally\footnote{Gauge kinetic functions and K\"ahler metric of the scalars are non-canonical.} coupled 
to four $U(1)$ gauge fields and three complex scalars. 
Microstates of spherically symmetric 4-dimensional BHs can be constructed in this theory by considering a multi-center system of intersecting 
D3-branes~\cite{Bena:2007kg,Gibbons:2013tqa,Bates:2003vx,Bianchi:2017bxl}. 
The metric reads
\begin{align} \label{4dsolution}
ds^2=-e^{2U} (dt+\omega)^2+e^{-2U}\sum_{i=1}^3 dx^2_i, 
\end{align}  
with
\begin{equation}
\label{def:expmin4U}
\begin{aligned}
e^{-4U} &= Z_1Z_2Z_3 V - \mu^2 V^2 \,,
\\
 *_3d\omega &=\ft12\left( VdW-WdV+K^IdL_I-L_IdK^I \right) \,,
 \end{aligned}
\end{equation}
and 
\begin{eqnarray}
 Z_I &=& L_I + \frac{\left|\epsilon_{IJK}\right|}{2}\frac{K^J K^K}{V}\,,\\
 \mu &=& \frac{W}{2} + \frac{L_I K^I}{2V} + 
{\left|\epsilon_{IJK}\right|}{}\frac{K^I K^J K^K}{6V^2}
\end{eqnarray}
where $\epsilon_{IJK}$ is the totally antisymmetric tensor, and
 $ \{ V, L_I, K^I, W \} $ are eight harmonic functions ($I=1,2,3$). As an ansatz, we take $N$-center harmonic functions  
 of the form
 \bea
 V &=& 1+\sum_{a=1}^N  {v_a\over |\vec{x}-\vec{x}_a|} \,,  \quad      L_I =1+ \sum_{a=1}^N   { {\ell}_{I,a} \over  
|\vec{x}-\vec{x}_a|}  \,,  \\ 
 K^I &=&  \sum_{a=1}^N  {k^I_{a}\over |\vec{x}-\vec{x}_a|} \,,  \quad \hspace{0.6cm} W =   \sum_{a=1}^N   { m_a \over  |\vec{x}-\vec{x}_a|} \,, 
\label{ansatz0}
 \eea
 with $\vec{x} =r (\sin\theta \, \cos\phi ,\sin\theta \, \sin\phi,\cos\theta)$\,. From a four-dimensional perspective these microstate geometries are singular at the centers but, for specific choice of charges and positions of the centers, they admit a regular horizonless five-dimensional uplift. Overall, the solution is regular, horizonless, and free of other pathologies (e.g., closed timelike curves). In general it carries four electric $Q_A=(Q_0,Q_I)$ and four magnetic $P^A=(P^0,P^I)$ charges.
For concreteness we shall focus on $3$-center ($N=3$) solutions and restrict ourselves to $4$-charge solutions by imposing $Q_0=P^I=0$, although the generalization is straightforward. We provide their explicit form in  Appendix~\ref{app:sol}. 
\begin{figure}
\center
\includegraphics[width=0.49\textwidth]{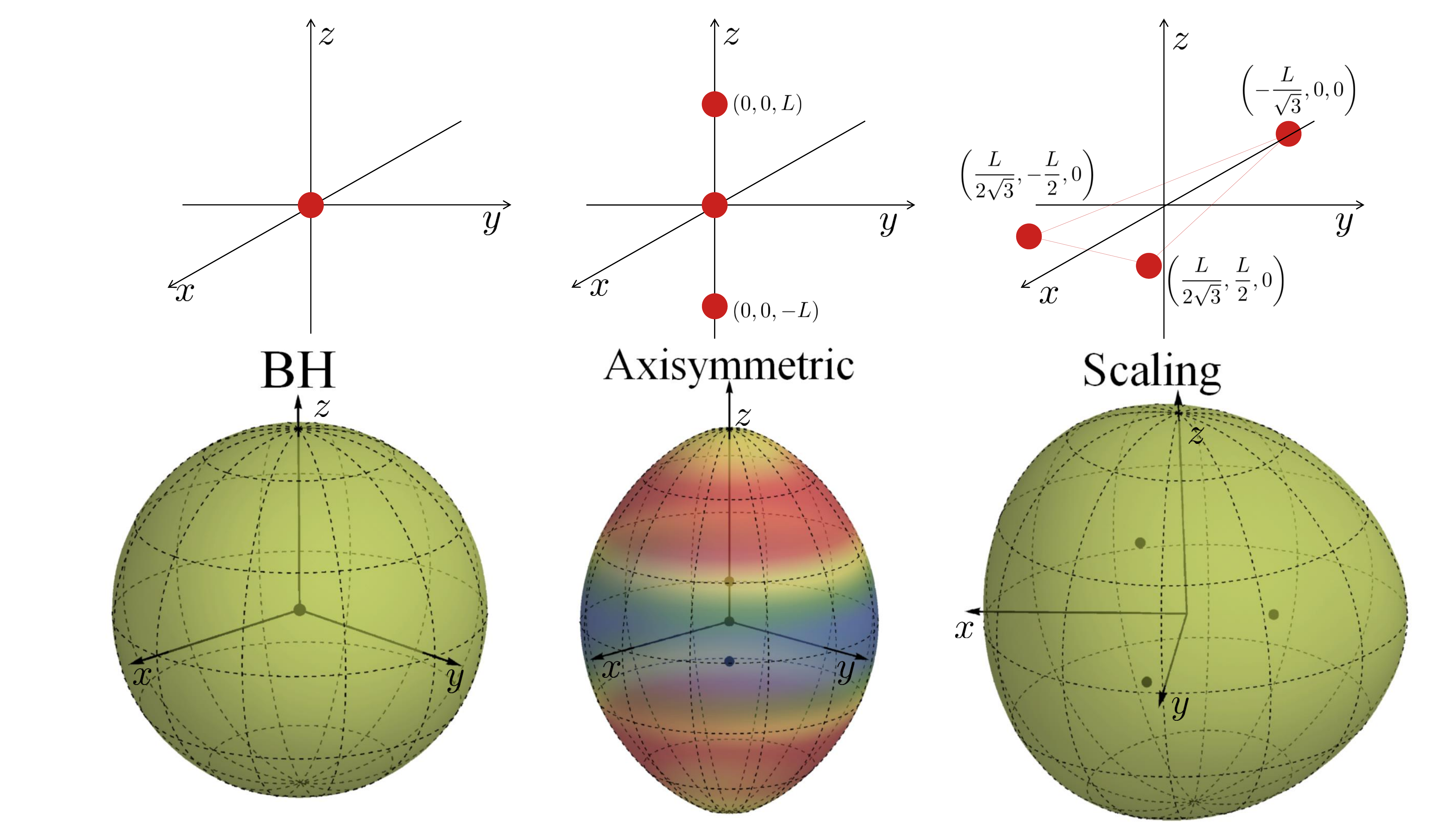} 
\caption{Top: Schematic representation of the multicenter microstate solutions considered in this work (see Appendix~\ref{app:sol} for details on their construction). The leftmost BH solution corresponds 
to 
the case in which the centers coincide. The middle solution is axially and equatorially symmetric, whereas the rightmost ``scaling'' solution breaks the axial symmetry.
Bottom: corresponding embedding diagram for each solution. The shape describes constant-$t$ and constant-$r$ surfaces of the metric deep down the fuzzball's throat. Deformations are related to the mass multipole moments, whereas the colors are weighted according to $g_{t\phi}$ to represent the leading current multipole moments (which vanish for the BH and the scaling solution)~\cite{Bianchi:2020bxa,Bianchi:2020miz}.
}
\label{fig:centers}
\end{figure}
The entropy of corresponding BH is explained by the huge parameter space of the solutions, 
which asymptotically reduce to the 4-charge BPS BH solution~\cite{Cvetic:1995uj}. The latter 
generalizes the extremal Reissner-Nordstr\"{o}m BH to the case of four different charges and three scalar fields. 
In the isotropic coordinates, the line element of the 4-charge BH reads~\cite{Cvetic:1995uj} 
\begin{align}
ds^{2}=-f(r)dt^{2}+f(r)^{-1}\left[
dr^{2}+r^{2}(d\theta^{2}+\sin^{2}\theta d\phi^{2}) \right]\,,\label{metricRN}
\end{align}
with $f(r)= \left( H_1 H_2 H_3 H_4\right)^{-1/2}$ and $H_A=1+{Q_A\over r}$.
The total mass is $M=\ft14 \left( Q_1+Q_2+Q_3+Q_4 \right)$ (with $Q_4=P^0$ in the cases we consider).
When $Q_{A}=Q$, the solution reduces to the extremal Reissner-Nordstr\"{o}m  BH with mass $M=Q$.

Although our method is general, for concreteness we shall consider some representative class of microstate geometries, 
depicted in Fig.~\ref{fig:centers}.
The first family comprises axisymmetric solutions with equatorial symmetry, i.e. reflection symmetric with respect to the equatorial plane, $\theta \to \pi-\theta$ (equivalently, $z \to -z$). 
They have three centers aligned along the $z$-axis, located at $\vec{x}_a = (0,0,z_a)$, with
\begin{equation}
z_1=L \,,\quad   z_2=0\,,\quad z_3=-L\,.
\end{equation}
This is a discrete three-parameter family of solutions with nonvanishing angular momentum, whose explicit form is provided in Appendix A 2. The size L is determined in terms of the charges of the centers.

The second family of $3$-center solutions we consider are those with $v_a=1$, four charges, and centers located on the vertices of a triangle. This is a five-parameter family which in general breaks both axial and 
equatorial symmetry~\cite{Bianchi:2020bxa,Bianchi:2020miz}.
The simplest element is the so-called ``scaling'' solution that corresponds to the three centers being located 
at the vertices of an equilateral triangle and has two free parameters --~related to the mass and the size of the triangle~-- while the angular momentum vanishes.
The explicit form of this solution is given in Appendix~\ref{app:scaling solution}.

In all cases, when the centers collapse to a single point, spherical symmetry is restored and the solution reduces to the extremal (non-rotating) BH.

\section{Linear response of BHs and microstates: prompt ringdown, QNMs, and echoes}
We now focus on the linear dynamics of a neutral massless scalar field propagating in the metrics described in the previous section. This is governed by the Klein-Gordon equation 
$\square\Phi(t,r,\theta,\phi)=0. $
Decomposing $\Phi$ in 
spherical harmonics, i.e. $\Phi=\sum_{lm}\Phi_{lm}(r,t)Y_{lm}(\theta,\phi)$, the (orbital) angular-momentum number is an integer
$l=0,1,2..$ and the azimuthal number $m$ is an integer such that $|m|\leq l$. When the metric is spherically symmetric (as in the BH case considered above), the azimuthal 
number is $(2l +1)$-fold degenerate and different $l$ modes are decoupled from each other. In the axisymmetric case, the degeneracy of 
$m$ is broken but $m$ is still a conserved quantum number: modes with different $m$ are decoupled. In the general case 
with no isometries, modes with different $l$ and $m$ can all mix with each others and it is more convenient to solve 
the Klein-Gordon equation directly as a $3+1$ evolution problem.

The prompt response of a compact object to some perturbation is universally described by the resonant excitation of its photon 
sphere, where (unstable) closed null orbits reside. For a BH spacetime, the photon-sphere modes 
coincide with the QNMs of the object and dominate the linear response~\cite{Cardoso:2016rao}. The QNMs are defined in 
the frequency domain as those complex eigenfrequencies, $\omega=\omega_R+i\omega_I$, which correspond to a solution 
that satisfies purely outgoing-wave boundary conditions at infinity and ``regularity'' conditions at the inner boundary. 
In the BH case, regularity at the horizon imposes purely ingoing-wave boundary conditions, whereas in the horizonless fuzzball case $\Phi$ must be regular at the origin
$r=0$. For each $(l,m)$ mode, there is a countably infinite number of QNMs identified 
by the overtone index $n=0,1,2,..$, with $n=0$ labelling the fundamental QNM dominating the linear response at late 
times. 

Crucially, if the spacetime is horizonless, the photon-sphere modes dominate only the initial ringdown until the 
perturbation has time to probe the inner boundary. Following the universal prompt ringdown, radiation can be reflected 
back and get quasi-trapped within the gravitational potential of the object, occasionally tunnelling to infinity and producing a 
series of repeated and modulated ``echoes''~\cite{Cardoso:2016rao,Cardoso:2016oxy,Mark:2017dnq,Correia:2018apm}. This transient regime interpolates 
between the prompt ringdown and the very late-time behavior, which is instead dominated by the long-lived modes of the 
horizonless compact object~\cite{Maggio:2020jml}.

\subsection{Analytical results in the geodesic approximation}\label{sec:geod}
There exists a tight relation between the ringdown modes of a compact object and some geodesic properties associated 
with the existence of an unstable photon sphere, as established in the eikonal 
limit~\cite{Ferrari:1984zz,Cardoso:2008bp}. In the static case, the real part of the QNM frequency is related to the 
(azimuthal) orbital frequency, whereas the imaginary part of the QNM corresponds to the Lyapunov exponent of the 
orbit~\cite{Cardoso:2008bp}. In the rotating case the relation between modes with generic $(l,m)$ and specific geodesic 
quantities is more involved~\cite{Yang:2012he}. Strictly speaking the geodesic approximation is valid when $l\gg1$ but 
it often works remarkably well also for smaller values of $l$~\cite{Cardoso:2016olt}.

\begin{table*}
\begin{center}
\begin{tabular}{||c|c|c|c|c|c|c|c|c||c|c||}
\hline
\hline
 $\kappa$& $M$ & $L/M$ & $J/M^2$ & $r_+/M$ & $r_-/M$ &  $M \omega_{\rm QNM,+}^{\rm fuzzball}$ & $M \omega_{\rm QNM,-}^{\rm fuzzball}$ & $M \omega_{\rm QNM}^{\rm BH}$ & $\Delta 
t_+/M$ & $\Delta t_-/M$ \\
 \hline
 3 & 13.75 & 0.1091 & 0.0714 & 0.7314 & 0.6759 & $0.6827-0.0767 i$ & $0.7324-0.0878 i$ & $0.6996-0.0871 i$ & 40.7  & 40.1\\
 4 & 24.25 & 0.0508 & 0.0544 & 0.7596 & 0.6490 & $0.6860-0.0841 i$ & $0.7372-0.0855 i$ & $0.7083-0.0859 i$ & 63.8  & 62.7\\
 5 & 37.75 & 0.0301 & 0.0439 & 0.7493 & 0.6379 & $0.6919-0.0852 i$ & $0.7380-0.0839 i$ & $0.7128-0.0851 i$ & 84.9 & 83.7\\
 6 & 54.25 & 0.0201 & 0.0367 & 0.7378 & 0.6345 & $0.6966-0.0854 i$ & $0.7373-0.0831 i$ & $0.7154-0.0847 i$ & 105.1 & 104.1\\
 7 & 73.75 & 0.0144 & 0.0315 & 0.7283 & 0.6343 & $0.7001-0.0853 i$ & $0.7362-0.0827 i$ & $0.7170-0.0844 i$ & 125.0 & 124.0\\
 8 & 96.25 & 0.0109 & 0.0276 & 0.7207 & 0.6352 & $0.7028-0.0852 i$ & $0.7351-0.0825 i$ & $0.7180-0.0841 i$ & 144.6 & 143.7\\
 9 & 121.75 & 0.0085 & 0.0246 & 0.7146 & 0.6367 & $0.7050-0.0851 i$ & $0.7341-0.0825 i$ & $0.7188-0.0840 i$ & 164.0 & 163.2\\
 10& 150.25 & 0.0069 & 0.0221 & 0.7097 & 0.6383 & $0.7067-0.0849 i$ & $0.7332-0.0824 i$ & $0.7193-0.0839 i$ & 183.3 & 182.6  \\
50& 3750 & 0.0003 & 0.0044 & 0.6746 & 0.6593 & $0.7188-0.0837 i$ & $0.7244-0.0830 i$ & $0.7216-0.0834 i$ & 942.2 & 942.1  \\
100& 15000 & 0.0001 & 0.0022 & 0.6706 & 0.6629 & $0.7103-0.0835 i$ & $0.7231-0.0832 i$ & $0.7217-0.0833 i$ & 1887.3 & 1887.2  \\
 \hline
 \hline
\end{tabular}
\end{center}
\caption{Summary of the ringdown features of an axisymmetric fuzzball geometry (with $\kappa_1=\kappa_2=\kappa_3=\kappa$) in the geodesic approximation for $l=m=\pm 2$. 
$M$ is the total mass, the ratio $L/M$ characterizes the distance of the centers, $J/M^2$ is the 
dimensionless angular momentum. The $\pm$ signs refers to co-rotating and counter-rotating orbits. $r_\pm/M$ are the dimensionless critical radii. 
$\Delta t_{\pm}$ is the echo time scale estimated using the WKB approximation [Eq.~\eqref{time_delay}] for co-rotating and counter-rotating orbits.
For the BH case, the QNMs are in agreement with an exact frequency-domain computation (see 
Appendix~\ref{app:FD}) within a few percent.}\label{tab:geod}
\end{table*}

Geodesic motion for a 
massless neutral particle moving in the spacetime given in Eq.~\eqref{4dsolution} can be described by the null 
Hamiltonian $\mathcal{H} =\frac{1}{2} g^{\mu\nu} P_\mu P_\nu=0$, where $P_\mu$ is the particle four-momentum. For simplicity we focus on the axisymmetric case, i.e. consider a stationary metric as in 
Eq.~\eqref{4dsolution}, where $\omega=\omega_\phi d\phi$.  The Hamiltonian in this case can be written as
\begin{align}
2\mathcal{H} = -e^{-2 U} E^2 + e^{2 U} \left(P_r^2+\frac{P_\theta^2 }{r^2}   + \frac{\left(P_\phi+\omega_\phi  E\right)^2}{
r^2\, \sin^2\theta}   \right)   \, \label{ham}
\end{align}
where $P_\phi$ and $E=-P_t$ are constants of motion, while $P_r$, $P_\theta$  vary along the trajectory. We notice that even assuming axial symmetry, $\omega_\phi$ and $U$ typically depend both on $r$ and $\theta$, so in general the radial and angular dynamics cannot be disentangled in simple terms. The situation improves if one further assumes equatorial symmetry.  For this choice $\dot{P}_\theta=-\partial_\theta {\cal H}=0$ at $\theta=\pi/2$, and a particle initially moving along the equator will remain on the plane.

 The null Hamiltonian  condition ${\cal H}=0$ follows from \eqref{ham} after setting $P_\theta=0$ and $\theta=\pi/2$. One finds  
  \be
  P_r^2- Q(r) =0 \label{pqr}
  \ee
  with radial effective potential
  \be
Q(r)=   - {  1 \over  r^2  }   \left[ P_\phi - b_+(r) E  \right]\left[ P_\phi - b_-(r) E \right] \label{qbb}
\ee
 and impact parameter functions
  \be
b_\pm(r) =-\omega_\phi(r) \pm r\,e^{-2U(r) } \,.
\ee
A particle falling from infinity will evolve according to Eq.~\eqref{pqr} till it reaches a turning point $r_*$, {\rm i.e.} a zero of $Q(r)$, and then bounces back to infinity. If the inversion point $r_*$ is a double root of $Q(r)$, the point cannot be reached in a finite time and the particle gets trapped forever orbiting around the mass center and approaching asymptotically a circular orbit (the light ring). This happens for a critical choice $E_c$   
 of the energy and of the radius $r_c$ obtained by solving the critical equations
 \be
 Q(E_c,r_c)=\partial_r Q(E_c,r_c)=0\,.
 \ee   
 Using Eq.~\eqref{qbb}, these equations can be written in the simple form
 \be
 b_c'(r_c) = 0   \quad , \quad
 E_c = {P_\phi \over b_c(r_c) } \label{brce}
 \ee
 where a prime denotes the radial derivative. Rotation produces an inner and an outer photon sphere radius $r_\pm$ obtained as the extrema of the $b_\pm(r)$ functions. 
The two signs distinguish between co-rotating and counter-rotating orbits
with respect to the angular momentum of the spinning microstate geometry.  

The QNM frequencies can be extracted from the WKB formula~\cite{Ferrari:1984zz,Cardoso:2008bp,Bianchi:2020des,Bianchi:2020yzr}
 \be
 {Q(r_c) \over \sqrt{ 2 Q''(r_c) } } ={\rm i} \left( n+\ft12 \right)\,,\label{wkb}
 \ee
 after replacing $E$ by the complex number $E=\omega_R+i\omega_I$ with a small imaginary part $\omega_I$. This equation can be solved at leading order in $\omega_I$ by taking $\omega_R=E_c$, with $E_c$ given by (\ref{brce}). Expanding then to linear order in $\omega_I$ 
  and plugging into (\ref{wkb}) one finds\footnote{\comment{Note that, with a slight abuse of notation, we shall indicate by $\omega_{\rm QNM}$ the modes computed through the WKB approximation. In the BH case, these modes coincide (in the $l\gg1$ limit) with the QNMs of a BH and we shall denote them as $\omega_{\rm QNM}^{\rm BH}$. In the fuzzball case these modes still describe the prompt ringdown as in the BH case and for consistency we shall denote by $\omega_{\rm QNM}^{\rm fuzzball}$.}}
 \be
\omega_{\rm QNM} =\frac{l +\frac{1}{2}}{|b_c| } -{\rm i} \lambda_c (2n+1)\,,
\ee 
 with $l=|m|=|P_\phi|$, for equatorial geodesics, and
\be
 \lambda_c =\left(  {\partial Q (r_c) \over \partial E}\right)^{-1}   \sqrt{ {Q''(r_c) \over 2}   } \,.
\ee

The above result simplifies considerably in the case of spherical symmetry. For example, for the $4$-charge BH with 
$Q_1=Q_2=M$, $Q_3=Q_4=\beta^2 M$, we get\footnote{For $\beta=0$, on has a 2-charge system with zero horizon area (`small BH').}
 \be
 \omega_{\rm QNM}^{\rm BH} M ={ l +\ft12\over  (1+\beta)^2} - {\rm i} (2n+1) {\sqrt{\beta/2}\over (1+\beta)^3}\,.
 \label{eq;QNM WKB}
 \ee

Interestingly, in the case of regular, horizonless objects the geodesic approximation can also capture the relevant timescales associated with echoes~\cite{Cardoso:2016rao,Cardoso:2016oxy,Abedi:2016hgu,Mark:2017dnq,Correia:2018apm,Pani:2018flj}. Geodesics with impact parameter $b=l/E$ along the equatorial plane are described by the radial equation  
 \be
   v_r(r,b)={dr\over dt} ={ {\partial {\cal H} \over \partial P_r} \over {\partial {\cal H} \over \partial P_t}}  
\approx {2 \sqrt{Q(r) } \over {\partial Q(r) \over \partial E}} \,,
   \ee 
 so the scattering time is given by
 \be
 \Delta t (b)= 2\int_{r_c}^{r_t} {dr\over v_r(r,b) }
 \label{time_delay}
 \ee
with $r_t$ the turning point of the orbit. 
The time delay as a function of $b$ has a minimum near b = 0.
minimal time delay provides an estimate of the delay between two subsequent echoes in the fuzzball geometry.

In Table~\ref{tab:geod} we provide a summary of the ringdown features in the geodesics approximation for a 
representative microstate geometry and compare them with the corresponding BH case. Although the method is generically 
valid for any axisymmetric spacetime, for concreteness we focus on the $3$-parameter family of axisymmetric 
fuzzball solutions presented in the previous section with $\kappa_1=\kappa_2=\kappa_3 = \kappa$, which is regular for any integer $\kappa\geq2$. 
Note that (equatorial) light rings exist in this geometry only when $\kappa \geq 3$. 

Some comments are in order. First, we note that the distance between the centers of the microstate geometry, 
$L/M$, monotonically decreases as $\kappa$ grows. Correspondingly the solution approaches the BH limit and the 
fuzzball photon-sphere QNMs coincide with those of the corresponding BH with the same mass. 
It is also interesting to note that the BH case ($L\to0$ or $\kappa\to\infty$) maximizes the real part of the QNM. This feature is analogous to the fact that the Lyapunov exponent of unstable null geodesics near the photon sphere is maximum for certain BH solutions~\cite{Bianchi:2020des}.
Furthermore, as $\kappa\gg1$ the time delays $\Delta t_{\pm}$ grows, showing that gravitational time dilation inside the fuzzball becomes larger in this 
limit. Based on these results, we would expect that the prompt ringdown of a fuzzball should be very similar to that of 
a BH, but extra features in the ringdown should appear on a timescale $\Delta t_{\pm}$, which is typically much longer
than the decay time, $\sim-1/\omega_I$, of the fundamental BH QNM.
In the next section we shall confirm these expectations by comparing the analytical approximation with fully numerical simulations.

\begin{figure*}[th]
\includegraphics[clip, width=3.5cm]{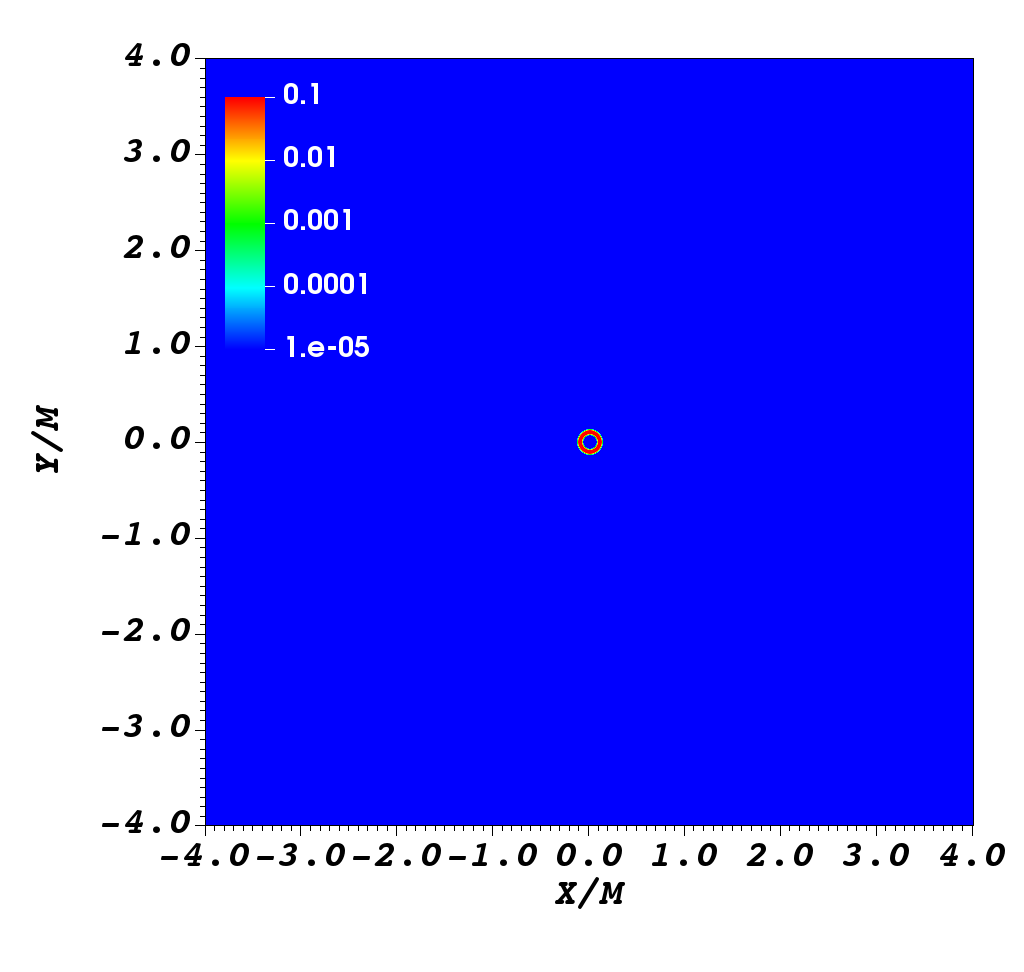}
\includegraphics[clip, width=3.5cm]{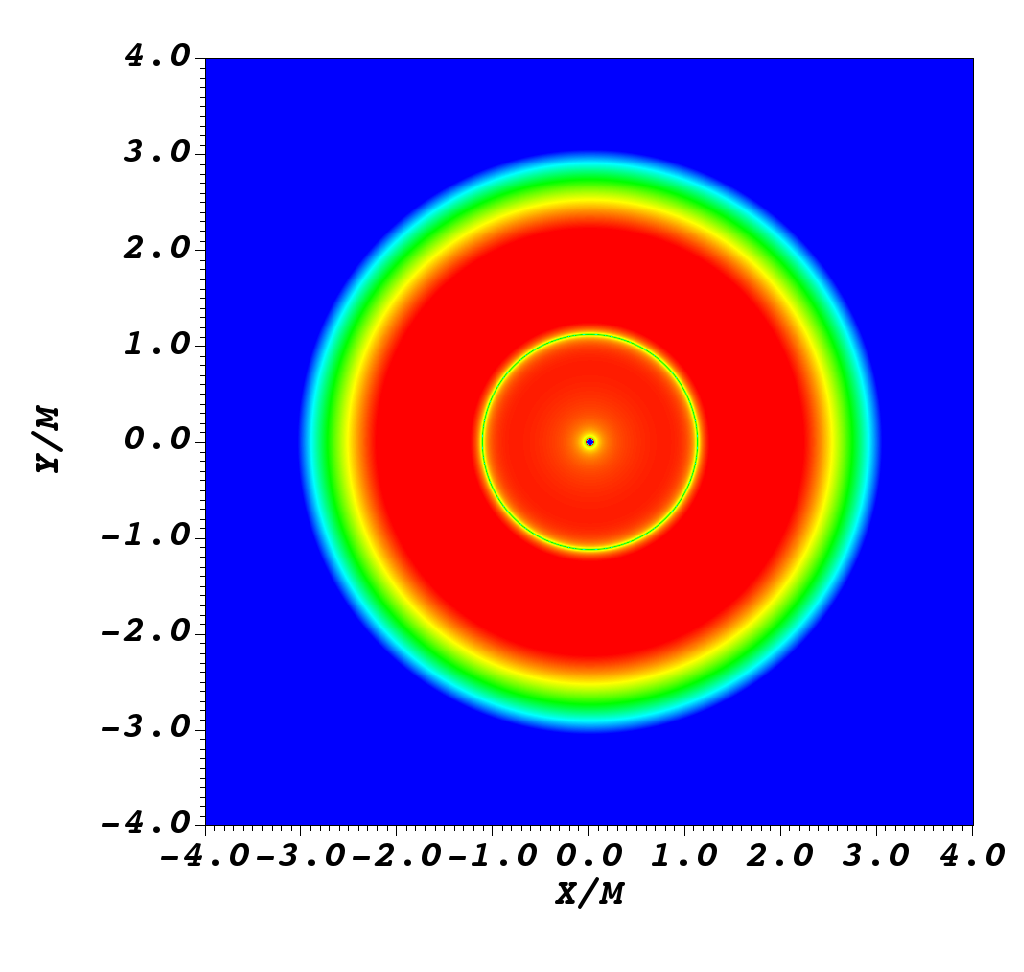}
\includegraphics[clip, width=3.5cm]{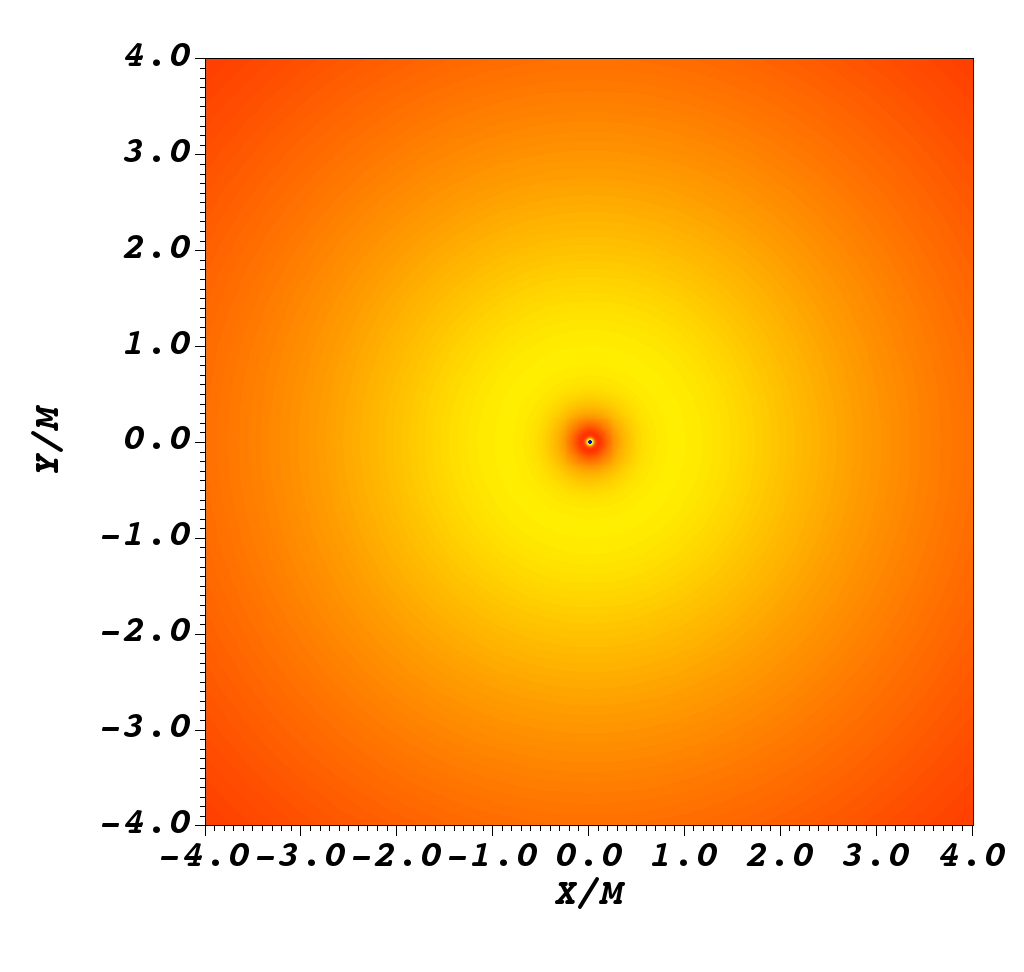}
\includegraphics[clip, width=3.5cm]{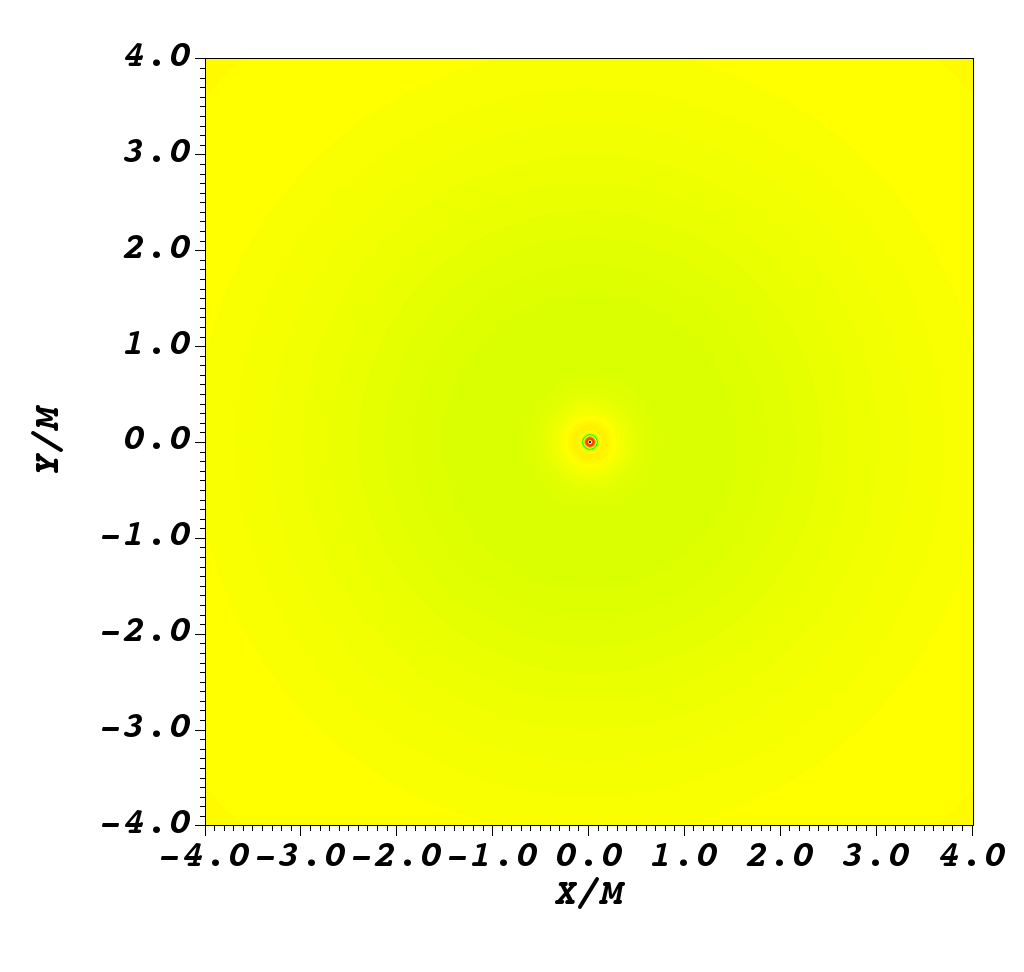}
\includegraphics[clip, width=3.5cm]{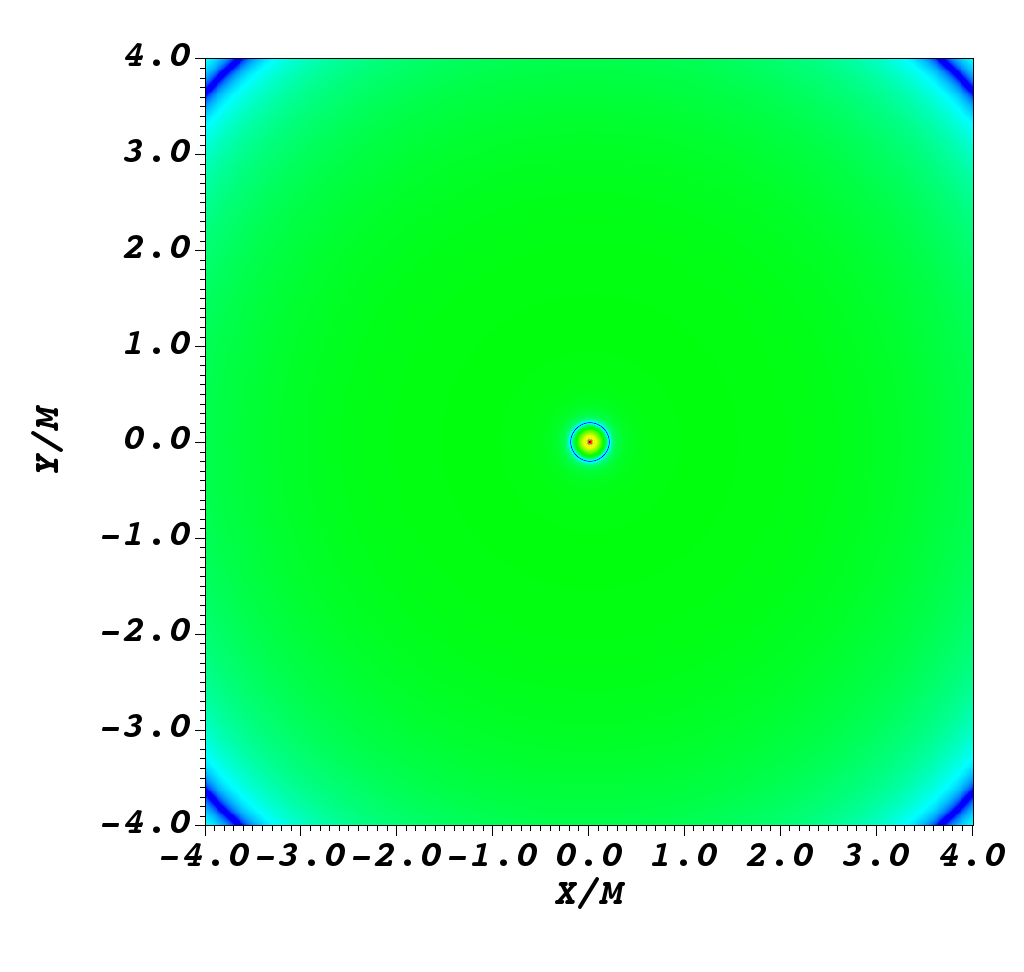}\\
\includegraphics[clip, width=3.5cm]{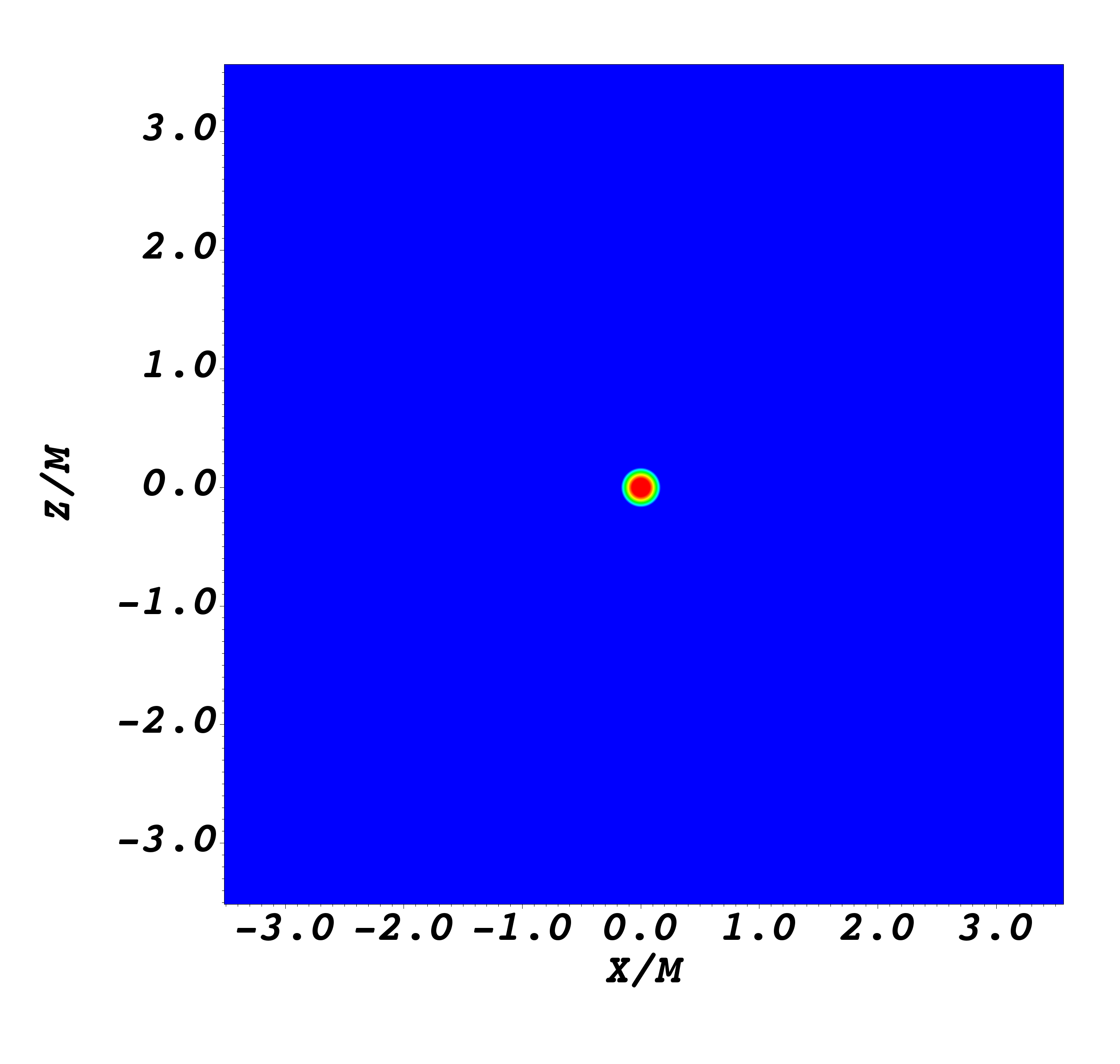}
\includegraphics[clip, width=3.5cm]{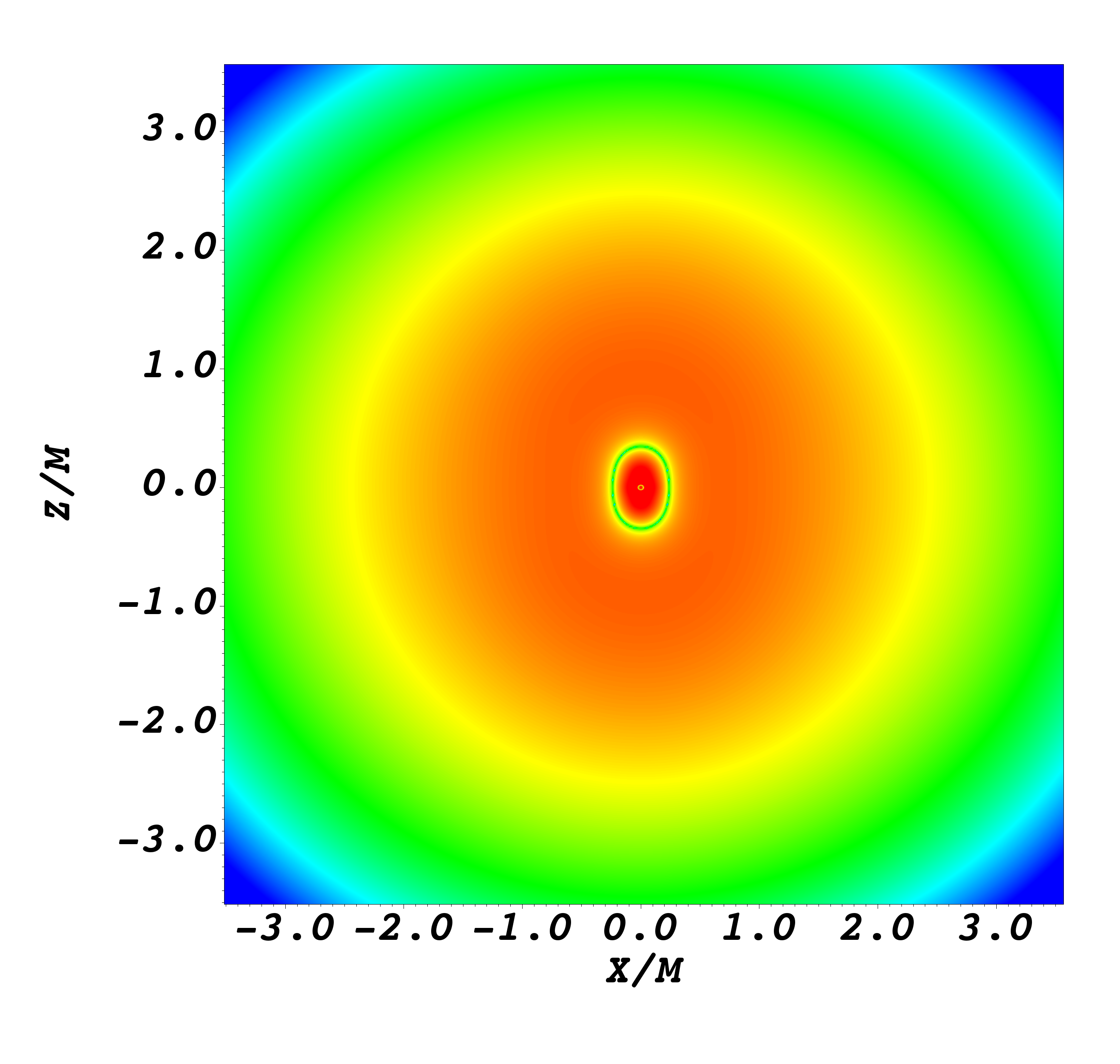}
\includegraphics[clip, width=3.5cm]{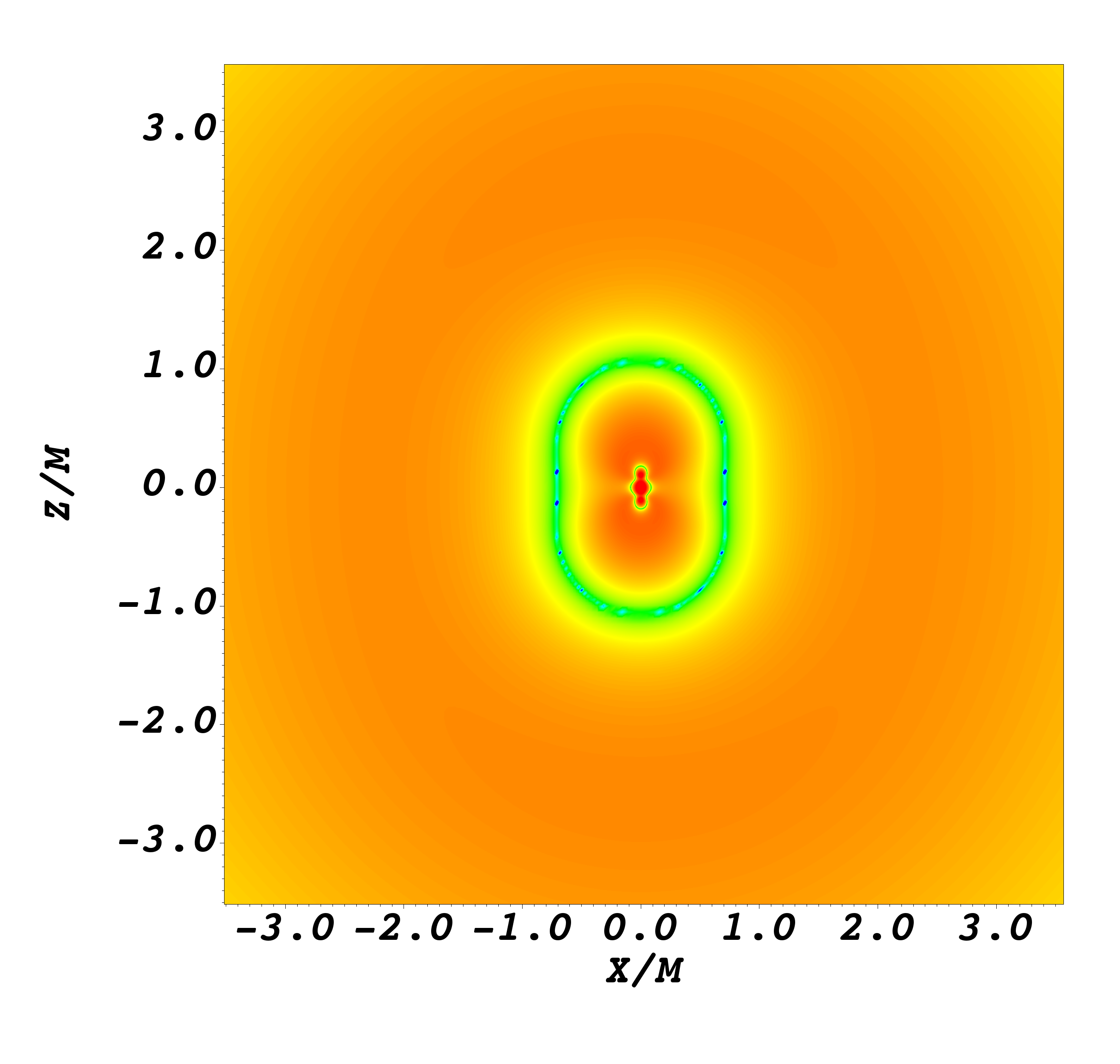}
\includegraphics[clip, width=3.5cm]{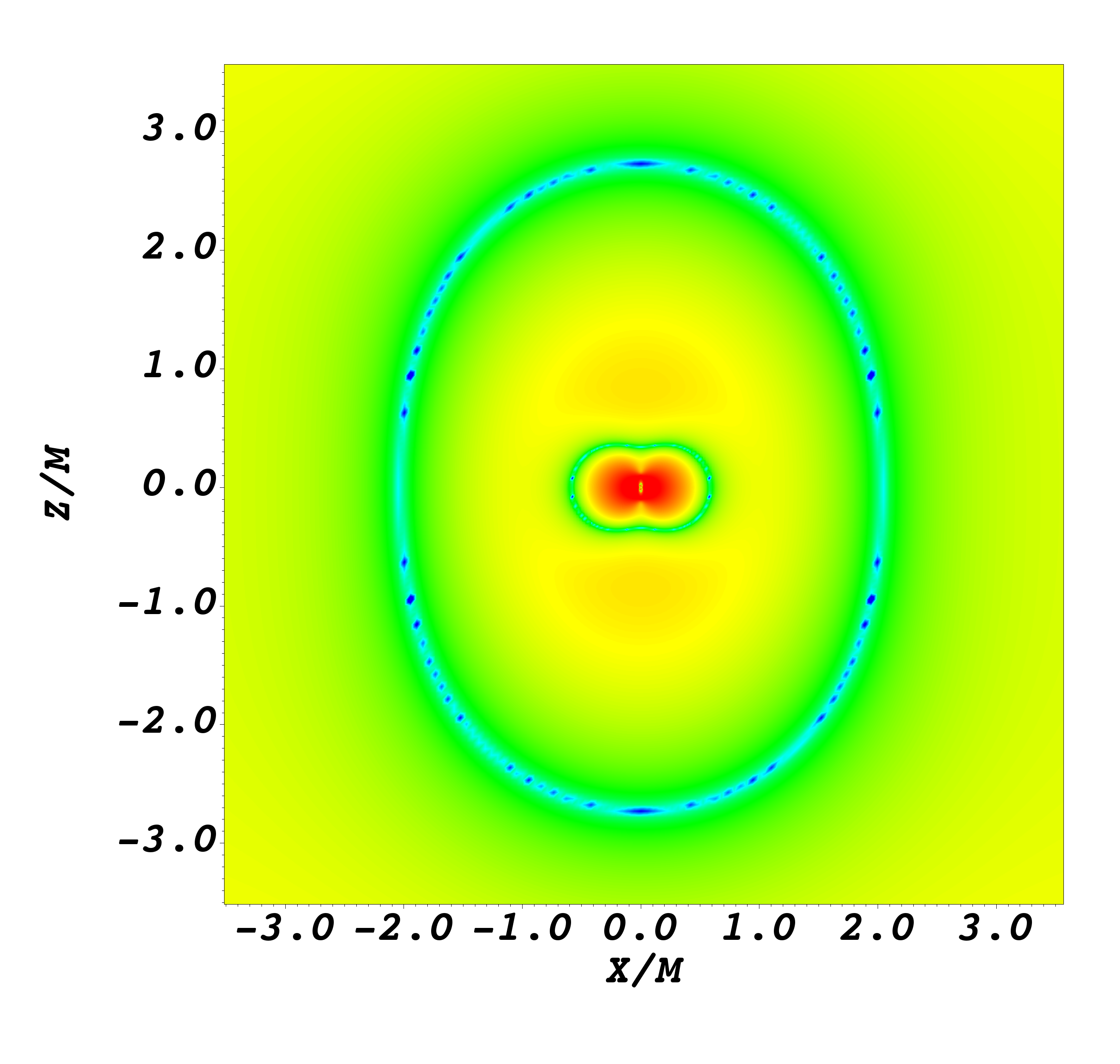}
\includegraphics[clip, width=3.5cm]{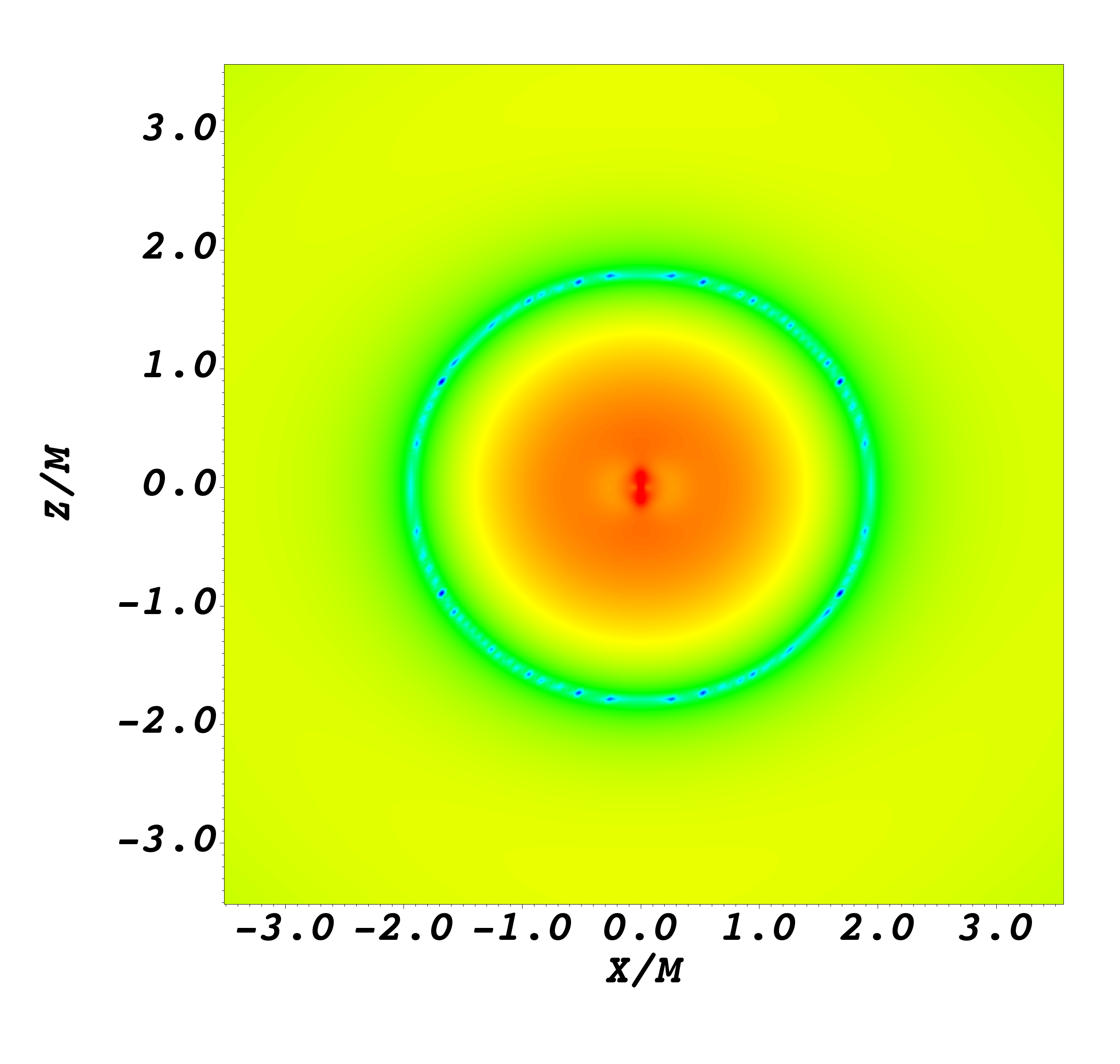}\\
\includegraphics[clip, width=3.5cm]{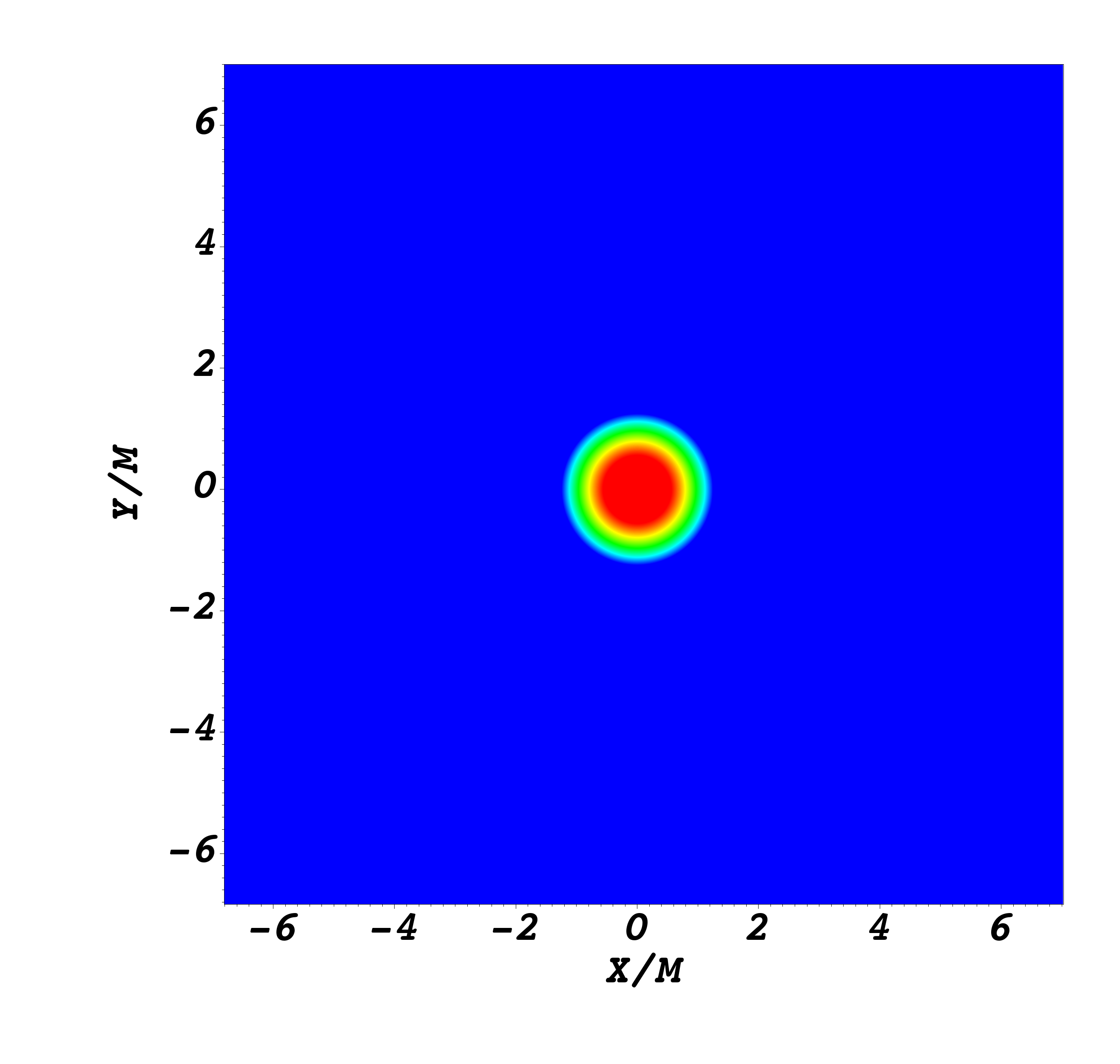}
\includegraphics[clip, width=3.5cm]{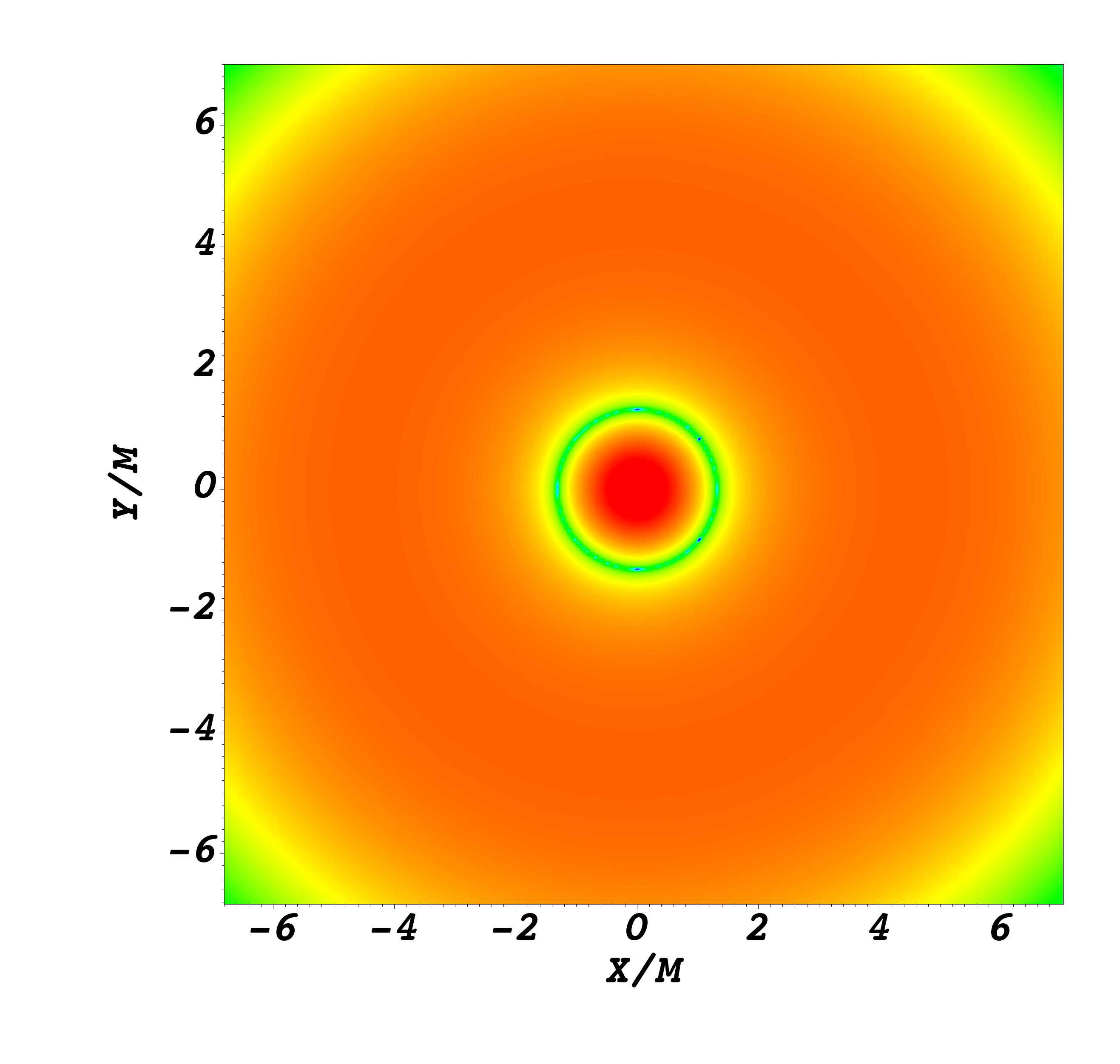}
\includegraphics[clip, width=3.5cm]{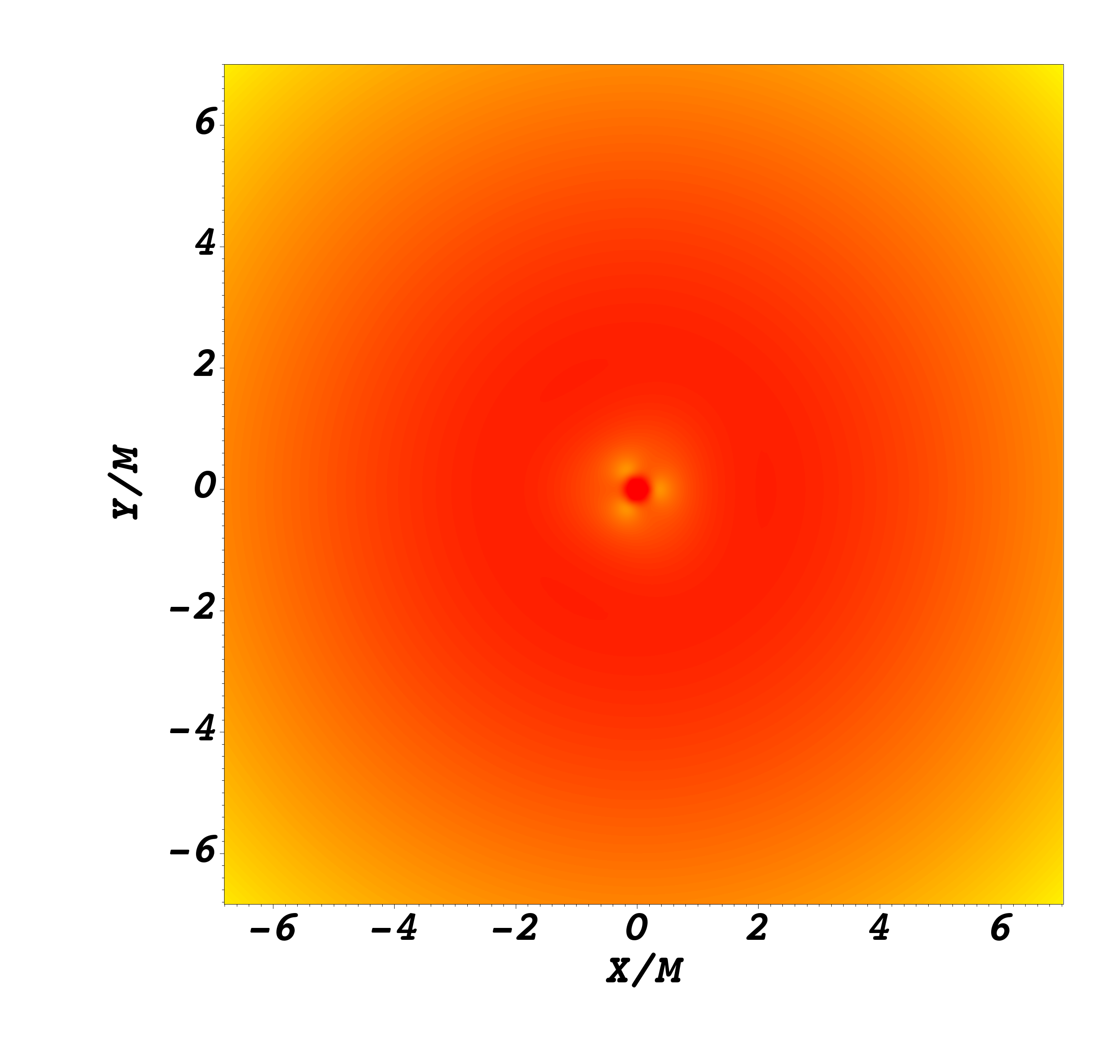}
\includegraphics[clip, width=3.5cm]{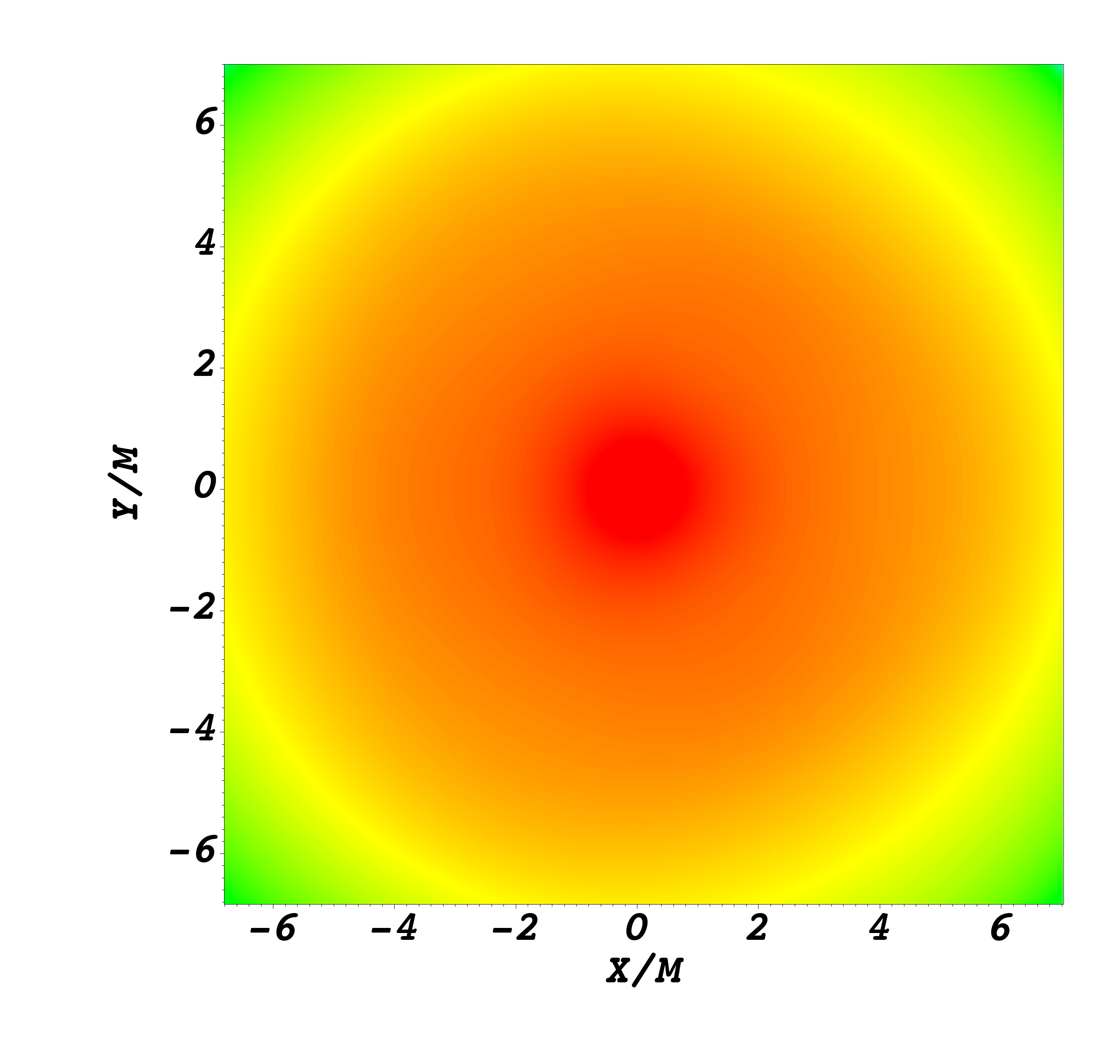}
\includegraphics[clip, width=3.5cm]{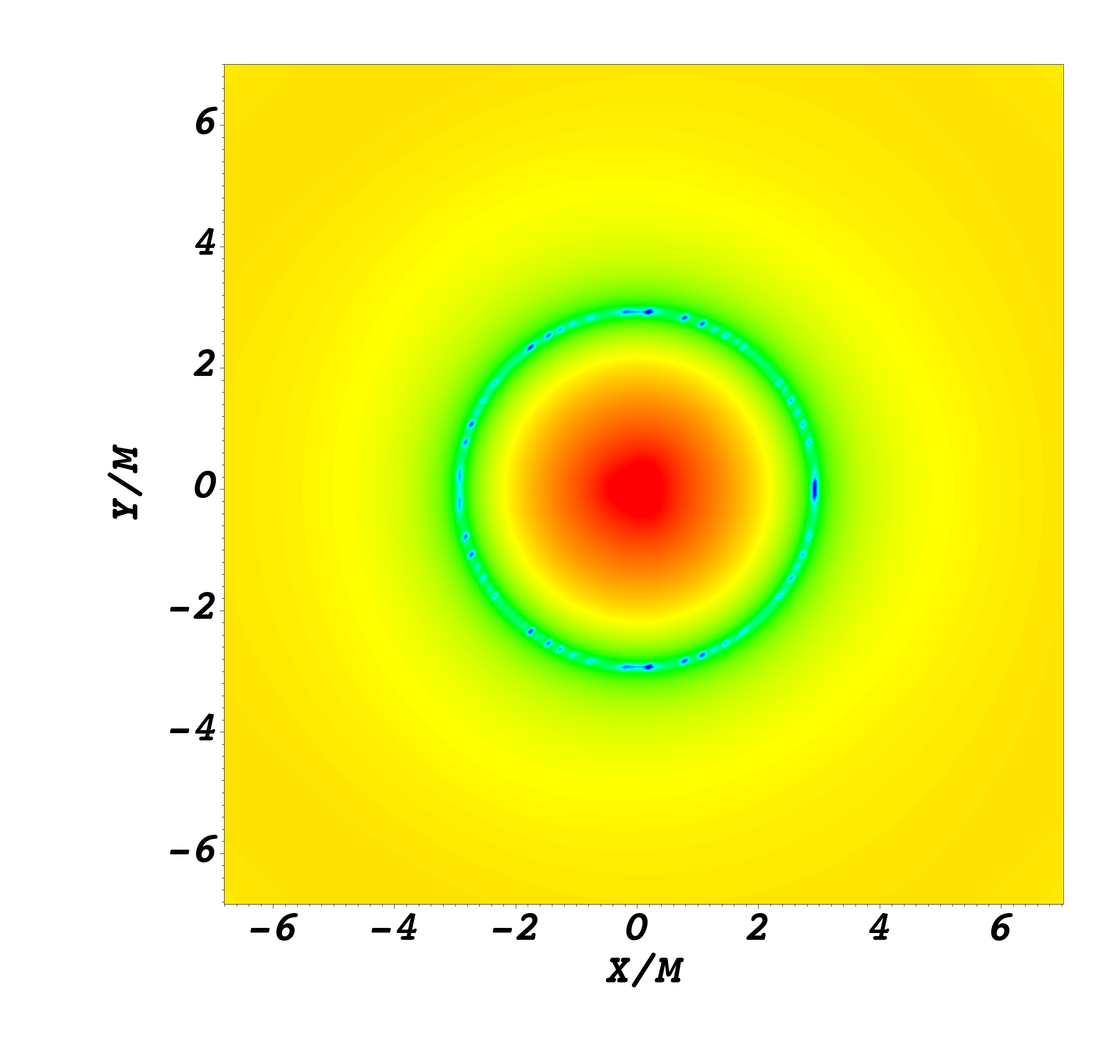}\\
\caption{Snapshots of the evolution of a scalar field around the geometries depicted in Fig.~\ref{fig:centers} and studied in the following, 
namely: $4$-charge BH (top panel) corresponding to left panel in Fig.~\ref{Graph_time_evolution_of_scalar_field_4charge_ID1_Amp1_w0_01_facwin0_001_M1_1_1_M2_0_9_epsilon1e-3_l0}, an axisymmetric fuzzball (middle panel) corresponding to Fig.~\ref{l0_m0_r80_SP_FuzzBall_3c_axisymmetric_analy_omega_wb_CaseI_n0_1_n1_3_n2_3_n3_3_P0_1_nu_1_Gauss_sigma0_5_MP_v5}, and a $3$-center scaling microstate (bottom 
panel) corresponding to bottom panel in Fig.~\ref{l0_m0_r30_SP_FuzzBall_3enters_scaling_analutic_omega_wb_lambda1_L1_analytic_omega_GaussianScalar_sigma1_MP_fac_reg_1}. 
From left to right, they are snapshots at $t/M=0,15,30,45,60$. Related movies are publicly available~\cite{webpage}.
\label{snapshot}}
\end{figure*}

\subsection{Numerical setup for time evolution}

Here, we explain the numerical set up for the time evolution of a scalar field in a microstate geometry. Owing to its spherical symmetry, the case of a $4$-charge BH is much simpler and can be studied with a $1+1$ evolution code or in the frequency domain, as discussed in Appendix~\ref{sec:Numerical code for 4-charge BH}.
However, since the general fuzzball spacetime does not have spatial isometries, 
the field equation for the scalar field is not separable.
We therefore evolve the scalar field using $3+1$ dimensional numerical simulations.
The numerical implementation is based on the Einstein Toolkit infrastructure~\cite{Loffler:2011ay,Zilhao:2013hia,EinsteinToolkit:2019_10}  
with mesh refinement provided by the Carpet package~\cite{Schnetter:2003rb,CarpetCode:web}, and multipatch infrastructure provided by Llama~\cite{Pollney_2011}.
The scalar field is evolved in the ScalarEvolve code, which has been previously used and tested 
in~\cite{Cunha:2017wao,Ikeda:2020xvt,Bernard:2019nkv}.
We implement a $4$-th order Runge-Kutta method, and the spatial derivatives are evaluated by $4$-th order finite differences.
The buffer zones between different refinement levels are evaluated by using $5$-th order interpolation in space and $2$-nd order interpolation in time.

Compared to the standard case of fields evolving on a BH spacetime, more resolution layers are needed. Indeed, the 
fuzzball spacetime has high curvature regions around the centers which must be resolved during the evolution and, at the 
same time, the asymptotic behavior of the field must be accurately extracted at large distances in order to get the 
ringdown signal. Furthermore, although regular from the five dimensional perspective, four-dimensional fuzzball 
geometries are singular at the centers and such singular behavior should be regularized to perform stable numerical 
simulations.
We regularize the geometry by replacing
\begin{equation}
 \frac{1}{|\vec{x}-\vec{x}_a|} \to \frac{{\rm erf}(|\vec{x}-\vec{x}_a|/\epsilon)}{|\vec{x}-\vec{x}_a|}
\end{equation}
in the harmonic functions $V$, $L_I$, $K^{I}$, and $W$. The error function smoothly 
interpolates between unity when $|\vec{x}-\vec{x}_a|\gg \epsilon$ and $\propto |\vec{x}-\vec{x}_a|$ when $|\vec{x}-\vec{x}_a|\ll\epsilon$.
Furthermore, the expressions of the one-form $\omega$ has a coordinate singularity on the axis for axisymmetric fuzzball 
solutions (see Eq.~\eqref{eq:define na nab, and dphiab}),
which can be regularized by the following replacement\footnote{Using two patches with $\omega_\pm = \omega \pm d\phi$ is not practical in numerical analyses.}:
\begin{eqnarray}
\frac{1}{1-(\vec{n}_{a}\cdot\vec{n}_{ab})^{2}}\to \frac{1}{1-(\vec{n}_{a}\cdot\vec{n}_{ab})^{2}+\epsilon}\,.
\end{eqnarray}

Details about code testing and the regularization procedure of spacetime are given in Appendix~\ref{app:code}.

Concerning the initial data, we have performed a variety of simulations with different initial profiles. We report here two representative cases.
The first type is a spherically symmetric Gaussian `shell':
\begin{eqnarray}
\Phi(0,x)&=&Ae^{
-\left(
\frac{r-r_{0}}{\sigma}
\right)^{2}}\,,
\label{eq:initial data momentarily static Gaussian}
\end{eqnarray}
whereas the second type is an $l=m=2$ profile:
\begin{align}
\Phi(0,x)&={\cal W}(r,r_{\rm max},r_{\rm min}){\rm Re}(Y_{2,2}(\theta,\phi))\frac{r}{\sigma}e^{-\frac{r-r_{0}}{\sigma}}\,,
\label{eq:l=2,m=2,ID}
\end{align}
where $\sigma$ and $r_{0}$ define the typical width and central value location of the initial profile, and ${\cal W}(r)$ is 
a window function, which is a $5$-th order smooth polynomial satisfying ${\cal W}(r_{\rm min},r_{\rm max},r_{\rm min})=0$ and 
${\cal W}(r_{\rm max},r_{\rm max},r_{\rm min})=1$. 
In both cases the initial scalar field is instantaneously static, i.e.
\begin{equation}
(\partial_{t}- \gamma^{i}\partial_{i})\Phi(0,x)=0\,,
\end{equation}
where $\gamma^i$ is the shift vector in the $3+1$ decomposition of the metric~\cite{Arnowitt:1962hi}.


\section{Numerical Results}
Here, we summarize the numerical results and discuss the scattering of a scalar wavepacket off fuzzball geometries.
Figure \ref{snapshot} shows some representative snapshots of our simulations on the $y=0$ plane. Related movies are publicly available~\cite{webpage}.

\begin{figure*}[tb]
\includegraphics[width=0.45\textwidth]{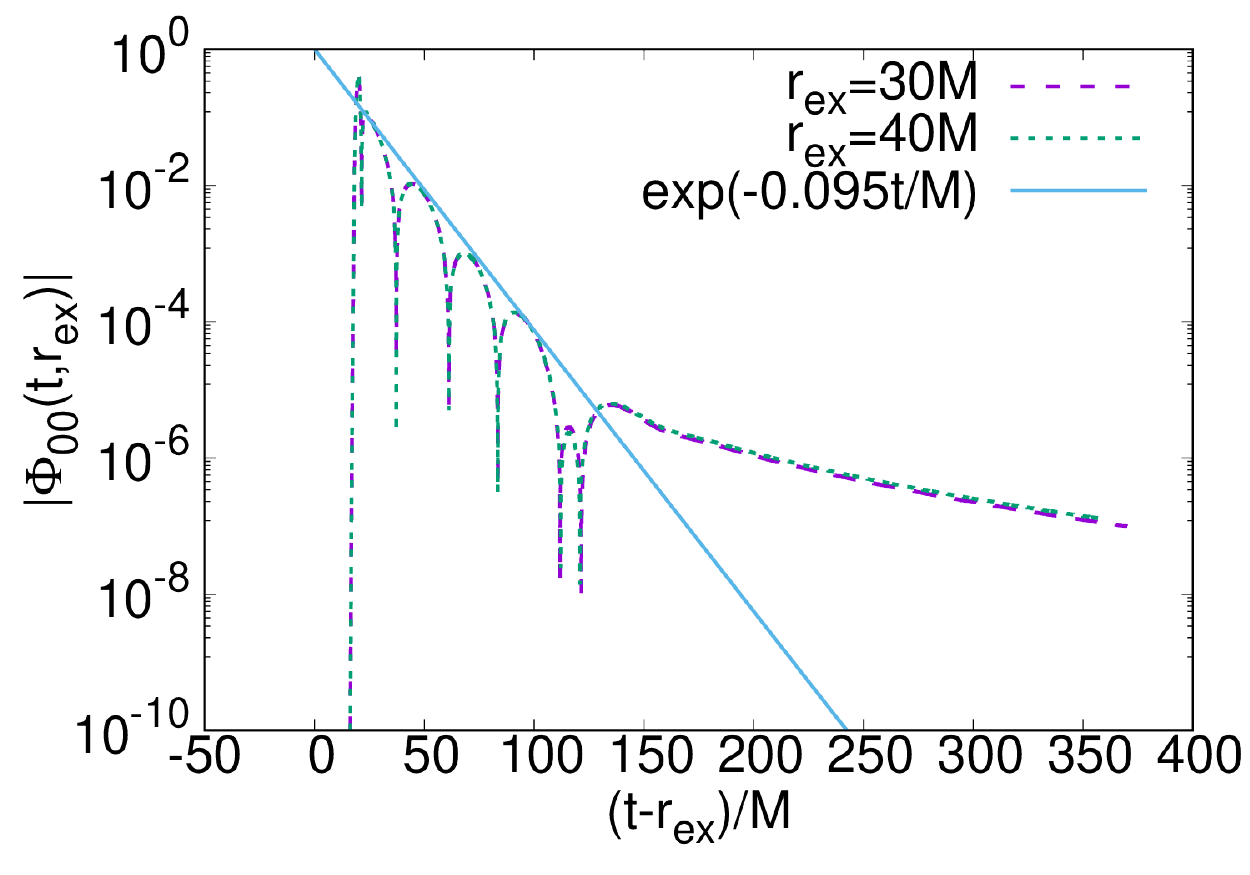}
\includegraphics[width=0.45\textwidth]{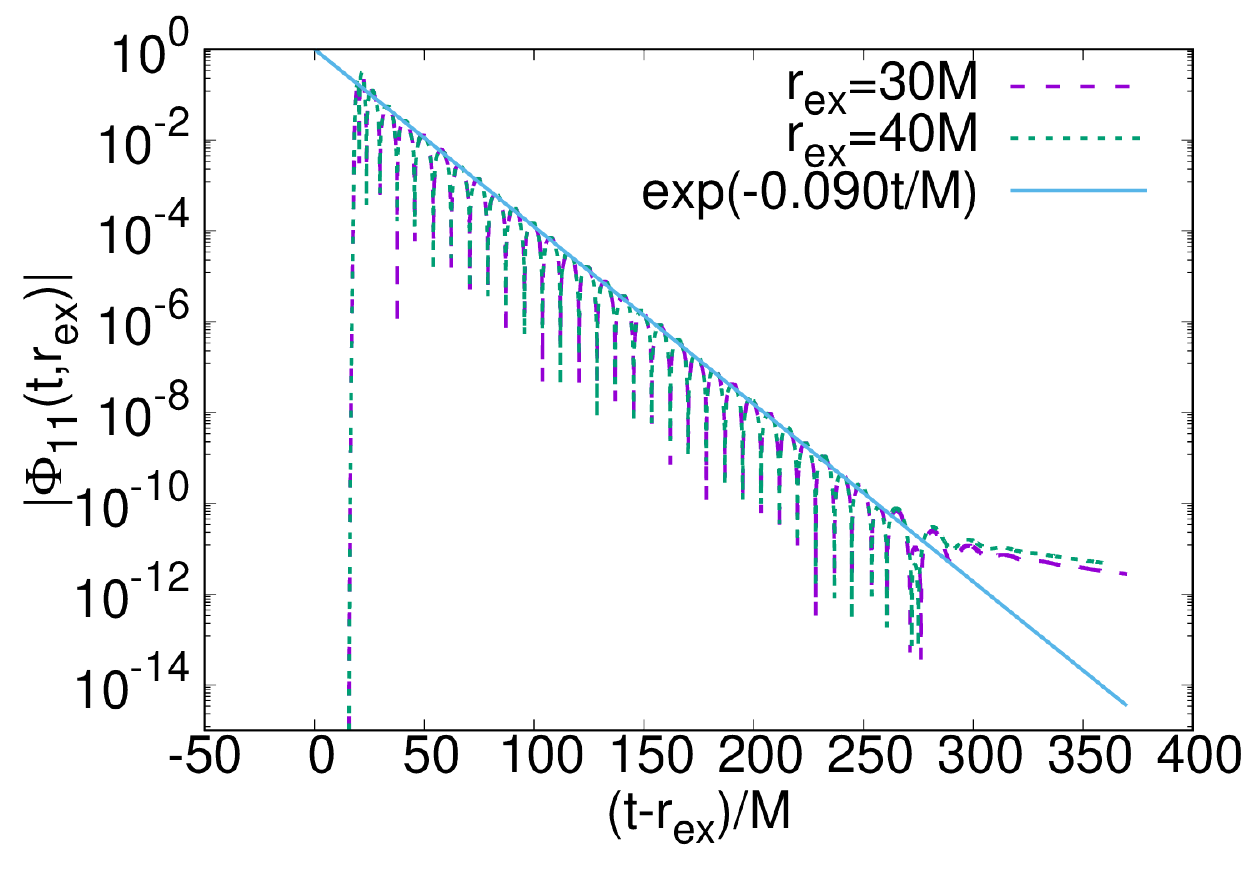}
\caption{
Time evolution of $(l,m)=(0,0)$ (left) and $(l,m)=(1,1)$ (right) multipole mode of a scalar field around a 4-charge BH 
solution with $(Q_{1}=Q_3,Q_{2}=Q_4)=(1.1,0.9)M$ at two different extraction radii.
For the initial data, we took a instantaneously static Gaussian centered at $1.1M$ in tortoise coordinates and with width $\sigma=0.01M$ (see Appendix~\ref{app:BH}). 
The general behavior is independent of the initial data.
\label{Graph_time_evolution_of_scalar_field_4charge_ID1_Amp1_w0_01_facwin0_001_M1_1_1_M2_0_9_epsilon1e-3_l0}}
\end{figure*}
\begin{figure*}[th]
\includegraphics[width=0.45\textwidth]{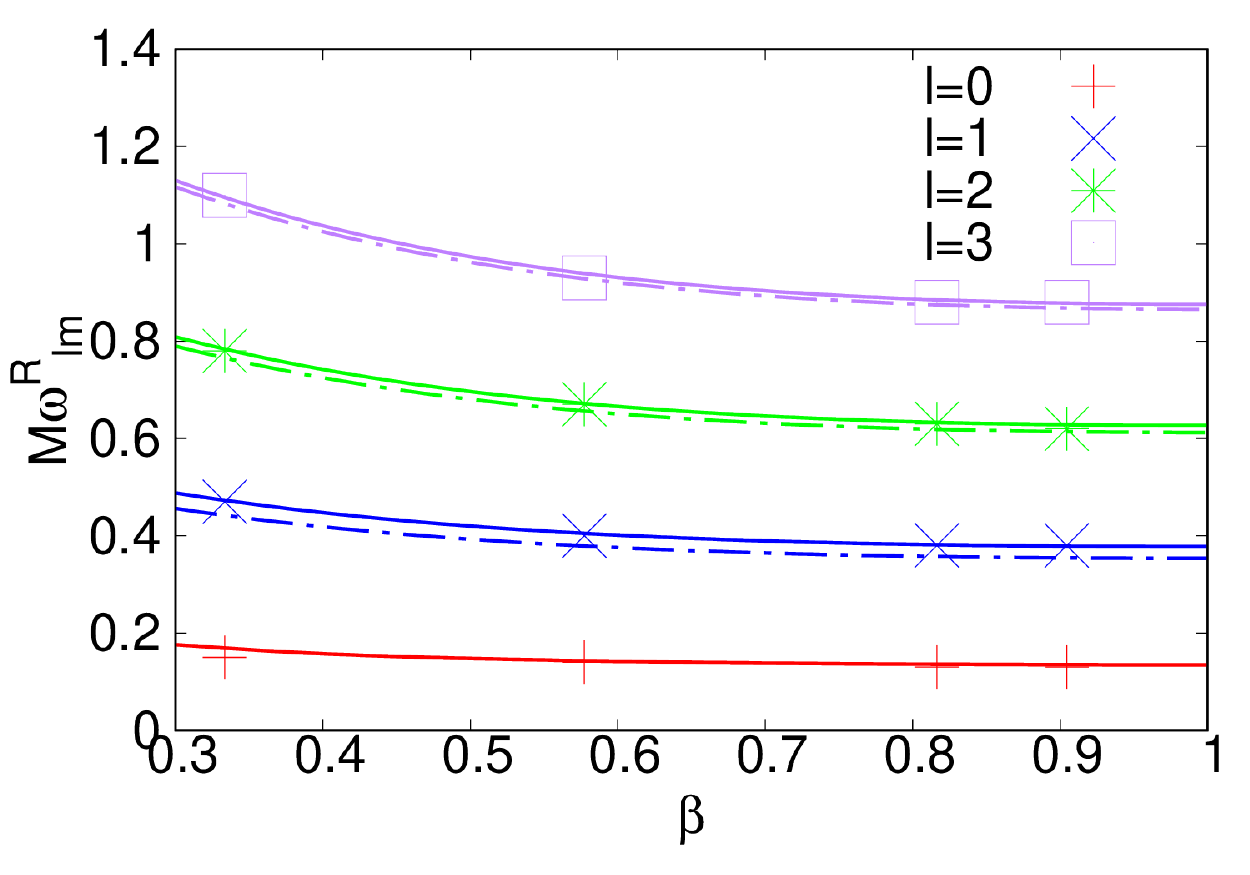}
\includegraphics[width=0.45\textwidth]{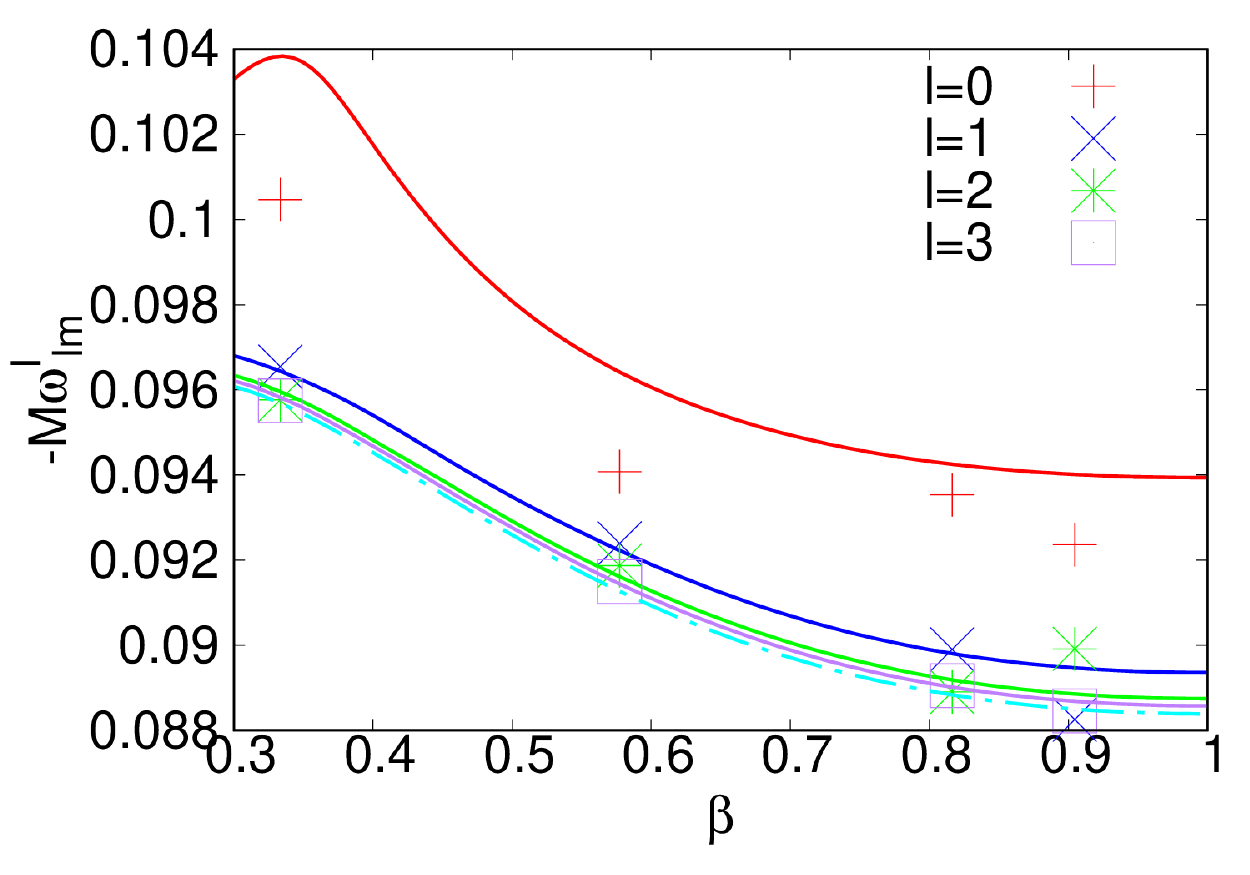}
\caption{
Real part (left panel) and imaginary part (right panel) of the fundamental QNMs of a $4$-charge BH with $Q_1=Q_3$ and
$Q_2=Q_4=\beta^2 Q_1$ as a function of $\beta$ for each $l$. Solid and 
dot-dashed curves refer to the exact frequency-domain computation (see Appendix~\ref{app:FD}) and the analytical result of the geodesic analysis (see 
Eq.~\ref{eq;QNM WKB}), respectively. Data points are the modes extracted from the time-domain waveform of our simulations.
\label{Graph_QNM_4chrage_real_QNM_plot-3_l0}}
\end{figure*}
\subsection{Ringdown of 4-charge BHs}
We evolve the scalar field around a 4-charge BH with $Q_{1}=Q_{3}$ and $Q_{2}=Q_{4}$ based on the formulation discussed 
in Appendix~\ref{sec:Numerical code for 4-charge BH},
and extract the field at fixed radius.
Figure~\ref{Graph_time_evolution_of_scalar_field_4charge_ID1_Amp1_w0_01_facwin0_001_M1_1_1_M2_0_9_epsilon1e-3_l0} shows 
the evolution of the scalar field with $l=0$ and $l=1$.
We observe the standard QNM ringing, wherein the signal dies off exponentially until producing the typical late-time 
power-law tail due to backscattering off the gravitational potential~\cite{Price:1971fb}.
Figure~\ref{Graph_QNM_4chrage_real_QNM_plot-3_l0} shows the QNM frequency as a function of the single 
dimensionless parameter of the metric, $\beta=\sqrt{Q_{3}/Q_{1}}$, with $\beta\to1$ being the extremal Reissner-Nordstr\"om limit.
The data points in Fig.~\ref{Graph_QNM_4chrage_real_QNM_plot-3_l0} are extracted from the simulations by fitting the time-domain waveform, whereas the dotted and solid curve respectively show
the QNMs given by the analytical geodesic approximation [Eq.\eqref{eq;QNM WKB}] and by an exact frequency-domain code (see Appendix~\ref{app:FD}).
The analytical geodesic result is in excellent agreement with the exact one (especially for large 
$l$), and both are in good agreement with the results extracted from the time domain.
The agreement of the imaginary part of the $l=0$ mode is less good (but still within a few percent) due to the smaller number of ringdown cycles before the power law (compare left and right panels of 
Fig.~\ref{Graph_time_evolution_of_scalar_field_4charge_ID1_Amp1_w0_01_facwin0_001_M1_1_1_M2_0_9_epsilon1e-3_l0}), which reduces the accuracy of the fit.
Since the BH QNMs extracted from the time evolution are in good agreement with the photon sphere QNMs computed with the geodesic approximation, and since the latter are independent of the object interior, 
we expect the prompt ringdown to 
be described by the photon 
sphere modes also in the fuzzball case~\cite{Cardoso:2016rao,Cardoso:2016oxy}.
In next subsection, we shall confirm this expectation and use it to distinguish the photon-sphere QNMs from other characteristic modes of the system.

\subsection{Ringdown and echoes of fuzzball microstates}

\begin{figure*}[tb]
\includegraphics[width=0.45\textwidth]{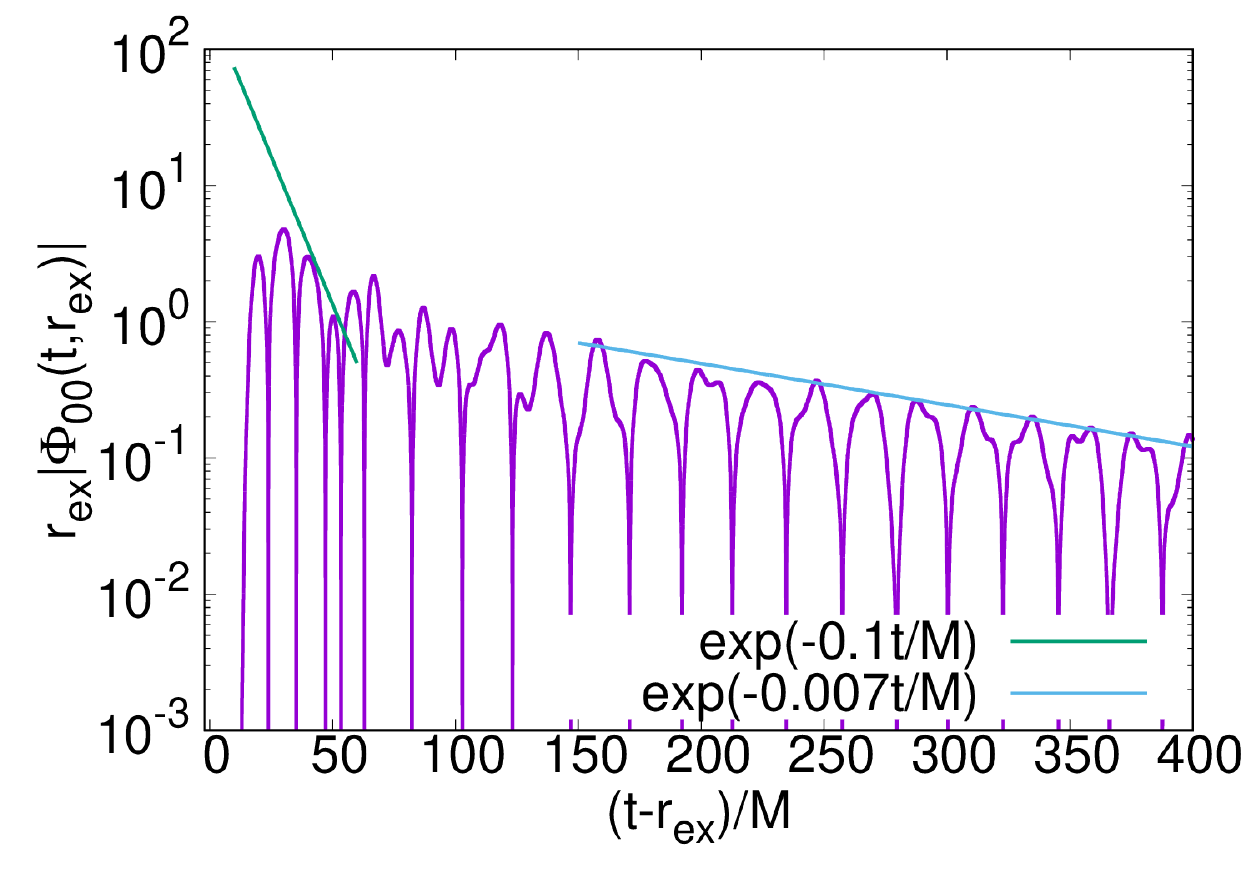}
\includegraphics[width=0.45\textwidth]{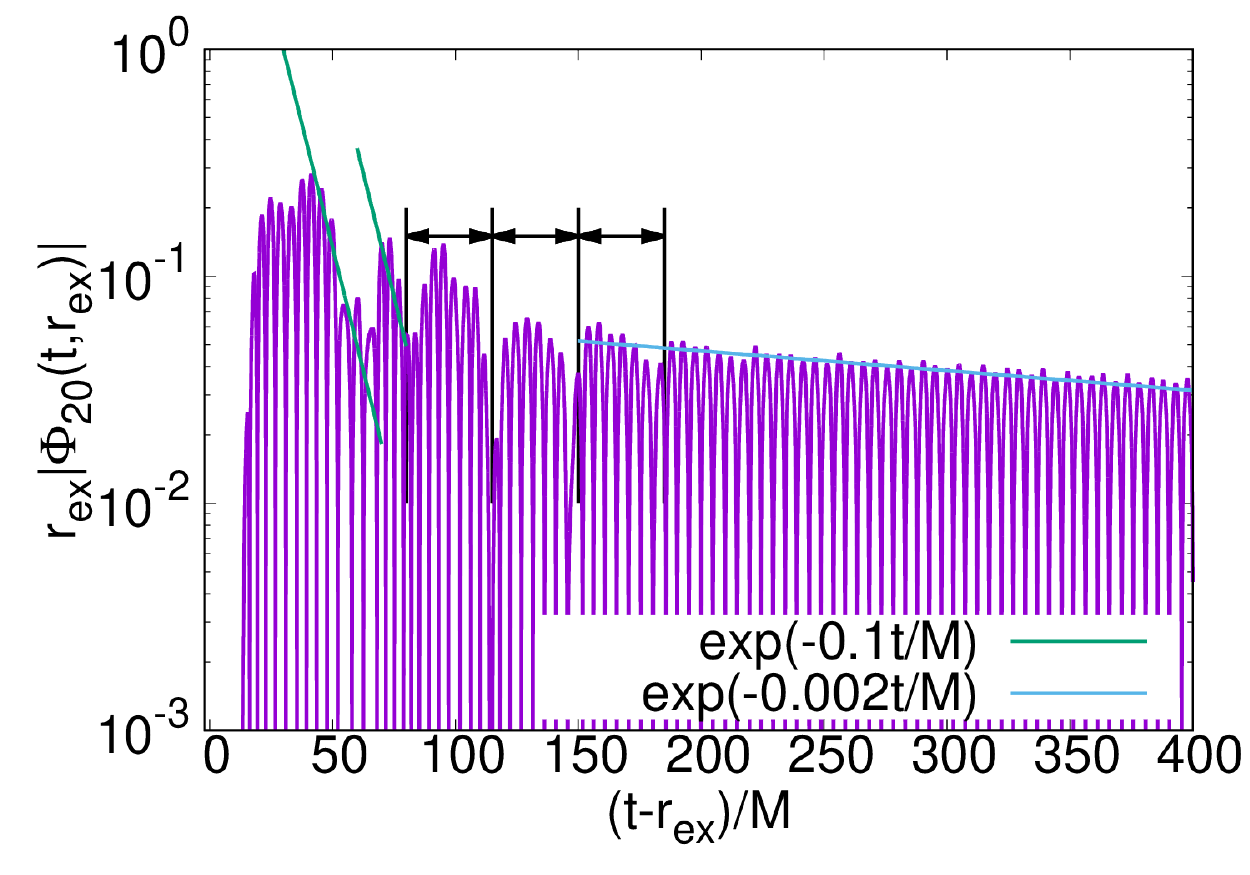}
\caption{
Time evolution of $l=m=0$ (left panel) and $l=2$, $m=0$ (right panel) multipole modes of the scalar field around 
axisymmetric fuzzball solution with $\kappa_1=\kappa_2=\kappa_3=\kappa=3$, and extracted at $r_{\rm ex}=23M$.
The evolution starts from instantaneously static spherically symmetric Gaussian profile [Eq.~\eqref{eq:initial data 
momentarily static Gaussian}] with $\sigma=0.036M$ and $r_{0}=0$.
\label{l0_m0_r80_SP_FuzzBall_3c_axisymmetric_analy_omega_wb_CaseI_n0_1_n1_3_n2_3_n3_3_P0_1_nu_1_Gauss_sigma0_5_MP_v5}
}
\end{figure*}

Let us start by discussing the case of axisymmetric fuzzball solutions.
Figure 
\ref{l0_m0_r80_SP_FuzzBall_3c_axisymmetric_analy_omega_wb_CaseI_n0_1_n1_3_n2_3_n3_3_P0_1_nu_1_Gauss_sigma0_5_MP_v5}
 shows the time evolution for initial data given in Eq.~\eqref{eq:initial data momentarily static Gaussian} with 
$\sigma=0.036M$ and $r_{0}=0$ around a $\kappa_1=\kappa_2=\kappa_3=\kappa=3$ axisymmetric fuzzball, taken as a representative example (the qualitative properties do not change for $\kappa>3$). 
The absence of spherical symmetry produces mode mixing: the initially spherical profile excites both spherical 
($l=m=0$) and quadrupolar ($l=2$, $m=0$) modes, where all dipolar ($l=1$) modes and the quadrupolar modes with 
$m\neq0$ vanish identically due to the symmetries of the metric and to the angular momentum 
composition rules. Hexadecupolar  ($l=4$, $m=0$) modes are also excited but are subleading and not shown in the plot. 

In the prompt-ringdown phase of both modes, we observe the standard damped oscillations with a decay rate 
of about $0.1M$, in agreement with the geodesic result for the photon-sphere modes (see Table~\ref{tab:geod} and Fig.~\ref{Graph_QNM_4chrage_real_QNM_plot-3_l0}).
At very late times, the decay is much slower and presumably corresponds to the long-lived QNMs of the fuzzball geometry. At 
intermediate times, we observe the typical echoes expected for horizonless 
geometries~\cite{Cardoso:2016rao,Cardoso:2016oxy,Abedi:2016hgu,Mark:2017dnq,Correia:2018apm}. Echoes are more evident for the ``induced'' quadrupolar modes compared to the spherical mode. The reason for this is twofold: i) the effective 
potential for modes with large angular momentum is higher, so generically modes with larger $l$ are more efficiently confined;
ii) the fuzzball's gravitational potential has a quadrupolar component that can trap $l=2$ modes more efficiently, especially for mostly equatorial $l=m=2$ modes discussed below. Therefore, for 
spherical modes, echoes could be absent or possibly 
buried in the low-frequency fuzzball fundamental QNM that dominates the late response.
For $l=2$, $m=0$, the typical time delay between echoes is roughly $\Delta t=35M$. Note, however, that it is difficult to identify accurately the echo delay time from the time evolution, especially due to the presence of multiple pulses. Indeed, our instantaneously static Gaussian pulse initially splits into an outgoing pulse and an ingoing pulse; the latter gets quickly reflected at $r=0$ and follows the outgoing pulse with a short delay~\cite{Testa:2018bzd}, which is approximately $\Delta T\approx 16 M$, as computed in the geodesic approximation for this solution. Therefore, we expect a complicate echo pattern from the modulation of this doublet. 

\begin{figure}[tb]
\includegraphics[width=0.45\textwidth]{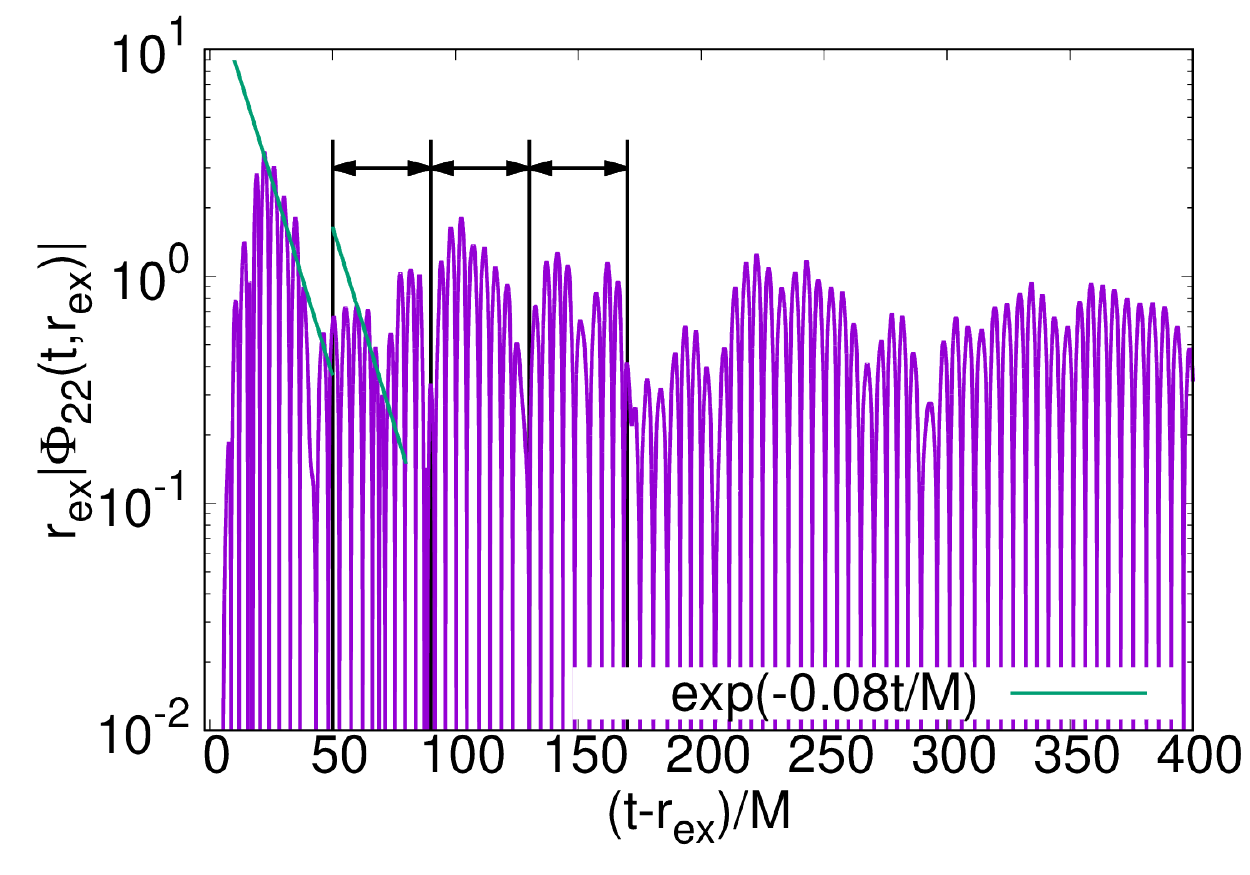}
\caption{
Same as in
Fig.~\ref{l0_m0_r80_SP_FuzzBall_3c_axisymmetric_analy_omega_wb_CaseI_n0_1_n1_3_n2_3_n3_3_P0_1_nu_1_Gauss_sigma0_5_MP_v5}
 but for the evolution of quadrupolar $l=m=2$ perturbations with initial data described by an instantaneously static 
$l=2,m=2$ Gaussian profile [Eq.~\eqref{eq:l=2,m=2,ID}] with $\sigma=0.036M$, $r_{0}=0.14M$, $r_{\rm max}=0.07M$, and 
$r_{\rm min}=0.04M$.
The extraction radius is $r_{\rm ex}=12M$. The prompt ringdown and the echo delay time are in agreement with the 
predictions in Table~\ref{tab:geod}.  
\label{l2_m2_r80_SP_FuzzBall_3enters_axisymmetric}}
\end{figure}

Note also that the echo delay time in Fig.~\ref{l0_m0_r80_SP_FuzzBall_3c_axisymmetric_analy_omega_wb_CaseI_n0_1_n1_3_n2_3_n3_3_P0_1_nu_1_Gauss_sigma0_5_MP_v5} cannot be directly compared with $\Delta t_\pm$ in Table~\ref{tab:geod}, since the latter is 
valid for equatorial motion that corresponds to $l=m$ modes.~\footnote{
However, preliminary analysis suggests that the timescale for $m\neq l$ geodesic motion is similar to the $l=m$ case.
Therefore, the echo time scale of left panel in Fig.\ref{l0_m0_r80_SP_FuzzBall_3c_axisymmetric_analy_omega_wb_CaseI_n0_1_n1_3_n2_3_n3_3_P0_1_nu_1_Gauss_sigma0_5_MP_v5} is compatible with the analytical approximation.
}
In order to perform such a comparison, in Fig.
\ref{l2_m2_r80_SP_FuzzBall_3enters_axisymmetric} we show the evolution starting from $l=2,m=2$ initial data 
[Eq.~\eqref{eq:l=2,m=2,ID}].
At early times, we observe the QNMs of the photon sphere (with a short decay time denoted by the green line), which 
is in good agreement with the estimation from the geodesic approximation and is indeed similar to the BH QNMs.
After the prompt ringdown, the expected echo pattern appears. The delay time is highlighted by black 
arrows and is approximately $40M$, also in good agreement with the time scale predicted by the geodesic 
analysis.
Some further interesting examples are presented in Appendix~\ref{app:extra}.

\comment{
We also note that the scaling of the delay time in the BH limit ($L\to0$) might be very different for microstates of extremal (i.e., zero-temperature) BHs, as opposed to their finite temperature counterparts. For example, for the extremal Reissner-Nordstrom metric in the Schwarzschild-like coordinates (in which the horizon is at $r=M$), the light-crossing time near the surface $r=M(1+\delta)$ scales as $\Delta t_+\sim 1/\delta$, as opposed to the $\Delta t_+\sim\log(\delta)$ behavior in the subextremal case. Interestingly, for more realistic configurations at finite temperature we expect the delay time grows much more slowly than in our extremal case, facilitating the appearance of echoes.}

Finally, let us focus on generic $3$-center microstate solutions. Although our method is general, for concreteness we 
focus on the scaling solution wherein the centers form an equilateral triangle (Fig.~\ref{fig:centers}). This solution 
is equatorially symmetric but breaks axial symmetry so the geodesic approximation discussed in Sec.~\ref{sec:geod} does 
not apply and we have to rely on numerical results only.

\begin{figure*}[tb]
\includegraphics[width=0.45\textwidth]{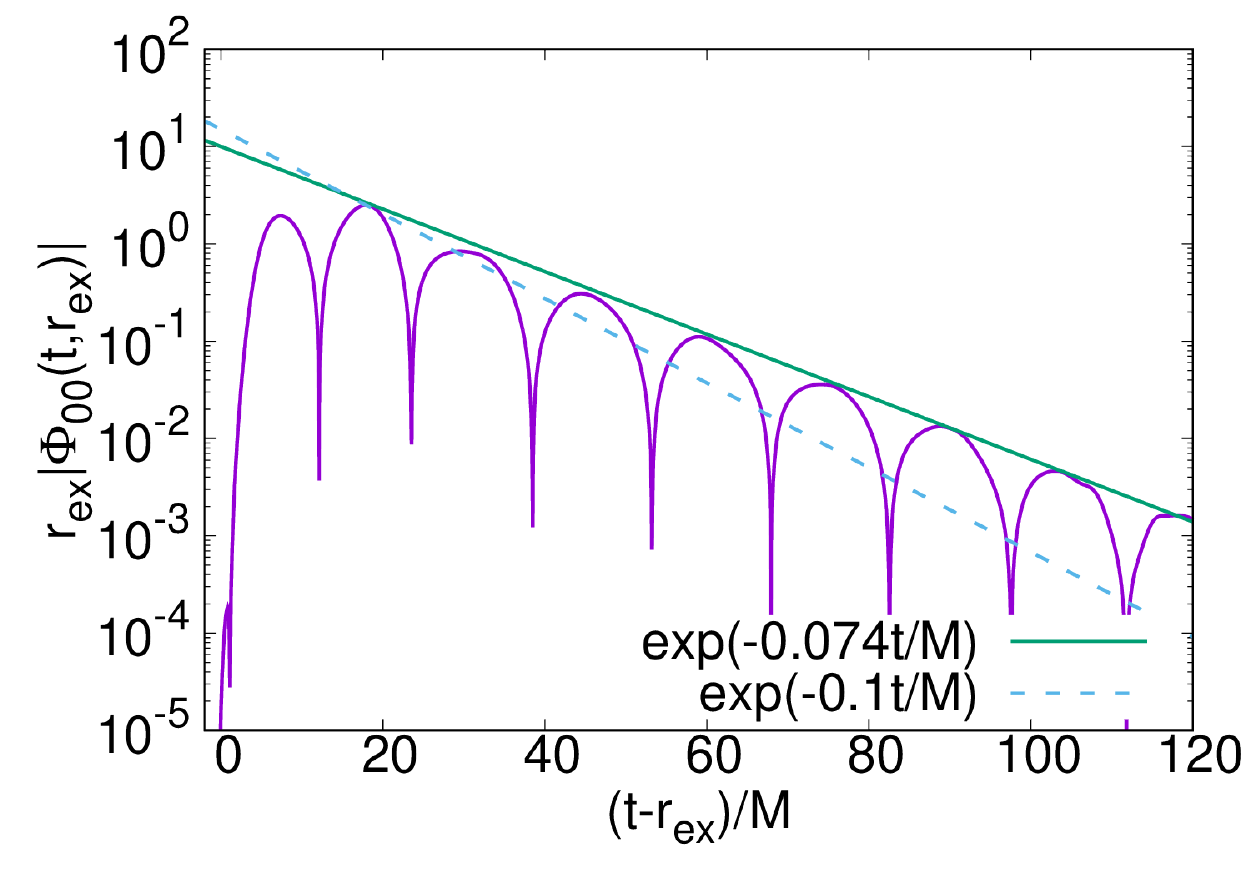}
\includegraphics[width=0.45\textwidth]{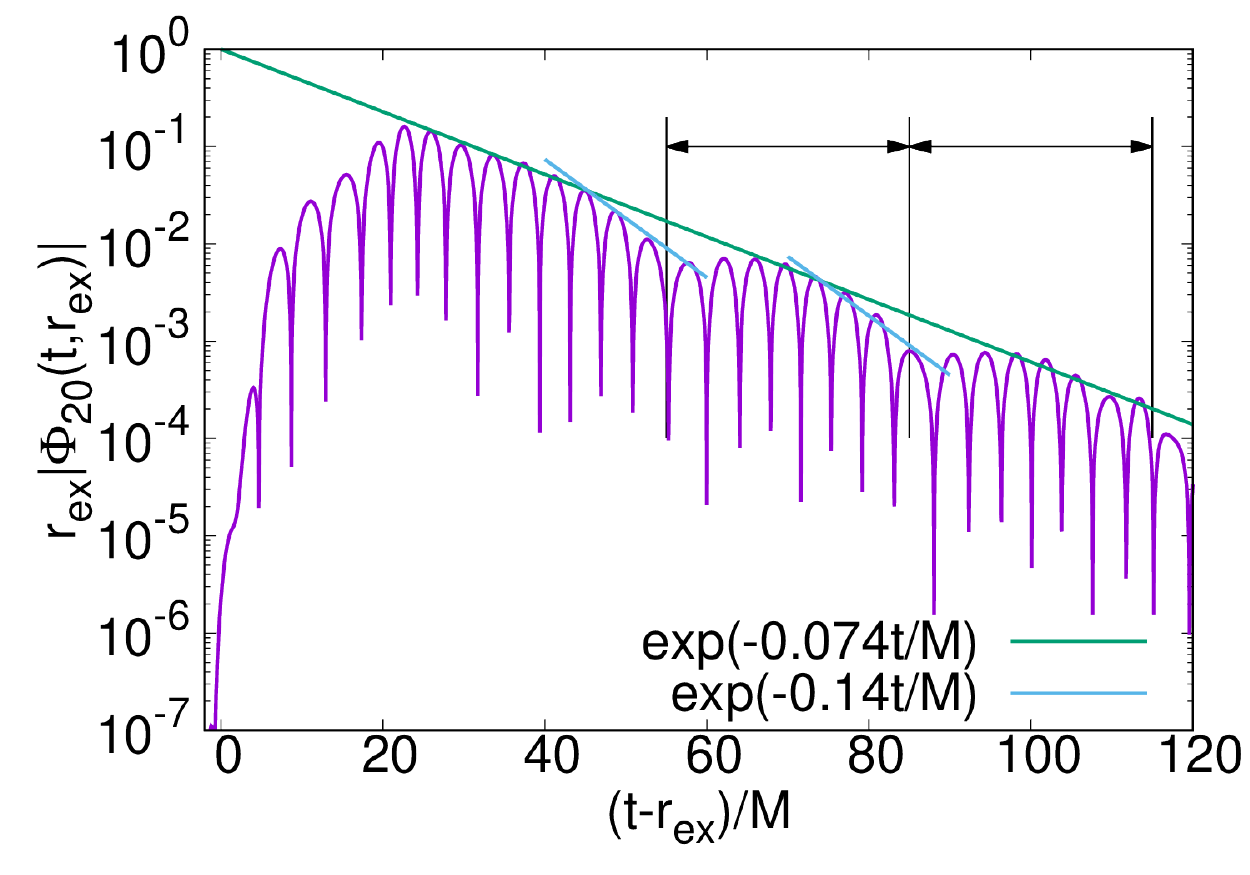}\\
\includegraphics[width=0.45\textwidth]{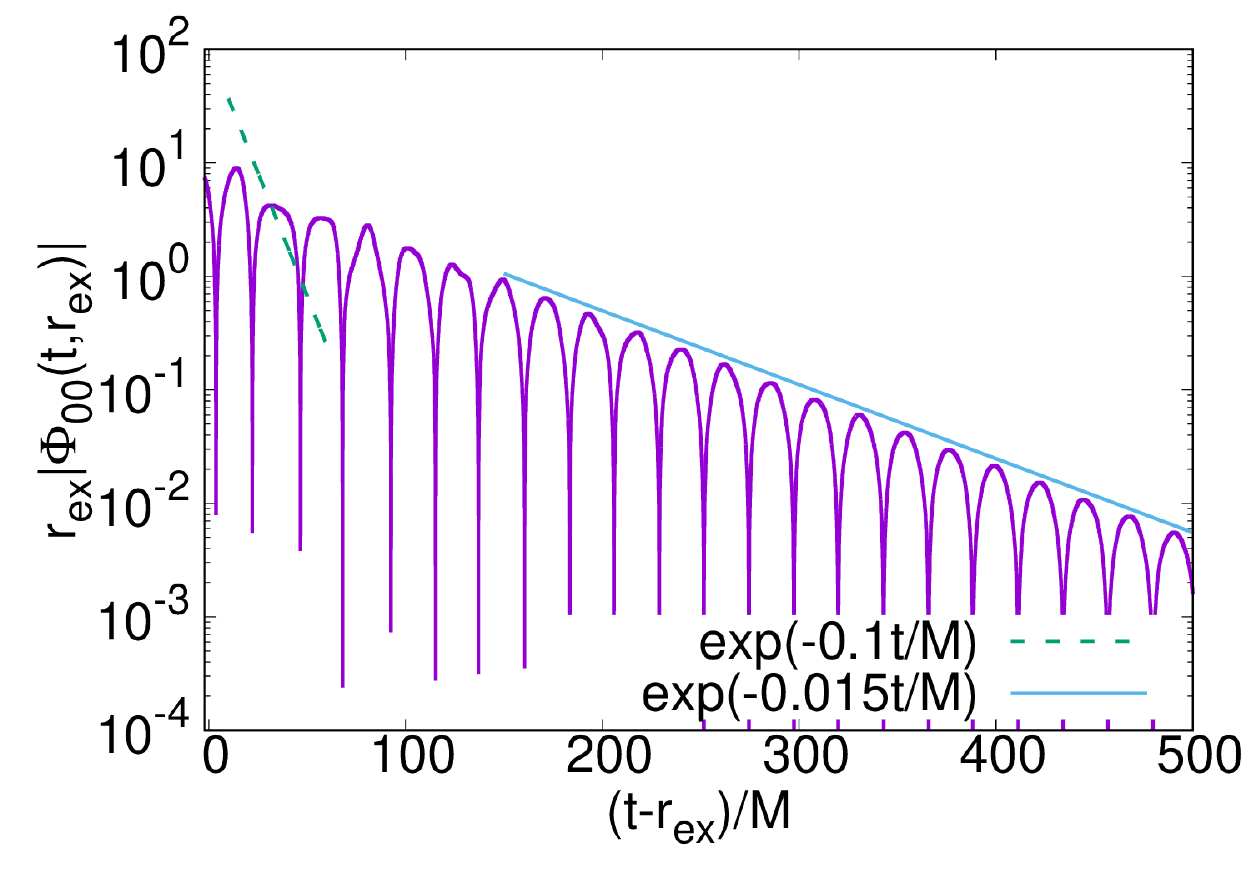}
\includegraphics[width=0.45\textwidth]{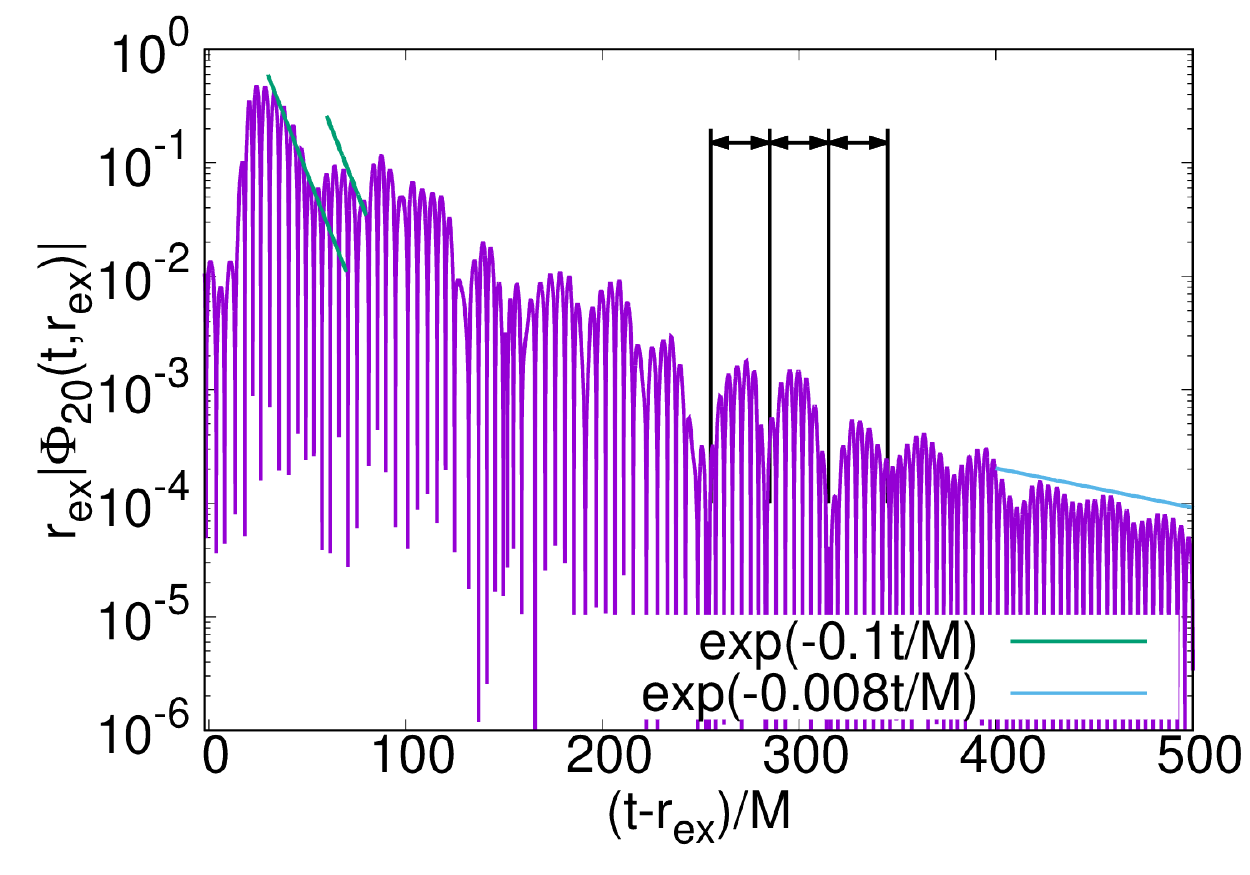}
\caption{
Time evolution of $l=0$ $m=0$ (left panel) and $l=2$, $m=0$ (right panel) multipole modes of the scalar field around 
scaling solution with $L=0.67M,\kappa=1$ (top panel), and  $L=0.27M,\kappa=2$ (bottom panel).
The evolution starts from instantaneously static spherically symmetric Gaussian profile [Eq.\ref{eq:initial data momentarily 
static Gaussian}] with $\sigma=0.67M$ (top panel) and $\sigma=0.27M$ (bottom panel).
The extraction radius is $r_{\rm ex}=20.0M$ (top panel) and $r_{\rm ex}=8.0M$ (bottom panel)
\label{l0_m0_r30_SP_FuzzBall_3enters_scaling_analutic_omega_wb_lambda1_L1_analytic_omega_GaussianScalar_sigma1_MP_fac_reg_1}}
\end{figure*}

Figure 
\ref{l0_m0_r30_SP_FuzzBall_3enters_scaling_analutic_omega_wb_lambda1_L1_analytic_omega_GaussianScalar_sigma1_MP_fac_reg_1} 
shows that the evolution of the scalar field starting from the spherically symmetric, instantaneously static, Gaussian 
initial data around two representative examples of scaling solution. The ringdown of this family of solutions shows the same 
qualitative features as in the axisymmetric case, in particular the echoes are more evident in the ``induced'' 
quadrupolar\footnote{Note that, even if the solution is not axisymmetric, the $l=2,\, m=\pm1$ modes are not excited by an initially spherical pulse. The reason is that the scaling solution only has 
$l=2,\, m=0$ and $l=3,\, m=\pm3$ multipole moments or higher~\cite{Bianchi:2020bxa,Bianchi:2020miz}. According to the standard angular-momentum sum rules, these multipoles cannot produce a 
mode with $m=1$ when coupled to a spherical ($l=m=0$) initial pulse.} mode $l=2,\, m=0$, whereas the spherical mode displays immediately the long-lived, low-frequency fuzzball QNMs, that 
have a much longer decay time relative to the corresponding photon-sphere quantity.
Although not shown, the evolution of $l=m=2$ initial data is similar to that presented in 
Fig.~\ref{l2_m2_r80_SP_FuzzBall_3enters_axisymmetric} for the axisymmetric solution.

\section{Discussion}
We have developed a general technique to study the linear response of a large family of regular, horizonless fuzzball 
geometries describing the microstates of extremal spherically-symmetric BHs. These microstates do not 
possess spatial isometries, which makes the problem dramatically more involved than in the BH case.

For the first time, we studied the evolution of scalar perturbations on these geometries by using a $3+1$ 
numerical-relativity code. Our method is generic and can be applied to any stationary fuzzball.
We unveiled the whole ringdown phenomenology studied in recent years for exotic compact 
objects~\cite{Cardoso:2019rvt,Maggio:2020jml}; in particular, we showed that the prompt ringdown of fuzzballs is 
associated with the photon-sphere modes and mimics the BH response, whereas the presence of 
echoes~\cite{Cardoso:2016rao,Cardoso:2016oxy,Abedi:2016hgu,Mark:2017dnq,Correia:2018apm} at late time is a \comment{smoking gun of some structure at the horizon scale and, in the particular model at hand, of the presence of the fuzzball's regular throat.}

\comment{We have performed several more simulations than those presented here and, in all cases, the perturbations always decay in time.
Although our analysis was not systematic, it provides strong numerical evidence that the fuzzball 
microstates under consideration are linearly stable, at least until our maximum simulation time, $t\approx1000M$. In addition, these geometries 
do not possess an ergoregion, so they are not plagued by the ergoregion instability of horizonless compact 
objects~\cite{1978CMaPh..63..243F,Cardoso:2007az,Chirenti:2008pf,Pani:2010jz,Cardoso:2008kj,Maggio:2017ivp, 
Maggio:2018ivz}.
On the other hand, it has been argued that linear perturbations in certain microstate geometries (with an ``evanescent'' ergoregion) decay logarithmically in time and could be prone to instabilities at the nonlinear level~\cite{Eperon:2016cdd} (see also \cite{Keir:2014oka,Cardoso:2014sna}). It would be interesting to check this result with numerical simulations for the solutions considered in this work. 
}

Although in viable astrophysical scenarios BHs are expected to be neutral, charged BH solutions to supergravity are a 
useful toy model to explore the properties of their corresponding microstates. The numerical analysis developed here 
does not rely on supersymmetry and can be directly applied to putative non-supersymmetric and neutral microstates, should 
the latter be found. In this case, we expect that all the features found in this work will be qualitatively the same.

A natural extension of our work is to consider other types of perturbations. While adapting our code to study a test 
vector field or any self-interacting (possibly massive) test fields is straightforward, considering metric 
perturbations is technically possible but much more challenging, since in these charged geometries the perturbations of the metric, gauge fields, and scalar fields are coupled to each other. However, 
also in this case we expect that the salient features of the ringdown will remain qualitatively the same.

A more promising avenue could be to embed this family of microstate geometries into an effective model such as the one based on the membrane paradigm developed in~\cite{Maggio:2020jml}. Our results 
can be used to calibrate the model and check if it reproduces the salient features of the fuzzball ringdown phenomenology. An important missing ingredient in current models is the fact that 
--~owing to the lack of spatial symmetries~-- different angular modes can be excited with comparable amplitude, even for highly symmetric initial data, as shown by our simulations.

On the longer run, the goal is to confront 
the predictions of the fuzzball scenario with gravitational-wave data, for example extending current ringdown tests~\cite{LIGOScientific:2019fpa,Maggio:2020jml,Abbott:2020jks} and echo 
searches~\cite{Abedi:2016hgu,Ashton:2016xff,Conklin:2017lwb,Westerweck:2017hus,Abedi:2018pst,
Conklin:2019fcs,Tsang:2019zra,Uchikata:2019frs,Abbott:2020jks} to include the predictions of this model, which are dramatically more complex than the toy models used so far to model 
gravitational-wave echoes.

On the theory side, our results also urge to address the ``measure problem'' with fuzzballs. As we have shown, in general the ringdown of a single microstate geometry is very peculiar and contains clear 
features that can be used to distinguish it from the standard BH ringdown. However, if a classical BH is described by a quantum superposition of microstates, then macroscopic observables (e.g. 
QNM frequencies and damping times) might be related to an average over the entire ensemble. Whether the signals should sum incoherently (possibly destroying the echo pattern) or what is the phase space and measure of this average remain open problems that should be urgently addressed if one wishes to test quantum-gravity effects near compact objects in the gravitational-wave era (see~\cite{Dimitrov:2020txx} for related discussion).

Studies of possible observational signatures of the fuzzball scenario are just in their infancy, but are acquiring considerable attention in the context of gravitational-wave and electromagnetic 
tests~\cite{Bena:2020see,Bianchi:2020bxa,Bena:2020uup,Bianchi:2020miz,Bacchini:2021fig} (see~\cite{Mayerson:2020tpn} for a review) that may also encode string corrections \cite{Addazi:2020obs, Aldi:2020qfu, Aldi:2021zhh} to memory effects \cite{Strominger:2014pwa}. We hope that our study on BH microstate spectroscopy could pave the way to further developments in this exciting field.

\begin{acknowledgments}
Numerical simulations have been made possible through a CINECA-INFN agreement, providing access to resources on MARCONI 
at CINECA, through the ``Baltasar Sete-Sois'' cluster at IST, and  through XC40 at YITP in Kyoto University.
This work also was granted access to the HPC resources of MesoPSL financed
by the Region Ile de France and the project Equip@Meso (reference ANR-10-EQPX-29-01) of the programme Investissements d'Avenir supervised
by the Agence Nationale pour la Recherche.
P.P. acknowledges financial support provided under the European Union's H2020 ERC, Starting 
Grant agreement no.~DarkGRA--757480. We also acknowledge support under the MIUR PRIN and FARE programmes (GW-NEXT, CUP:~B84I20000100001), and 
from the Amaldi Research Center funded by the MIUR program ``Dipartimento di Eccellenza'' (CUP: 
B81I18001170001).
The authors would like to acknowledge networking support by the GWverse COST Action
CA16104, ``Black holes, gravitational waves and fundamental physics.''
M.B., D.C., A.G. and J.F.M. would like to acknowledge partial support by Grant ID1202 ``Strong Interactions: from Lattice QCD to Strings, Branes and Holography'' within the `Beyond Borders 2019' scheme of the University of Roma ``Tor Vergata''.
\end{acknowledgments}

\appendix
\begin{widetext}
\section{Fuzzball microstate geometries} \label{app:sol}
In this appendix we provide details on the fuzzball microstate geometries considered in the main text.

\subsection{General multicenter fuzzball solutions}

 We are interested in solutions of the form~\eqref{4dsolution} admitting a regular horizonless five-dimensional uplift with metric
\begin{equation}
ds_5^2 = -(Z_1\, Z_2\, Z_3)^{-{2\over 3}} \left[dt+\mu(d\Psi+w_0) +w\right]^2 + (Z_1\, Z_2\, Z_3)^{1\over 3}\left[ V^{-1}  (d\Psi+w_0)^2+ V d\vec x^2 \right]  
\end{equation}
and $*_3 dw_0=dV$. It can be seen that the resulting metric reduces to $\mathbb{R}$ times a Gibbons-Hawking space near the centers if $\mu$ vanishes and $Z_I$ remain finite near the centers. The 
origin of the Gibbons-Hawking space can display orbifold singularities if $|v_a|\neq 1$ but the latter can be easily regularized by splitting the centers (see~\cite{Bena:2007kg} for a review). 
The finiteness of $Z_I$ is ensured by setting
\begin{equation}
\label{fuzz-rel}
\ell_{I,a}=-{1\over 2} \frac{|\epsilon_{IJK}|  k_a^J k_a^K }{v_a} \,,
\quad 
,
\quad 
m_a = \frac{k_a^1 k_a^2 k_a^3 }{v_a^2}
\end{equation}
while the vanishing of $\mu$ leads to the so called ``bubble equations"~\cite{Bena:2007kg}
\begin{equation}
\label{bubble-general}
\sum_{b = 1}^N \frac{\Gamma_{ab}}{r_{ab}} = \Lambda_a  \,,
\end{equation}
where $r_{ab}=|\vec{x}_a-\vec{x}_b|$, 
\begin{eqnarray}
\Gamma_{ab} =  v_a v_b \prod_{I=1}^3\left( \frac{k_a^I}{v_a} - \frac{k_b^I}{v_b} \right) 
\qquad , \qquad 
\Lambda_a  =   \sum_{I=1}^3 k_a^I - \frac{ k_a^1 k_a^2 k_a^3}{v_a^2}    \,.
\end{eqnarray}
Notice that Eq.~\eqref{bubble-general} gives $N-1$ equations for the distances $r_{ab}$ and one equation for the coefficients $k_a^I$, namely
\begin{equation}
\sum_{a = 1}^N \Lambda_a = 0\,.
\end{equation}
In order to remove singularities and closed time-like curves one should impose the following conditions 
\begin{equation}
e^{2U} >0\,,
\quad 
Z_I V > 0\,.   \label{regularitycond}
\end{equation}
Note that the first condition prevents the existence of an ergoregion and therefore these geometries are free of the ergoregion instability that exists for horizonless ultracompact objects~\cite{1978CMaPh..63..243F,Cardoso:2007az,Chirenti:2008pf,Pani:2010jz,Cardoso:2008kj,Maggio:2017ivp, 
Maggio:2018ivz} as well as for non-supersymmetric fuzzballs of the so-called JMaRT family~\cite{JMaRT, Cardoso:2005gj, Bianchi:2019lmi}.

The explicit expression for the one-form $\omega$ can be derived from Eq.~\eqref{def:expmin4U} by making use of the bubble equations (\ref{bubble-general}) and relations (\ref{fuzz-rel}). One finds
\begin{equation}
\omega = \frac{1}{4}\sum_{a,b}^{N}\Gamma_{ab} \,\omega_{a b}\,,
\end{equation}
with $\omega_{a b}= \vec{\omega}_{ab}{\cdot}d\vec{x}$ a solution of 
\begin{eqnarray}
\vec{\nabla}\times\vec{\omega}_{ab}&=&\frac{1}{|\vec{x}-\vec{x}_{a}|}\vec{\nabla}\frac{1}{|\vec{x}-\vec{x}_{b}|}-\frac{1}{|\vec{x}-\vec{x}_{b}|}\vec{\nabla}\frac{1}{|\vec{x}-\vec{x}_{a}|}\,.
 \end{eqnarray}
The solution can be written as
\begin{equation}
\omega_{a b} = \frac{(\vec{n}_a+\vec{n}_{a b}){\cdot}(\vec{n}_b-\vec{n}_{a b})}{r_{a b}}\,d\phi_{a b}\,,
\end{equation}
with 
\begin{equation}
\vec{n}_a = \frac{\vec{x}-\vec{x}_a}{|\vec{x}-\vec{x}_{a}|}\,,
\quad
\vec{n}_{a b} = \frac{\vec{x}_a-\vec{x}_b}{r_{a b}}\,,
\quad
d\phi_{a b} = \frac{(\vec{n}_a\times \vec{n}_{a b})\cdot d\vec{x}}{|\vec{x}-\vec{x}_{a}| \left[1-(\vec{n}_a\cdot \vec{n}_{a b})^2\right]}\,.
\label{eq:define na nab, and dphiab}
\end{equation}
The solution carry in general four electric, $Q_A=(Q_0,Q_I)$, and four magnetic, $P_A=(P^0,P^I)$, charges respectively given by
\begin{eqnarray}
P^0 &=& \sum_a v_a \quad ,\quad Q_I=\sum_a \ell_{I,a} = -{1\over 2} \sum_a \frac{|\epsilon_{IJK}|  k_a^J k_a^K }{v_a} \,,\nn\\
 P^I &=& \sum_a k^I_a \quad , \quad Q_0 = \sum_a m_a = \sum_a \frac{k_a^1 k_a^2 k_a^3 }{v_a^2}\,.
\end{eqnarray}

\subsection{Axisymmetric three-center solutions with $v_a \neq 1$}\label{app:ss_axi}
 
 We consider first a family of axisymmetric solutions with equatorial symmetry $z \to -z$ ($\theta \to \pi-\theta)$ . 
 We align the three centers along the $z$-axis
 \begin{equation}
\begin{aligned}
\vec{x}_a &= (0,0,z_a) \quad, \quad  z_1=L \qquad, \qquad    z_2=0\qquad , \qquad z_3=-L \,,
\end{aligned}
\end{equation}
and take the charges to be
\begin{equation}
\begin{aligned}
v_1 & =v_3=-1 \quad, \quad v_2= 3 \quad, \quad 
k^I{}_a &= \left(
\begin{array}{ccc}
 -\kappa_1 		& ~~0~~	& \kappa_1 \\
 -\kappa_2 		& ~~0~~	& \kappa_2 \\
 -\kappa_3 		& ~~0~~	& \kappa_3 \\  
\end{array}
\right)
\end{aligned}
\end{equation}
This is therefore a $3$-parameter family of solutions spanned by the integers $\kappa_1$, $\kappa_2$, and $\kappa_3$.
For this case, $\Gamma_{ab}$ and $\Lambda_a$ take the simple form
\begin{equation}
\begin{aligned}
\Gamma_{ab} &= \kappa_1 \kappa_2 \kappa_3 \left(
\begin{array}{ccc}
 	0 & -{v_2\over v_1^2} 	&  -{8\over v_1} \\
 {v_2\over v_1^2}  		& ~~0~~	& -{v_2\over v_1^2}  \\
{8\over v_1}		& {v_2\over v_1^2} 	& 0 \\  
\end{array}
\right) \qquad, \qquad 
 \Lambda_{a} = \kappa_1 \kappa_2 \kappa_3 \left(
\begin{array}{c}
 	 -{{q}} +{1  \over v_1^2} 	  \\
0 \\
	 {{q}} -{1 \over v_1^2} 	  \\  
\end{array}
\right)\,,
\end{aligned}
\end{equation}
with ${{q}}={\kappa_1+\kappa_2+\kappa_3\over \kappa_1 \kappa_2 \kappa_3}$.
The bubble equations are solved for
\be
L = {4\,v_1+v_2 \over {{q}} \,v_1^2-1}\,,
\ee
whereas the one-form $\omega =\omega_\phi d\phi$  takes the form
 \begin{equation}
\omega_\phi = \sum_{b>a}^N\frac{\Gamma_{ab}}{2r_a r_b\left(z_a-z_b\right) }\left[\left(r_a+z_a-r \cos\theta\right) \left(r_b-z_b+r \cos\theta\right)-r^2 \sin ^2\theta\right]
\end{equation}
with $N=3$ and 
\begin{equation}
r_a =| \vec{x}-\vec{x}_a|\,,
\end{equation}
so that
 \be
  r_1^2= r^2+L^2-2\, r\, L\, \cos\theta \quad, \quad r^2_2 = r^2 \quad, \quad r_3^2= r^2+L^2+2 \,r\, L\,\cos\theta \,.
\end{equation}


For completeness, we list the harmonic functions and other relevant quantities. The eight harmonic functions are
\begin{equation}
\begin{aligned}
&V= 1+v_1\left[\frac{1}{r_1}+\frac{1}{r_3}\right]+\frac{v_2}{r_2} \quad, \quad
W= \frac{ \kappa_1 \kappa_2 \kappa_3}{v_1^2 }\left[-\frac{1}{r_1}+ \frac{1}{r_3}\right] \\
&L_1= 1{-} \frac{\kappa_2 \kappa_3}{v_1} \left[\frac{1}{r_1}{+}\frac{1}{r_3}\right] \quad, \quad
L_2= 1{-} \frac{ \kappa_1 \kappa_3}{v_1}\left[\frac{1}{r_1}+\frac{1}{r_3}\right]  \quad, \quad
L_3= 1{-} \frac{\kappa_1 \kappa_2}{v_1} \left[\frac{1}{r_1}{+}\frac{1}{r_3}\right] \\
&K_1= {-}\kappa_1 \left[\frac{1}{r_1}   {-}\frac{1}{r_3} \right] \quad, \quad
K_2= \kappa_2 \left[{-}\frac{1}{r_1}{+}\frac{1}{r_3}\right] \quad, \quad
K_3= \kappa_3 \left[ - \frac{1}{r_1}  {+}\frac{1}{r_3} \right]\,.
\end{aligned}
\end{equation}
The nonvanishing charges are (remind $v_{1}=v_{3}=-1,v_{2}=3$)
\begin{equation}\label{charge1}
P^0=2v_1+v_2 \quad, \quad Q_1 = -\frac{2 \kappa_2 \kappa_3}{v_1} \quad, \quad
Q_2 = -\frac{2 \kappa_1\kappa_3   }{v_1   }  \quad, \quad
Q_3 = -\frac{ 2  \kappa_1 \kappa_2}{v_1} 
\end{equation}
To ensure that the charges are positive, we assume
\begin{equation}
  \kappa_1,\kappa_2,\kappa_3 \geq 1 \quad, \quad v_1<0 \quad, \quad  v_2+2v_1>0\,.
\end{equation}
 The mass $M$ and entropy $S$ of the corresponding BH read
\begin{equation}
\begin{aligned}\label{ms1}
M &= \frac{P^0{+}Q_1{+}Q_2{+}Q_3}{4} =\frac{ v_1(v_2+2v_1) - 2 (\kappa_1 \kappa_2 +\kappa_2 \kappa_3+  \kappa_1 \kappa_3) }{4 v_1} \,,\\
S^2 &\propto \lim_{r \to 0} r^4 e^{-4U} =-(v_2+2v_1) \frac{8 \kappa_1^2 \kappa_2^2 \kappa_3^2 }{v_1^3}  \,,
\end{aligned}
\end{equation}
and are positive by construction. 
The total angular momentum of the solution reads $\vec{J}= J \vec{e}_z$, 
with
\be
J = - \kappa_1 \kappa_2 \kappa_3 {4 v_1 +v_2 \over 2 v_1^2}.
\ee
Notice that under the following scaling
\begin{equation}
v_a \to \gamma v_a \quad, \quad \kappa_i \to \gamma \kappa_i\,,
\end{equation}
all the charges scale as
\begin{equation}
Q_I \to \gamma Q_I \quad, \quad
P^0 \to \gamma P^0 \quad, \quad
M \to \gamma M \quad, \quad
J   \to \gamma^2 J \quad, \quad
S   \to \gamma^2 S\,.
\end{equation}
This is an overall scaling, therefore dimensionless quantities such as $M \omega_{\rm QNM}$ are not affected by it. This implies that any given solution can be scaled to a solution with arbitrarily 
large charges and mass.

Finally, the $4$-charge BH associated to the fuzzball solution is obtained by bringing the three centers to the origin. In this case harmonic functions become
\begin{equation}
\begin{aligned}
&V= 1+\frac{v_2+2v_1}{r} \quad, \quad
 L_1= 1{+} \frac{ 2 \kappa_2 \kappa_3}{ v_1 r}  \quad, \quad
L_2=1{+} \frac{ 2 \kappa_1 \kappa_3}{ v_1 r} \quad, \quad
L_3= 1{+} \frac{ 2 \kappa_1 \kappa_2}{v_1 r} \\
&K^I=  {{W}}=0\,,
\end{aligned}
\end{equation}
leading to
\begin{equation}
\begin{aligned}
e^{-4U} &= L_1L_2L_3 V \qquad ,\qquad \omega = 0
 \end{aligned}
\end{equation}
The mass, charges, and entropy are given by Eqs.~\eqref{charge1}-\eqref{ms1}, whereas the BH is nonspinning, $J=0$.

\subsection{Three-center solutions with $v_a=1$} 
 
 The second family of solutions we consider are those with three centers,  $v_a=1$ and four charges $Q_0=P^I=0$.   
  The general solution to the bubble equation  is described by four integers $\kappa_a$ specifying the $k_a^I$-matrix to be of the form
 \be
 k^I{}_a = \left(
\begin{array}{ccc}
 -\kappa_1 \kappa_2 &  -\kappa_1 \kappa_3  &  \kappa_1(\kappa_2+\kappa_3) \\
  \kappa_3 &    \kappa_2  &  -\kappa_2-\kappa_3 \\
   -\kappa_4  &  \kappa_4  & 0 \\  
\end{array}
\right).  
 \ee
 and therefore
 \be
\begin{aligned}
V &= 1+\sum_{a=1}^3 {1\over r_a}\,,   \quad   {{W}}=  \kappa_1 \,\kappa_2 \,\kappa_3\, \kappa_4  \left( { 1\over 
r_1}-{1\over r_2}\right)
\\
 L_1 &=1+  \kappa_4 \left( {\kappa_3 \over r_1}-{\kappa_2\over r_2} \right)\,,  \quad  L_2 =1+  \kappa_1 \kappa_4 
\left( 
 -{\kappa_2\over r_1}+{\kappa_3\over r_2} \right)
\\
 L_3 &=1+  \kappa_1 \left( {\kappa_2\kappa_3 \over r_1}+{\kappa_2 \kappa_3\over r_2} +{(\kappa_2+\kappa_3)^2 \over r_3} 
\right)\,, \quad
K^1 =  \kappa_1 \left( -{\kappa_2 \over r_1}-{\kappa_3\over r_2}+{\kappa_2+\kappa_3\over r_3} \right)\\
  K^2 &=  {\kappa_3 \over r_1}+{\kappa_2 \over r_2}-{\kappa_2+\kappa_3 \over r_3}\,,  \quad 
  K^3 =  \kappa_4 \left( -{1\over r_1}+{1\over r_2} \right)
\end{aligned}
 \ee
 again with $r_a =| \vec{x}-\vec{x}_a|$
and $\kappa_i$ some arbitrary integers. The solution describes a microstate of a Reissner-Nordstr\"om BH with a magnetic charge $P^0$ and three electric charges 
$Q_I$ given by 
\be
P^0 {\,=\,} 3 \quad ,\quad 
Q_1 {\,=\,} \kappa_4(\kappa_3{\,-\,}\kappa_2)\quad ,\quad
Q_2 {\,=\,} \kappa_1\kappa_4(\kappa_3 {\,-\,}\kappa_2) \quad ,\quad  
Q_3 {\,=\,} \kappa_1(\kappa_2^2{\,+\,}4\kappa_2\kappa_3{\,+\,}\kappa_3^2) \,. \\
\ee
The bubble equations  constrain the distances $r_{ab}=| \vec{x}_a-\vec{x}_b|$ between the 
centers to be related by
 \bea
r_{12}&=& 
\frac{2\kappa_1\kappa_4(\kappa_2-\kappa_3)^2r_{23}}{
\kappa_1\kappa_4(2\kappa_2^2+5\kappa_2\kappa_3+2\kappa_3^2)+(\kappa_2+\kappa_4-\kappa_1\kappa_3+\kappa_1\kappa_2\kappa_3
\kappa_4)r_{23}}\nn\\
r_{13}&=& 
\frac{\kappa_1\kappa_4(2\kappa_2+\kappa_3)(\kappa_2+2\kappa_3)r_{23}}{
\kappa_1\kappa_4(2\kappa_2^2+5\kappa_2\kappa_3+2\kappa_3^2)-(\kappa_1-1)(\kappa_2+\kappa_3)r_{23}}.
\label{bubble2!!!}
\eea

\subsubsection{Scaling solution}\label{app:scaling solution}

 The simplest solution in this class is the so-called scaling solution, corresponding to the choice
 \be
 \kappa_1=1 \qquad , \qquad \kappa_2= 0 \qquad ,\qquad \kappa_3=\kappa_4=\kappa\,.
\ee
For this choice one finds
\begin{align}
P^0=&\,3
\quad, \quad 
Q_I=(\kappa^2,\kappa^2,\kappa^2)
\quad, \quad 
M = \frac{3(1{\,+\,}\kappa^2)}{4}
\quad, \quad 
J=0  \,.
\end{align}
In particular, note that the solution is non-spinning. The centers are located at the vertices of an equilateral triangle. 
\begin{eqnarray}
(x_{1},y_{1},z_{1})&=&\left(\frac{L}{2\sqrt{3}},\frac{L}{2},0\right)\,,\quad \quad 
(x_{2},y_{2},z_{2})=\left(\frac{L}{2\sqrt{3}},-\frac{L}{2},0\right)\quad, \quad 
(x_{3},y_{3},z_{3})=\left(-\frac{L}{\sqrt{3}},0,0\right)\,.
\end{eqnarray}

\subsubsection{Another axisymmetric solution}

Although not considered in the main text, for completeness we provide here another axisymmetric solution.
In this case axial symmetry is found for the choice $\kappa_2=0$ and $\kappa_4=\kappa_1 \kappa_3$, i.e.
    \begin{equation}
 k^I{}_a = \kappa_3\left(
\begin{array}{ccc}
 0 &  -\kappa_1  &  \kappa_1 \\
  1 &    0  &  -1 \\
   -\kappa_1  &  \kappa_1  & 0 \\  
\end{array}
\right) \,.
 \end{equation}
The centers are located at
\be
\vec{x}_a = (0,0,z_a) \quad, \quad  z_1=L \qquad, \qquad    z_2=0\qquad , \qquad z_3=-L\,,
\ee
with
\begin{equation}
L = \frac{\kappa_1^2\kappa_3^2}{\kappa_1-1}\,.
\end{equation}
The mass, charges, angular momentum and entropy of the solution read
\begin{equation}
\begin{aligned}
P^0  &= 3 \,,
\quad 
Q_1 {\,=\,} \kappa_1\kappa_3^2\, ,
\quad
Q_2 {\,=\,} \kappa_1^2\kappa_3^2\, ,
\quad  
Q_3 {\,=\,} \kappa_1\kappa_3^2 \,,\\
M &= \frac{3+\kappa_1\kappa_3^2(2+\kappa_3)}{4}\,,
\quad
\vec J = -\frac{1}{2}\kappa_1^2\kappa_3^3\,\vec{e}_z \qquad , \qquad S = 3\kappa_1^4\kappa_3^6\,.
\end{aligned}
\end{equation}

\section{Numerical implementation for 4-charge BH \label{sec:Numerical code for 4-charge BH}}\label{app:BH}

\subsection{Time-domain analysis}\label{app:TD}

To evolve the scalar field in a spherically symmetric BH, it is convenient to define an effective potential for different angular modes. The isotropic coordinates in Eq.~\eqref{metricRN} can 
be transformed into areal radius coordinates as:
\begin{align}
ds^{2}&=-f(r)dt^{2}
+g(r)\left(
\frac{2{{\varrho}}}{r^{2}g'(r)+2rg(r)}
\right)^{2}d{{\varrho}}^{2}
+{{\varrho}}^{2}d^{2}\Omega\,,
\end{align}
where ${{\varrho}}^{2}=g(r)r^{2}$.
In the general case of four independent charges, the analytical expression for ${{\varrho}}(r)$ is cumbersome.
If we assume $Q_{1}=Q_{3}$ and $Q_{2}=Q_{4}$ for simplicity, then the horizon areal radius is ${{\varrho}}_{\rm H}=\sqrt{Q_{1}Q_{2}}$, and the metric reduces to
\begin{eqnarray}
ds^{2}=-{{B}}({{\varrho}})^{2}dt^{2}+{{A}}({{\varrho}})^{2}d{{\varrho}}^{2}+{{\varrho}}^{2}d^{2}\Omega\,,
\end{eqnarray}
where
\begin{eqnarray}
{{B}}({{\varrho}})&=&\frac{\sqrt{(Q_{1}-Q_{2})^{2}+4{{\varrho}}^{2}}-(Q_{1}+Q_{2})}{2{{\varrho}}}\,,\nonumber\\
{{A}}({{\varrho}})&=&\frac{4{{\varrho}}^{2}}{(Q_{1}-Q_{2})^{2}+4{{\varrho}}^{2}-(Q_{1}+Q_{2})\sqrt{(Q_{1}-Q_{2})^{2}+4{{\varrho}}^{2}}}\,.\nonumber\\
\end{eqnarray}
Since the space-time is spherically symmetric, 
we can separate the angular dependence of the field as
\begin{eqnarray}
\Phi&=&\sum_{l,m}\frac{\sigma_{lm}(t,{{\varrho}})}{{{\varrho}}}Y_{lm}(\theta,\phi)\,,
\end{eqnarray}
where $Y_{lm}(\theta,\phi)$ are the spherical harmonics.
The tortoise coordinate ${{\varrho}}_{\ast}$ is defined as $\frac{d{{\varrho}}_{\ast}}{d{{\varrho}}}=\frac{{{A}}}{{{B}}}$, which yields
\begin{align}
{{\varrho}}_{\ast}=&\frac{1}{2}\sqrt{(Q_{1}-Q_{2})^{2}+4{{\varrho}}^{2}}+\frac{2Q_{1}Q_{2}}{Q_{1}+Q_{2}-\sqrt{(Q_{1}-Q_{2})^{2}+4{{\varrho}}^{2}}}&+(Q_{1}+Q_{2})\ln\left(
Q_{1}+Q_{2}-\sqrt{(Q_{1}-Q_{2})^{2}+4{{\varrho}}^{2}}
\right)\,.
\end{align}
Finally, the evolution equation for $\sigma_{lm}(t,R)$ in tortoise coordinates is governed by
\begin{eqnarray}
-\frac{\partial^{2}}{\partial t^{2}}\sigma_{lm}+\frac{\partial^{2}}{\partial {{\varrho}}_{\ast}^{2}}\sigma_{lm}-V_{\rm eff}({{\varrho}})\sigma_{lm}=0
\label{eq:scalar field in tortoise coordinate}
\end{eqnarray}
where the effective potential reads
\begin{align}
V_{\rm eff}({{\varrho}}_{\ast})&={{B}}^{2}\left(\frac{l(l+1)}{{{\varrho}}^{2}}+ \frac{1}{{{\varrho}}{{A}}^{2}}\left(
\frac{\partial_{{{\varrho}}}{{B}}}{{{B}}}-\frac{\partial_{{{\varrho}}}{{A}}}{{{A}}}
\right)
\right)\,.
\end{align}
We integrate Eq.~\eqref{eq:scalar field in tortoise coordinate} numerically with an ingoing boundary condition at ${{\varrho}}_{\ast}\to -\infty$ (i.e., 
$\partial_t\sigma_{lm}\sim\partial_{{{\varrho}}_\ast}\sigma_{lm}$), and outgoing boundary condition at ${{\varrho}}_{\ast}\to\infty$ (i.e., $\partial_t\sigma_{lm}\sim-\partial_{{{\varrho}}_\ast}\sigma_{lm}$).
\par
For the initial data for simulation we choose an instantaneously static Gaussian initial data,
\begin{eqnarray}
\sigma_{lm}(0,{{\varrho}}_{\ast})&=&{{a}}e^{-\left(
\frac{{{\varrho}}_{\ast}-{{\varrho}}_{0}}{\sigma}
\right)^{2}}\,,\\
\dot{\sigma}_{lm}(0,{{\varrho}}_{\ast})&=&0\,,
\end{eqnarray}
where ${{a}}$, ${{\varrho}}_{0}$, and $\sigma$ are the initial amplitude, position, and width of the Gaussian pulse, respectively. Note that our initial data are given for each $(l,m)$ mode separately, since the 
modes are decoupled in the spherically symmetric background.
We used a C++ numerical code to integrate Eq.~\eqref{eq:scalar field in tortoise coordinate} using a 4th-order Runge-Kutta method, where spatial derivatives are evaluated by 4th-order 
finite difference.

\subsection{Frequency-domain analysis}\label{app:FD}
In the spherically-symmetric case it is straightforward to integrate the Klein-Gordon equation as an eigenvalue problem in order to compute directly the QNMs of the $4$-charge BH. 

In the particular case $Q_{1}=Q_{3}$ and $Q_{2}=Q_{4}$, Eq.~\eqref{eq:scalar field in tortoise coordinate} reduces to a single ordinary differential equation in a Schroedinger-like form by 
assuming $\sigma_{lm}=\Psi(\varrho)e^{-i\omega t}$. Boundary conditions are then imposed ($\Psi\to e^{\pm i\omega {\varrho_\ast}}$ as $\varrho_\ast\to\pm\infty$) in order to compute the complex eigenvalues $\omega$. 
In the general case, $r^4_H =  Q_{1} Q_{2} Q_{3} Q_{4}$, the same method can be implemented by integrating the Klein-Gordon equation on the metric~\eqref{metricRN} using directly isotropic coordinates. 
By decomposing the field as $\Phi=\sum_{lm}\frac{{{F}}(r)}{r}Y_{lm}e^{-i\omega t}$, the Klein-Gordon equation takes a particularly simple form,
\begin{equation}
 {{F}}''+\left(\omega^2\frac{(Q_1+r)(Q_2+r)(Q_3+r)(Q_4+r)}{r^4}-\frac{l(l+1)}{r^2}\right){{F}}=0\,.
\end{equation}
In both cases, we computed the fundamental eigenfrequencies $\omega$ using an extended version of the direct integration method~\cite{Chandrasekhar:1975zza,Pani:2013pma}. Following \cite{Aminov:2020yma}, the QNMs  can also be found `analytically' by means of the `quantum' Seiberg-Witten curve of ${\cal N}=2$ super Yang-Mills theory with $SU(2)$ gauge group and RG-invariant scale $\Lambda$, coupled to $N_f=2$ hypermultiplet doublets with masses $m_{1,2}$, after the identifications ${\Lambda} = +2i\hbar\omega r_H$, 
${m_1} = {\hbar\over 2} - 2i\hbar \omega \sum_A Q_A$, ${m_2} = - {\hbar\over 2} - 2i \hbar \omega  \sum_A {r_H^2 \over Q_A}$  and ${u} = \hbar^2\left[(l+{1\over 2})^2 + \omega^2 (2 r_H^2 - \sum_{A<B}   Q_AQ_B)\right]$ \cite{QNMvsSWwip}.
 
\section{Code tests and regularization procedure for fuzzball simulations} \label{app:code}
In this appendix we provide some technical details of our code.
First of all, since microstate geometries are cumbersome, it is useful to check the analytical solutions.
For this purpose, in the left panel of Fig.~\ref{Graph_converngece}, we check the numerical convergence of residual in the differential form of Eq.~\eqref{def:expmin4U} for different resolutions. The 
residual shows convergence with fourth-order accuracy, as expected since the spatial derivatives are evaluated by fourth-order-accurate finite difference stencils.
\begin{figure*}[tb]
\includegraphics[width=0.45\textwidth]{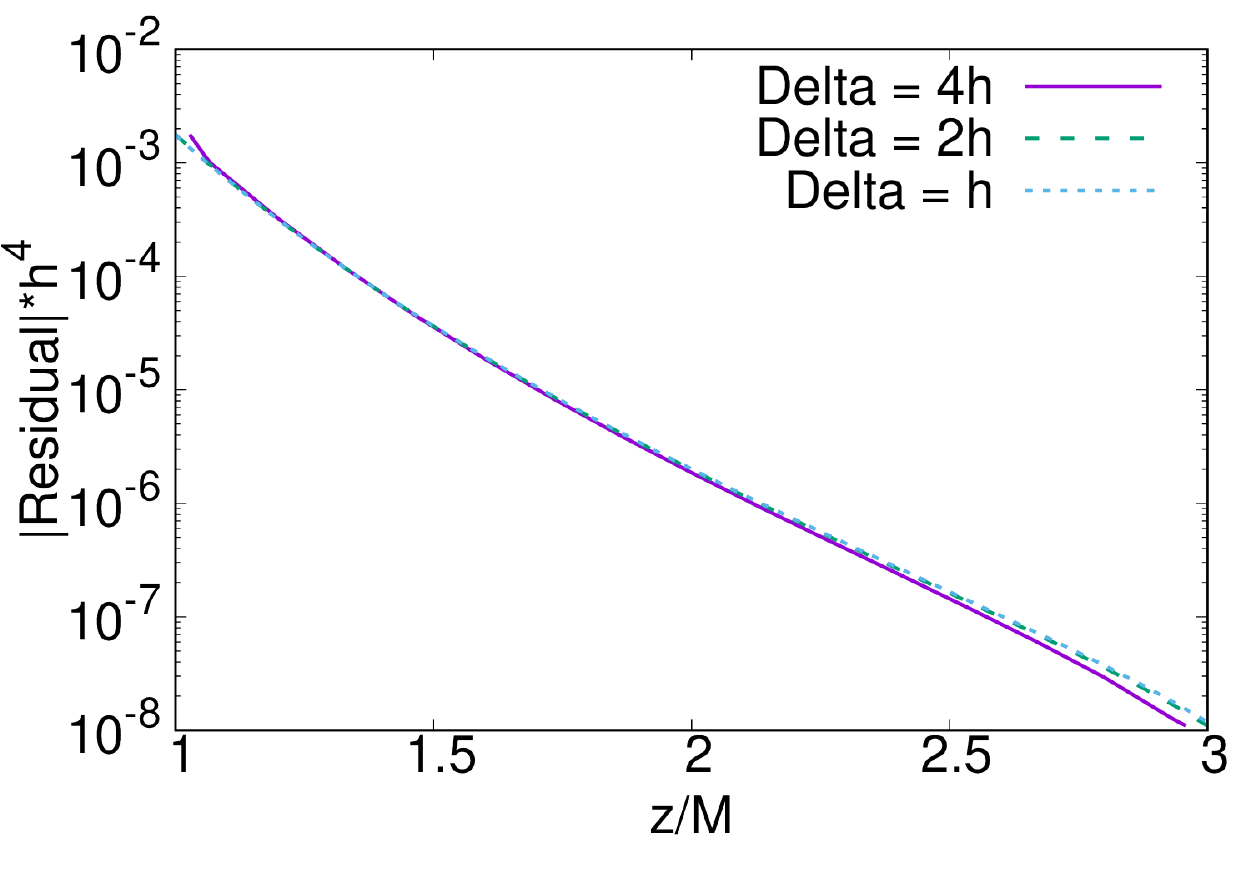}
\includegraphics[width=0.45\textwidth]{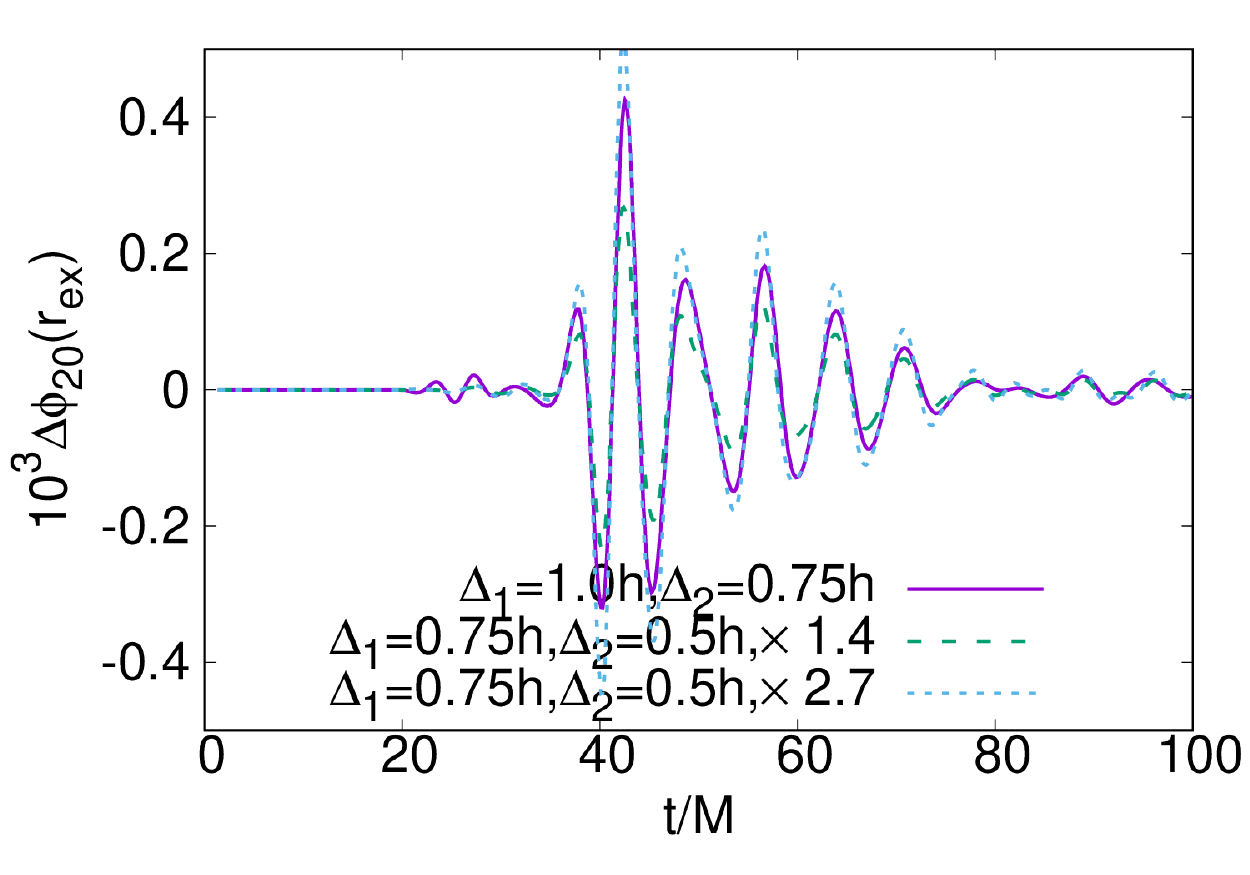}
\caption{Convergence of the residual of the differential form in Eq.~\eqref{def:expmin4U} on $z$-axis for different resolutions $h$ (left panel), and convergence analysis of the $l=2$, $m=0$ 
multipole mode of the scalar field at $r_{\rm ex}=20M/3$ 
(right panel).
\label{Graph_converngece}}
\end{figure*}

The right panel of Fig.~\ref{Graph_converngece} shows the evolution of the difference $\Delta\Phi_{20}$ between the field $\Phi_{20}$ at different resolutions $\Delta_{1}$ and 
$\Delta_{2}$ for a simulation starting from instantaneously static Gaussian initial data with $\sigma=0.33M$ around the microstate scaling solution with $\kappa=1$, and $L=0.67M$.
In our code, the spatial derivatives are approximated with fourth-order-accurate finite difference stencils, and the time integration is performed by a 4th-order Runge-Kutta method.
The boundary of the mesh refinement boundary is interpolated with second order and fifth order accuracy in time and space.
We define the convergence factor 
\begin{align}
Q_{n}=\frac{\Phi_{20}(\Delta_{c})-\Phi_{20}(\Delta_{m})}{\Phi_{20}(\Delta_{m})-\Phi_{20}(\Delta_{h})}=\frac{\Delta_{c}^{n}-\Delta_{m}^{n}}{\Delta_{m}^{n}-\Delta_{h}^{n}}\,,
\end{align}
where $\Delta_{c}=\Delta$, $\Delta_{m}=0.75\Delta$, $\Delta_{h}=0.5\Delta$, and $n$ is expected convergence factor.
As shown in the right panel of Fig.~\ref{Graph_converngece} the numerical evolutions converge between second- and third-order accuracy, as expected.

Finally, we discuss the regularization procedure to resolve the singularities of the $4$-dimensional fuzzball geometry near the centers. In Fig.~\ref{l0_m0_r30_plot_SP_FuzzBall_3enters_scaling_analutic_omega_wb_lambda1_L1_analytic_omega_GaussianScalar_sigma1_MP_fac_reg_1_diff_repsilon} we show an example of evolution for different values of the regularization parameter $\epsilon$. While the initial response is independent of $\epsilon$ --~since it is related to the properties of the photon sphere and not to the region near the centers~-- the behavior at later times depends on $\epsilon$ if the latter parameter is not sufficiently small. The two smallest values of $\epsilon$ give the same evolution for $l=0$, whereas small finite-$\epsilon$ differences arise at late times for $l=2$. Such differences become smaller as $\epsilon\to0$ and, anyway, do not affect the overall structure of the signal. While using even smaller values of $\epsilon$ is computationally very expensive, our simulations indicate that for sufficiently small $\epsilon$ the evolution is regular and smoothly convergent.

\begin{figure*}[tb]
\includegraphics[width=0.45\textwidth]{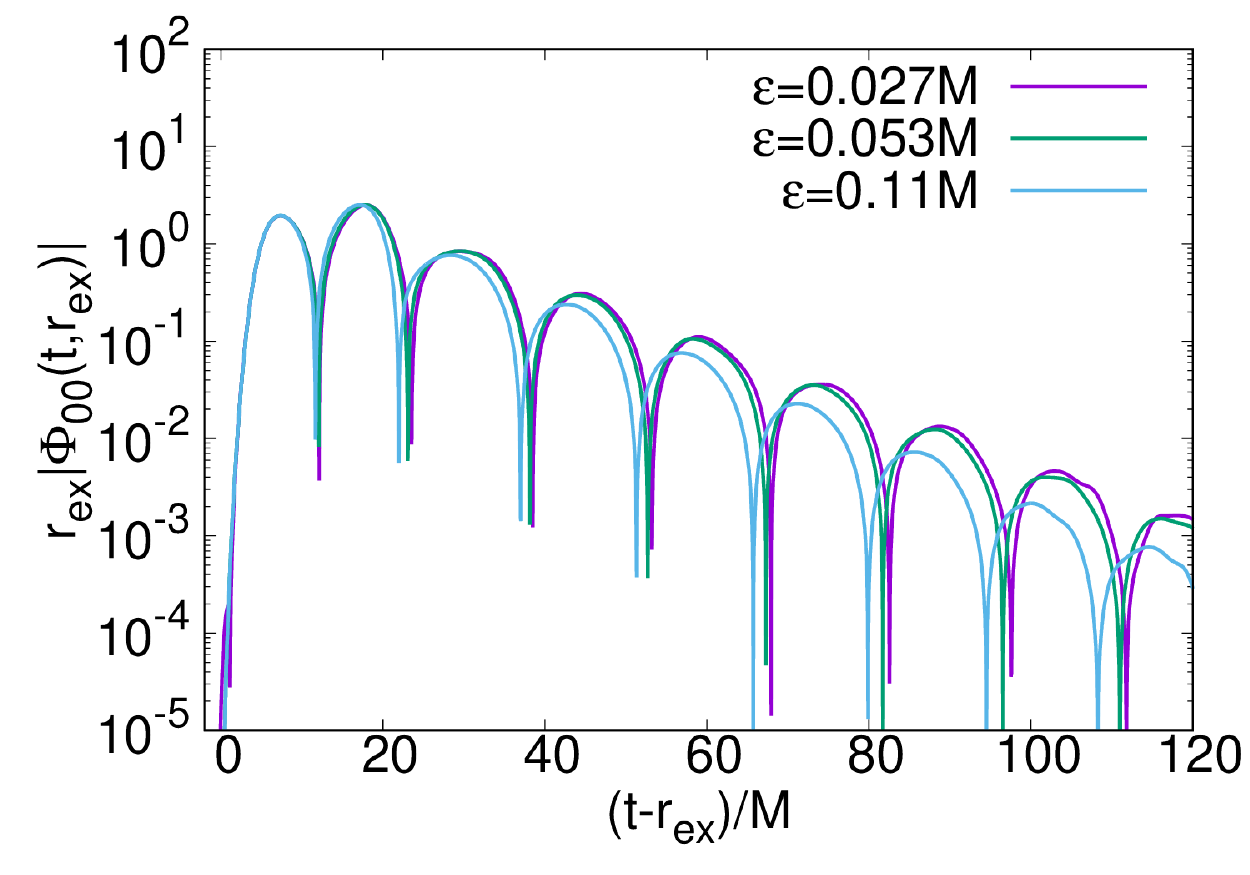}
\includegraphics[width=0.45\textwidth]{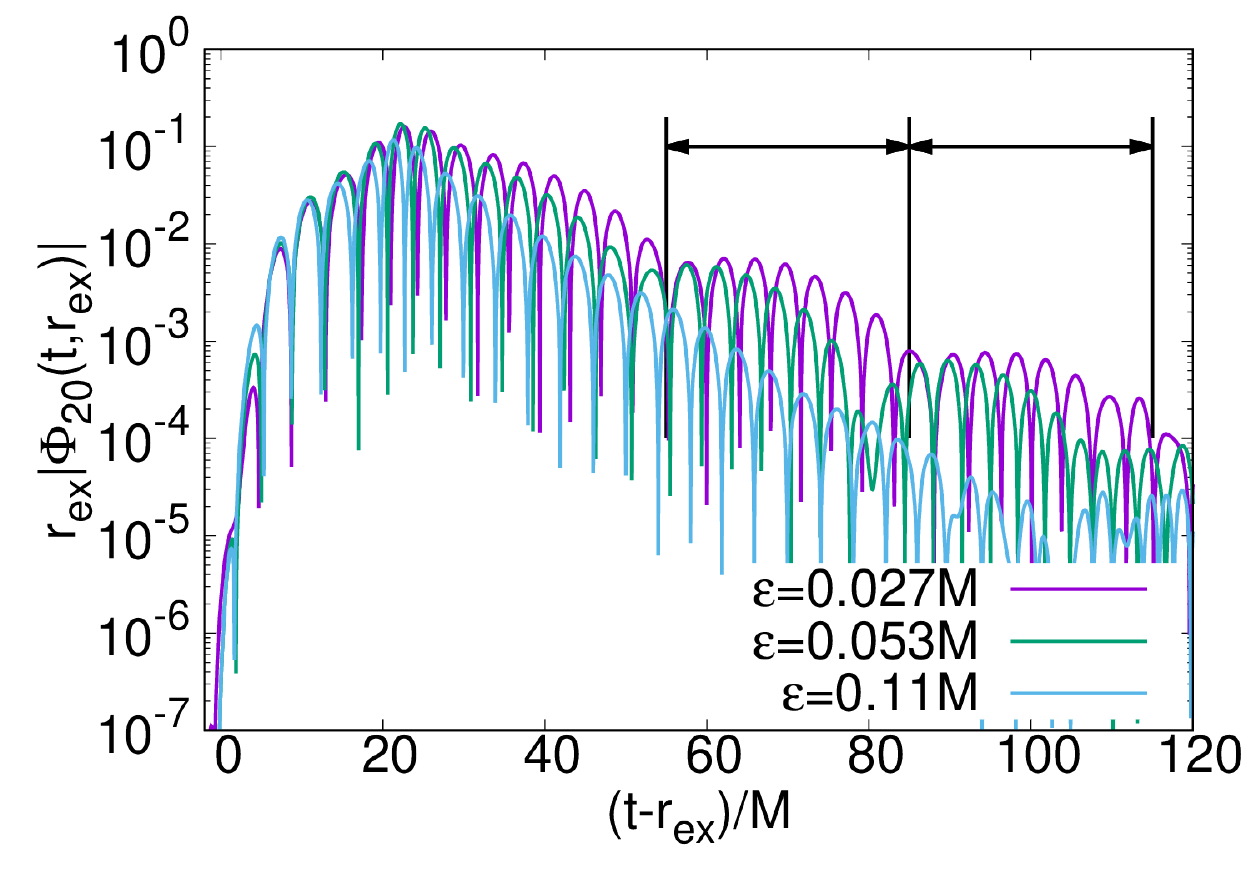}
\caption{
Time evolution of $l=0$ $m=0$ (left panel) and $l=2$, $m=0$ (right panel) multipole modes of the scalar field around scaling solution with $L=0.67M,\kappa=1$.
The evolution starts from instantaneously static spherically symmetric Gaussian profile Eq.\ref{eq:initial data momentarily static Gaussian} with $\sigma=0.33M$.
The extraction radius is $r_{\rm ex}=20.0M$.
\label{l0_m0_r30_plot_SP_FuzzBall_3enters_scaling_analutic_omega_wb_lambda1_L1_analytic_omega_GaussianScalar_sigma1_MP_fac_reg_1_diff_repsilon}}
\end{figure*}

\comment{Note also that the singularities at the centers are problematic only if they coincide with some of the numerical grid points, otherwise a natural cut-off is set by the grid numerical resolution. We have adopted this strategy to probe much smaller values of the cut-off, confirming the generic behavior discussed above, although for very small values of the cut-off reaching numerical convergence requires very high resolution.}

\section{Extra simulations}\label{app:extra}

For completeness, in this appendix we consider two further examples of evolution in the axisymmetric microstate geometry with $\kappa_1=\kappa_2=\kappa_3=\kappa$. In particular, in the left and right panels of Fig.~\ref{fig:n2n4} we consider the evolution of the $l=m=2$ initial data on a solution with $\kappa=2$ and $\kappa=4$, respectively. The former case is interesting because, according to our geodesic analysis, it does not possess a light ring on the equatorial plane. Correspondingly, we do not observe echoes in this case since radiation confined near the equator cannot be efficiently trapped. On the other hand, the standard prompt ringdown shown in the left panel could be identified with a nontrivial geodesic structure outside the equatorial plane for this geometry.
Finally, in the case $\kappa=4$ there is an equatorial light ring which can efficiently confine radiation. Furthermore, in this case the fuzzball's throat is deeper and the associated echo delay time is about $64M$ (see Table~\ref{tab:geod}). This is in qualitative agreement with the timescales shown in the right panel of Fig.~\ref{fig:n2n4}, confirming that the physical picture drawn from the geodesics approximation is reliable.

\begin{figure*}[tb]
\includegraphics[width=0.45\textwidth]{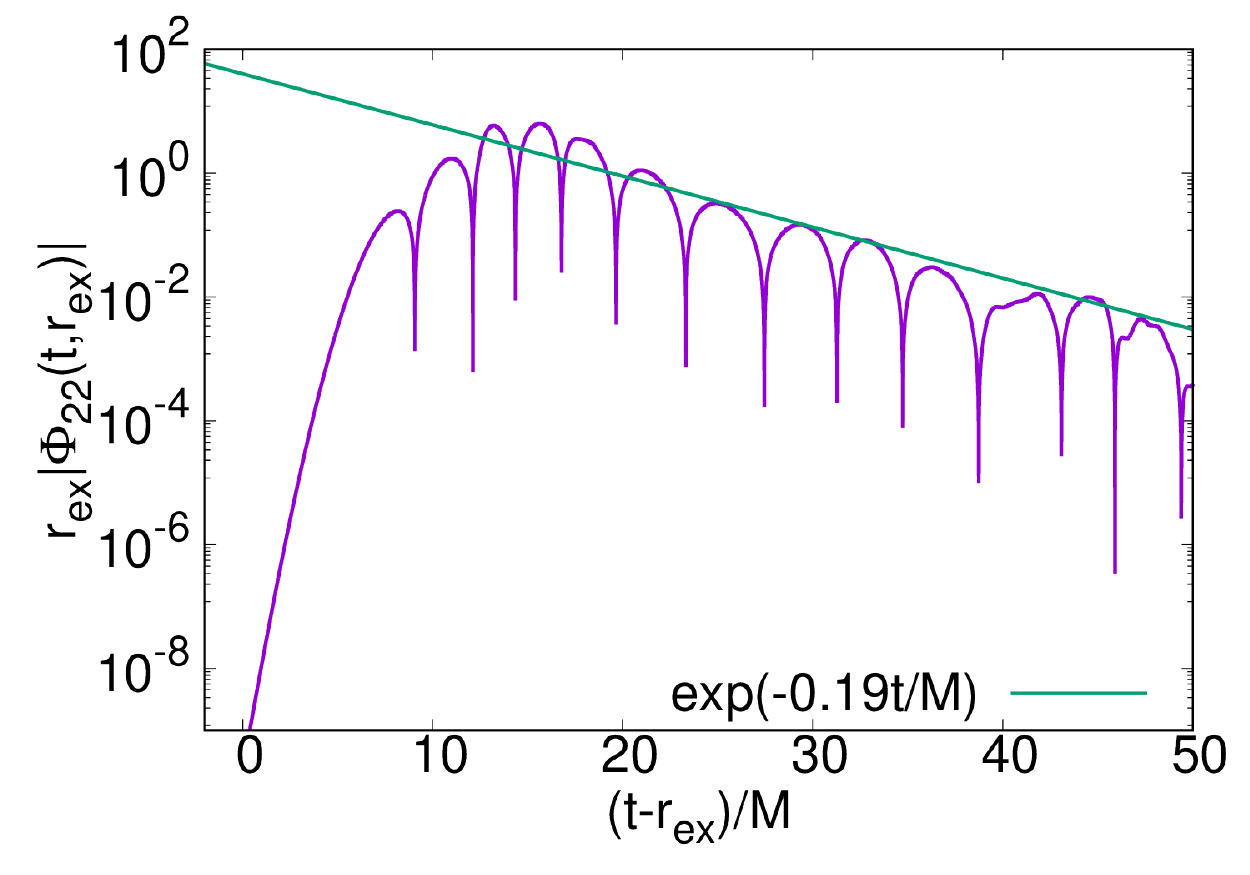}
\includegraphics[width=0.45\textwidth]{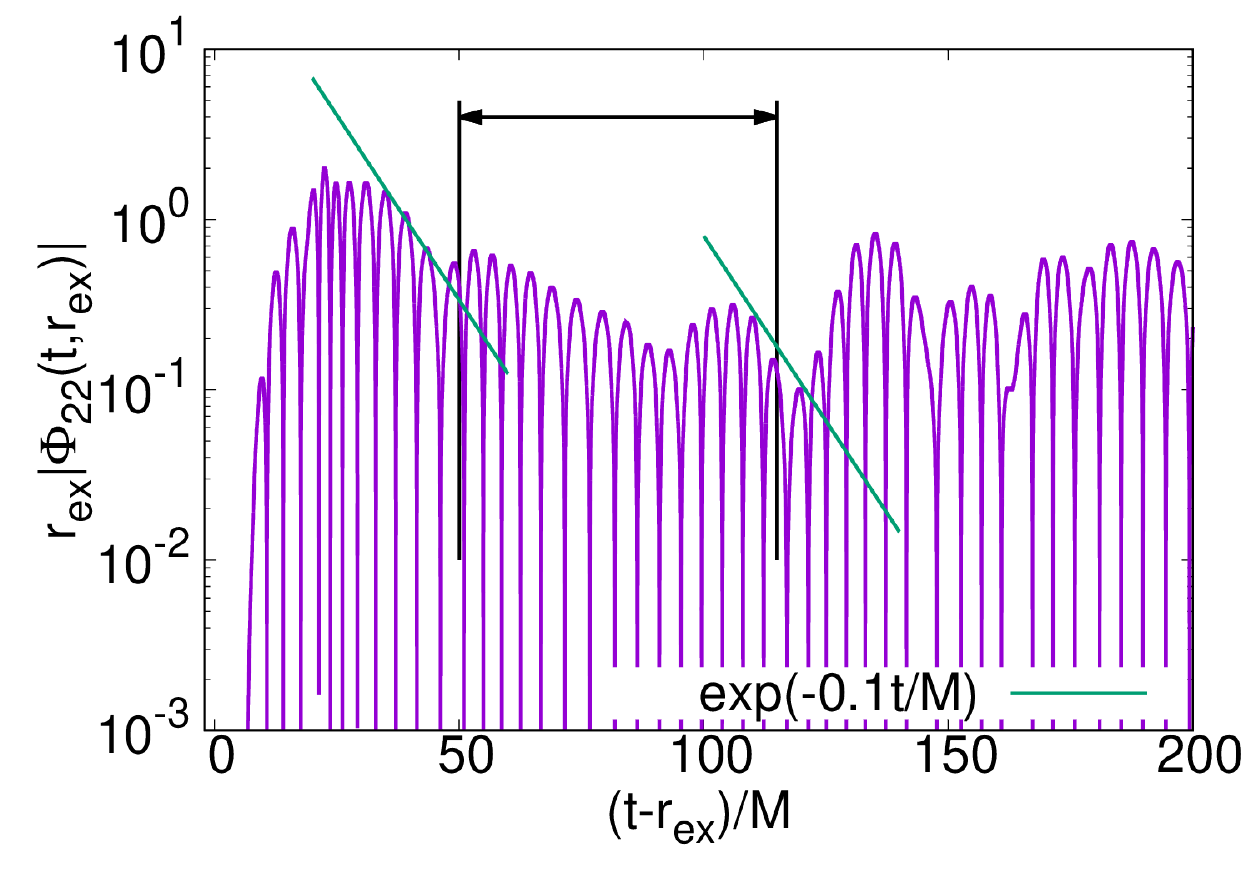}
\caption{Evolution of an instantaneously static $l=m=2$ initial Gaussian profile on the axisymmetric microstate geometry with $\kappa_1=\kappa_2=\kappa_3=\kappa$.
Left panel: $\kappa=2$; in this case the absence of an equatorial light ring implies the absence of echoes at late time. Right panel: $\kappa=4$; also in this case the echo delay time is consistent with the geodesics prediction in Table~\ref{tab:geod}.}
\label{fig:n2n4}
\end{figure*}

\end{widetext}

\bibliographystyle{utphys}
\bibliography{References}

\providecommand{\href}[2]{#2}\begingroup\raggedright\begin{thebibliography}{100}

\bibitem{Carter71}
B.~Carter, ``Axisymmetric black hole has only two degrees of freedom,''
  \href{http://dx.doi.org/10.1103/PhysRevLett.26.331}{{\em Phys. Rev. Lett.}
  {\bfseries 26} (Feb, 1971) 331--333}.
  \url{http://link.aps.org/doi/10.1103/PhysRevLett.26.331}.

\bibitem{Robinson:1975bv}
D.~Robinson, ``{Uniqueness of the Kerr black hole},''
\href{http://dx.doi.org/10.1103/PhysRevLett.34.905}{{\em Phys. Rev. Lett.}
  {\bfseries 34} (1975) 905--906}.

\bibitem{Heusler:1998ua}
M.~Heusler, ``{Stationary black holes: Uniqueness and beyond},'' {\em Living
  Rev. Relativity} {\bfseries 1} no.~6, (1998) .
\url{http://www.livingreviews.org/lrr-1998-6}.

\bibitem{Chrusciel:2012jk}
P.~T. Chrusciel, J.~L. Costa, and M.~Heusler, ``{Stationary Black Holes:
  Uniqueness and Beyond},'' {\em Living Rev. Relativ.} {\bfseries 15} (2012) 7,
\href{http://arxiv.org/abs/1205.6112}{{\ttfamily arXiv:1205.6112 [gr-qc]}}.

\bibitem{Holzhey:1991bx}
C.~F.~E. Holzhey and F.~Wilczek, ``{Black holes as elementary particles},''
  \href{http://dx.doi.org/10.1016/0550-3213(92)90254-9}{{\em Nucl. Phys. B}
  {\bfseries 380} (1992) 447--477},
  \href{http://arxiv.org/abs/hep-th/9202014}{{\ttfamily arXiv:hep-th/9202014}}.

\bibitem{Hawking:1973uf}
S.~Hawking and G.~Ellis, \href{http://dx.doi.org/10.1017/CBO9780511524646}{{\em
  {The Large Scale Structure of Space-Time}}}.
\newblock Cambridge Monographs on Mathematical Physics. Cambridge University
  Press, 2, 2011.

\bibitem{Brill:1972xj}
D.~R. Brill, P.~L. Chrzanowski, C.~Martin~Pereira, E.~D. Fackerell, and J.~R.
  Ipser, ``{Solution of the scalar wave equation in a kerr background by
  separation of variables},''
  \href{http://dx.doi.org/10.1103/PhysRevD.5.1913}{{\em Phys. Rev. D}
  {\bfseries 5} (1972) 1913--1915}.

\bibitem{Teukolsky:1972my}
S.~A. Teukolsky, ``{Rotating black holes - separable wave equations for
  gravitational and electromagnetic perturbations},''
\href{http://dx.doi.org/10.1103/PhysRevLett.29.1114}{{\em Phys. Rev. Lett.}
  {\bfseries 29} (1972) 1114--1118}.

\bibitem{TheLIGOScientific:2016pea}
{\bfseries LIGO Scientific, Virgo} Collaboration, B.~P. Abbott {\em et~al.},
  ``{Binary Black Hole Mergers in the first Advanced LIGO Observing Run},''
  \href{http://dx.doi.org/10.1103/PhysRevX.6.041015}{{\em Phys. Rev. X}
  {\bfseries 6} no.~4, (2016) 041015},
  \href{http://arxiv.org/abs/1606.04856}{{\ttfamily arXiv:1606.04856 [gr-qc]}}.
  [Erratum: Phys.Rev.X 8, 039903 (2018)].

\bibitem{Vishveshwara:1970cc}
C.~V. Vishveshwara, ``{Stability of the schwarzschild metric},''
\href{http://dx.doi.org/10.1103/PhysRevD.1.2870}{{\em Phys. Rev.} {\bfseries
  D1} (1970) 2870--2879}.

\bibitem{ChandraBook}
S.~Chandrasekhar, {\em {The mathematical theory of black holes}}.
\newblock 1985.

\bibitem{1980ApJ...239..292D}
S.~{Detweiler}, ``{Black holes and gravitational waves. III - The resonant
  frequencies of rotating holes},''
  \href{http://dx.doi.org/10.1086/158109}{{\em \apj} {\bfseries 239} (July,
  1980) 292--295}.

\bibitem{Dreyer:2003bv}
O.~Dreyer, B.~J. Kelly, B.~Krishnan, L.~S. Finn, D.~Garrison, and
  R.~Lopez-Aleman, ``{Black hole spectroscopy: Testing general relativity
  through gravitational wave observations},''
  \href{http://dx.doi.org/10.1088/0264-9381/21/4/003}{{\em Class. Quant. Grav.}
  {\bfseries 21} (2004) 787--804},
\href{http://arxiv.org/abs/gr-qc/0309007}{{\ttfamily arXiv:gr-qc/0309007
  [gr-qc]}}.

\bibitem{Berti:2005ys}
E.~Berti, V.~Cardoso, and C.~M. Will, ``{On gravitational-wave spectroscopy of
  massive black holes with the space interferometer LISA},''
  \href{http://dx.doi.org/10.1103/PhysRevD.73.064030}{{\em Phys. Rev.}
  {\bfseries D73} (2006) 064030},
\href{http://arxiv.org/abs/gr-qc/0512160}{{\ttfamily arXiv:gr-qc/0512160
  [gr-qc]}}.

\bibitem{Kokkotas:1999bd}
K.~D. Kokkotas and B.~G. Schmidt, ``{Quasinormal modes of stars and black
  holes},'' \href{http://dx.doi.org/10.12942/lrr-1999-2}{{\em Living Rev. Rel.}
  {\bfseries 2} (1999) 2},
\href{http://arxiv.org/abs/gr-qc/9909058}{{\ttfamily arXiv:gr-qc/9909058
  [gr-qc]}}.

\bibitem{Berti:2009kk}
E.~Berti, V.~Cardoso, and A.~O. Starinets, ``{Quasinormal modes of black holes
  and black branes},''
  \href{http://dx.doi.org/10.1088/0264-9381/26/16/163001}{{\em Class. Quant.
  Grav.} {\bfseries 26} (2009) 163001},
\href{http://arxiv.org/abs/0905.2975}{{\ttfamily arXiv:0905.2975 [gr-qc]}}.

\bibitem{Isi:2019aib}
M.~Isi, M.~Giesler, W.~M. Farr, M.~A. Scheel, and S.~A. Teukolsky, ``{Testing
  the no-hair theorem with GW150914},''
  \href{http://dx.doi.org/10.1103/PhysRevLett.123.111102}{{\em Phys. Rev.
  Lett.} {\bfseries 123} no.~11, (2019) 111102},
  \href{http://arxiv.org/abs/1905.00869}{{\ttfamily arXiv:1905.00869 [gr-qc]}}.

\bibitem{Giesler:2019uxc}
M.~Giesler, M.~Isi, M.~A. Scheel, and S.~Teukolsky, ``{Black Hole Ringdown: The
  Importance of Overtones},''
  \href{http://dx.doi.org/10.1103/PhysRevX.9.041060}{{\em Phys. Rev. X}
  {\bfseries 9} no.~4, (2019) 041060},
  \href{http://arxiv.org/abs/1903.08284}{{\ttfamily arXiv:1903.08284 [gr-qc]}}.

\bibitem{TheLIGOScientific:2016src}
{\bfseries LIGO Scientific, Virgo} Collaboration, B.~P. Abbott {\em et~al.},
  ``{Tests of general relativity with GW150914},''
  \href{http://dx.doi.org/10.1103/PhysRevLett.116.221101,
  10.1103/PhysRevLett.121.129902}{{\em Phys. Rev. Lett.} {\bfseries 116}
  no.~22, (2016) 221101}, \href{http://arxiv.org/abs/1602.03841}{{\ttfamily
  arXiv:1602.03841 [gr-qc]}}.
[Erratum: Phys. Rev. Lett.121,no.12,129902(2018)].

\bibitem{LIGOScientific:2019fpa}
{\bfseries LIGO Scientific, Virgo} Collaboration, B.~Abbott {\em et~al.},
  ``{Tests of General Relativity with the Binary Black Hole Signals from the
  LIGO-Virgo Catalog GWTC-1},''
  \href{http://dx.doi.org/10.1103/PhysRevD.100.104036}{{\em Phys. Rev. D}
  {\bfseries 100} no.~10, (2019) 104036},
  \href{http://arxiv.org/abs/1903.04467}{{\ttfamily arXiv:1903.04467 [gr-qc]}}.

\bibitem{Berti:2016lat}
E.~Berti, A.~Sesana, E.~Barausse, V.~Cardoso, and K.~Belczynski,
  ``{Spectroscopy of Kerr black holes with Earth- and space-based
  interferometers},''
  \href{http://dx.doi.org/10.1103/PhysRevLett.117.101102}{{\em Phys. Rev.
  Lett.} {\bfseries 117} no.~10, (2016) 101102},
  \href{http://arxiv.org/abs/1605.09286}{{\ttfamily arXiv:1605.09286 [gr-qc]}}.

\bibitem{Berti:2015itd}
E.~Berti {\em et~al.}, ``{Testing General Relativity with Present and Future
  Astrophysical Observations},''
  \href{http://dx.doi.org/10.1088/0264-9381/32/24/243001}{{\em Class. Quant.
  Grav.} {\bfseries 32} (2015) 243001},
\href{http://arxiv.org/abs/1501.07274}{{\ttfamily arXiv:1501.07274 [gr-qc]}}.

\bibitem{Abbott:2020jks}
{\bfseries LIGO Scientific, Virgo} Collaboration, R.~Abbott {\em et~al.},
  ``{Tests of General Relativity with Binary Black Holes from the second
  LIGO-Virgo Gravitational-Wave Transient Catalog},''
  \href{http://arxiv.org/abs/2010.14529}{{\ttfamily arXiv:2010.14529 [gr-qc]}}.

\bibitem{LambShift}
W.~E. Lamb and R.~C. Retherford, ``Fine structure of the hydrogen atom by a
  microwave method,'' \href{http://dx.doi.org/10.1103/PhysRev.72.241}{{\em
  Phys. Rev.} {\bfseries 72} (Aug, 1947) 241--243}.
  \url{https://link.aps.org/doi/10.1103/PhysRev.72.241}.

\bibitem{Mathur:2009hf}
S.~D. Mathur, ``{The Information paradox: A Pedagogical introduction},''
  \href{http://dx.doi.org/10.1088/0264-9381/26/22/224001}{{\em Class. Quant.
  Grav.} {\bfseries 26} (2009) 224001},
  \href{http://arxiv.org/abs/0909.1038}{{\ttfamily arXiv:0909.1038 [hep-th]}}.

\bibitem{Penrose:1969pc}
R.~Penrose, ``{Gravitational collapse: The role of general relativity},'' {\em
  Riv. Nuovo Cim.} {\bfseries 1} (1969) 252--276.
[Gen. Rel. Grav.34,1141(2002)].

\bibitem{Wald:1997wa}
R.~M. Wald,
  \href{http://dx.doi.org/10.1007/978-94-017-0934-7_5}{``{Gravitational
  collapse and cosmic censorship},''} in {\em {Black Holes, Gravitational
  Radiation and the Universe: Essays in Honor of C.V. Vishveshwara}},
  pp.~69--85.
\newblock 1997.
\newblock
\href{http://arxiv.org/abs/gr-qc/9710068}{{\ttfamily arXiv:gr-qc/9710068
  [gr-qc]}}.
\newblock

\bibitem{Penrose_CCC}
R.~Penrose, ``{Singularities of Spacetime (in Theoretical Principles in
  Astrophysics and Relativity)},'' in {\em {Chicago University Press, Chicago,
  1978 217 P.}}
\newblock 1978.

\bibitem{Bekenstein}
J.~D. Bekenstein, ``{Black holes and entropy},'' {\em Physical Review D}
  {\bfseries 7} no.~8, (1973) 2333.

\bibitem{Hawking:1976de}
S.~W. Hawking, ``{Black Holes and Thermodynamics},''
\href{http://dx.doi.org/10.1103/PhysRevD.13.191}{{\em Phys. Rev.} {\bfseries
  D13} (1976) 191--197}.

\bibitem{Hawking:1974sw}
S.~Hawking, ``{Particle Creation by Black Holes},''
  \href{http://dx.doi.org/10.1007/BF02345020}{{\em Commun. Math. Phys.}
  {\bfseries 43} (1975) 199--220}. [Erratum: Commun.Math.Phys. 46, 206 (1976)].

\bibitem{Lunin:2001jy}
O.~Lunin and S.~D. Mathur, ``{AdS / CFT duality and the black hole information
  paradox},'' \href{http://dx.doi.org/10.1016/S0550-3213(01)00620-4}{{\em Nucl.
  Phys. B} {\bfseries 623} (2002) 342--394},
  \href{http://arxiv.org/abs/hep-th/0109154}{{\ttfamily arXiv:hep-th/0109154}}.

\bibitem{Lunin:2002qf}
O.~Lunin and S.~D. Mathur, ``{Statistical interpretation of Bekenstein entropy
  for systems with a stretched horizon},''
  \href{http://dx.doi.org/10.1103/PhysRevLett.88.211303}{{\em Phys. Rev. Lett.}
  {\bfseries 88} (2002) 211303},
  \href{http://arxiv.org/abs/hep-th/0202072}{{\ttfamily arXiv:hep-th/0202072}}.

\bibitem{Mathur:2005zp}
S.~D. Mathur, ``{The Fuzzball proposal for black holes: An Elementary
  review},'' \href{http://dx.doi.org/10.1002/prop.200410203}{{\em Fortsch.
  Phys.} {\bfseries 53} (2005) 793--827},
  \href{http://arxiv.org/abs/hep-th/0502050}{{\ttfamily arXiv:hep-th/0502050}}.

\bibitem{Mathur:2008nj}
S.~D. Mathur, ``{Fuzzballs and the information paradox: A Summary and
  conjectures},'' \href{http://arxiv.org/abs/0810.4525}{{\ttfamily
  arXiv:0810.4525 [hep-th]}}.

\bibitem{Bena:2015bea}
I.~Bena, S.~Giusto, R.~Russo, M.~Shigemori, and N.~P. Warner, ``{Habemus
  Superstratum! A constructive proof of the existence of superstrata},''
  \href{http://dx.doi.org/10.1007/JHEP05(2015)110}{{\em JHEP} {\bfseries 05}
  (2015) 110}, \href{http://arxiv.org/abs/1503.01463}{{\ttfamily
  arXiv:1503.01463 [hep-th]}}.

\bibitem{Bena:2016agb}
I.~Bena, E.~Martinec, D.~Turton, and N.~P. Warner, ``{Momentum Fractionation on
  Superstrata},'' \href{http://dx.doi.org/10.1007/JHEP05(2016)064}{{\em JHEP}
  {\bfseries 05} (2016) 064}, \href{http://arxiv.org/abs/1601.05805}{{\ttfamily
  arXiv:1601.05805 [hep-th]}}.

\bibitem{Bena:2016ypk}
I.~Bena, S.~Giusto, E.~J. Martinec, R.~Russo, M.~Shigemori, D.~Turton, and
  N.~P. Warner, ``{Smooth horizonless geometries deep inside the black-hole
  regime},'' \href{http://dx.doi.org/10.1103/PhysRevLett.117.201601}{{\em Phys.
  Rev. Lett.} {\bfseries 117} no.~20, (2016) 201601},
  \href{http://arxiv.org/abs/1607.03908}{{\ttfamily arXiv:1607.03908
  [hep-th]}}.

\bibitem{Bena:2017xbt}
I.~Bena, S.~Giusto, E.~J. Martinec, R.~Russo, M.~Shigemori, D.~Turton, and
  N.~P. Warner, ``{Asymptotically-flat supergravity solutions deep inside the
  black-hole regime},'' \href{http://dx.doi.org/10.1007/JHEP02(2018)014}{{\em
  JHEP} {\bfseries 02} (2018) 014},
  \href{http://arxiv.org/abs/1711.10474}{{\ttfamily arXiv:1711.10474
  [hep-th]}}.

\bibitem{Bianchi:2017bxl}
M.~Bianchi, J.~F. Morales, L.~Pieri, and N.~Zinnato, ``{More on microstate
  geometries of 4d black holes},''
  \href{http://dx.doi.org/10.1007/JHEP05(2017)147}{{\em JHEP} {\bfseries 05}
  (2017) 147}, \href{http://arxiv.org/abs/1701.05520}{{\ttfamily
  arXiv:1701.05520 [hep-th]}}.

\bibitem{Bena:2017upb}
I.~Bena, D.~Turton, R.~Walker, and N.~P. Warner, ``{Integrability and
  Black-Hole Microstate Geometries},''
  \href{http://dx.doi.org/10.1007/JHEP11(2017)021}{{\em JHEP} {\bfseries 11}
  (2017) 021}, \href{http://arxiv.org/abs/1709.01107}{{\ttfamily
  arXiv:1709.01107 [hep-th]}}.

\bibitem{Strominger:1996sh}
A.~Strominger and C.~Vafa, ``{Microscopic origin of the Bekenstein-Hawking
  entropy},'' \href{http://dx.doi.org/10.1016/0370-2693(96)00345-0}{{\em Phys.
  Lett. B} {\bfseries 379} (1996) 99--104},
  \href{http://arxiv.org/abs/hep-th/9601029}{{\ttfamily arXiv:hep-th/9601029}}.

\bibitem{Horowitz:1996ay}
G.~T. Horowitz, J.~M. Maldacena, and A.~Strominger, ``{Nonextremal black hole
  microstates and U duality},''
  \href{http://dx.doi.org/10.1016/0370-2693(96)00738-1}{{\em Phys. Lett. B}
  {\bfseries 383} (1996) 151--159},
  \href{http://arxiv.org/abs/hep-th/9603109}{{\ttfamily arXiv:hep-th/9603109}}.

\bibitem{Maldacena:1997de}
J.~M. Maldacena, A.~Strominger, and E.~Witten, ``{Black hole entropy in M
  theory},'' \href{http://dx.doi.org/10.1088/1126-6708/1997/12/002}{{\em JHEP}
  {\bfseries 12} (1997) 002},
  \href{http://arxiv.org/abs/hep-th/9711053}{{\ttfamily arXiv:hep-th/9711053}}.

\bibitem{Bianchi:2017sds}
M.~Bianchi, D.~Consoli, and J.~Morales, ``{Probing Fuzzballs with Particles,
  Waves and Strings},'' \href{http://dx.doi.org/10.1007/JHEP06(2018)157}{{\em
  JHEP} {\bfseries 06} (2018) 157},
  \href{http://arxiv.org/abs/1711.10287}{{\ttfamily arXiv:1711.10287
  [hep-th]}}.

\bibitem{Bianchi:2018kzy}
M.~Bianchi, D.~Consoli, A.~Grillo, and J.~F. Morales, ``{The dark side of
  fuzzball geometries},'' \href{http://dx.doi.org/10.1007/JHEP05(2019)126}{{\em
  JHEP} {\bfseries 05} (2019) 126},
  \href{http://arxiv.org/abs/1811.02397}{{\ttfamily arXiv:1811.02397
  [hep-th]}}.

\bibitem{Bena:2018mpb}
I.~Bena, E.~J. Martinec, R.~Walker, and N.~P. Warner, ``{Early Scrambling and
  Capped BTZ Geometries},''
  \href{http://dx.doi.org/10.1007/JHEP04(2019)126}{{\em JHEP} {\bfseries 04}
  (2019) 126}, \href{http://arxiv.org/abs/1812.05110}{{\ttfamily
  arXiv:1812.05110 [hep-th]}}.

\bibitem{Bena:2019azk}
I.~Bena, P.~Heidmann, R.~Monten, and N.~P. Warner, ``{Thermal Decay without
  Information Loss in Horizonless Microstate Geometries},''
  \href{http://dx.doi.org/10.21468/SciPostPhys.7.5.063}{{\em SciPost Phys.}
  {\bfseries 7} no.~5, (2019) 063},
  \href{http://arxiv.org/abs/1905.05194}{{\ttfamily arXiv:1905.05194
  [hep-th]}}.

\bibitem{Bena:2020uup}
I.~Bena and D.~R. Mayerson, ``{Black Holes Lessons from Multipole Ratios},''
  \href{http://arxiv.org/abs/2007.09152}{{\ttfamily arXiv:2007.09152
  [hep-th]}}.

\bibitem{Bianchi:2020des}
M.~Bianchi, A.~Grillo, and J.~F. Morales, ``{Chaos at the rim of black hole and
  fuzzball shadows},'' \href{http://dx.doi.org/10.1007/JHEP05(2020)078}{{\em
  JHEP} {\bfseries 05} (2020) 078},
  \href{http://arxiv.org/abs/2002.05574}{{\ttfamily arXiv:2002.05574
  [hep-th]}}.

\bibitem{Bena:2020see}
I.~Bena and D.~R. Mayerson, ``{Multipole Ratios: A New Window into Black
  Holes},'' \href{http://dx.doi.org/10.1103/PhysRevLett.125.221602}{{\em Phys.
  Rev. Lett.} {\bfseries 125} no.~22, (2020) 22},
  \href{http://arxiv.org/abs/2006.10750}{{\ttfamily arXiv:2006.10750
  [hep-th]}}.

\bibitem{Bianchi:2020miz}
M.~Bianchi, D.~Consoli, A.~Grillo, J.~F. Morales, P.~Pani, and G.~Raposo,
  ``{The multipolar structure of fuzzballs},''
  \href{http://dx.doi.org/10.1007/JHEP01(2021)003}{{\em JHEP} {\bfseries 01}
  (2021) 003}, \href{http://arxiv.org/abs/2008.01445}{{\ttfamily
  arXiv:2008.01445 [hep-th]}}.

\bibitem{Bianchi:2020yzr}
M.~Bianchi, D.~Consoli, A.~Grillo, and J.~F. Morales, ``{Light rings of
  five-dimensional geometries},''
  \href{http://arxiv.org/abs/2011.04344}{{\ttfamily arXiv:2011.04344
  [hep-th]}}.

\bibitem{Mayerson:2020tpn}
D.~R. Mayerson, ``{Fuzzballs and Observations},''
  \href{http://dx.doi.org/10.1007/s10714-020-02769-w}{{\em Gen. Rel. Grav.}
  {\bfseries 52} no.~12, (2020) 115},
  \href{http://arxiv.org/abs/2010.09736}{{\ttfamily arXiv:2010.09736
  [hep-th]}}.

\bibitem{Cardoso:2019rvt}
V.~Cardoso and P.~Pani, ``{Testing the nature of dark compact objects: a status
  report},'' \href{http://dx.doi.org/10.1007/s41114-019-0020-4}{{\em Living
  Rev. Rel.} {\bfseries 22} no.~1, (2019) 4},
  \href{http://arxiv.org/abs/1904.05363}{{\ttfamily arXiv:1904.05363 [gr-qc]}}.

\bibitem{Maggio:2020jml}
E.~Maggio, L.~Buoninfante, A.~Mazumdar, and P.~Pani, ``{How does a dark compact
  object ringdown?},''
  \href{http://dx.doi.org/10.1103/PhysRevD.102.064053}{{\em Phys. Rev. D}
  {\bfseries 102} no.~6, (2020) 064053},
  \href{http://arxiv.org/abs/2006.14628}{{\ttfamily arXiv:2006.14628 [gr-qc]}}.

\bibitem{Cardoso:2016rao}
V.~Cardoso, E.~Franzin, and P.~Pani, ``{Is the gravitational-wave ringdown a
  probe of the event horizon?},''
  \href{http://dx.doi.org/10.1103/PhysRevLett.117.089902,
  10.1103/PhysRevLett.116.171101}{{\em Phys. Rev. Lett.} {\bfseries 116}
  no.~17, (2016) 171101}, \href{http://arxiv.org/abs/1602.07309}{{\ttfamily
  arXiv:1602.07309 [gr-qc]}}.
[Erratum: Phys. Rev. Lett.117,no.8,089902(2016)].

\bibitem{Cardoso:2016oxy}
V.~Cardoso, S.~Hopper, C.~F.~B. Macedo, C.~Palenzuela, and P.~Pani,
  ``{Gravitational-wave signatures of exotic compact objects and of quantum
  corrections at the horizon scale},''
  \href{http://dx.doi.org/10.1103/PhysRevD.94.084031}{{\em Phys. Rev. D}
  {\bfseries 94} no.~8, (2016) 084031},
  \href{http://arxiv.org/abs/1608.08637}{{\ttfamily arXiv:1608.08637 [gr-qc]}}.

\bibitem{Barausse:2014tra}
E.~Barausse, V.~Cardoso, and P.~Pani, ``{Can environmental effects spoil
  precision gravitational-wave astrophysics?},''
  \href{http://dx.doi.org/10.1103/PhysRevD.89.104059}{{\em Phys. Rev.}
  {\bfseries D89} no.~10, (2014) 104059},
\href{http://arxiv.org/abs/1404.7149}{{\ttfamily arXiv:1404.7149 [gr-qc]}}.

\bibitem{Holdom:2016nek}
B.~Holdom and J.~Ren, ``{Not quite a black hole},''
  \href{http://dx.doi.org/10.1103/PhysRevD.95.084034}{{\em Phys. Rev.}
  {\bfseries D95} no.~8, (2017) 084034},
\href{http://arxiv.org/abs/1612.04889}{{\ttfamily arXiv:1612.04889 [gr-qc]}}.

\bibitem{Conklin:2017lwb}
R.~S. Conklin, B.~Holdom, and J.~Ren, ``{Gravitational wave echoes through new
  windows},'' \href{http://dx.doi.org/10.1103/PhysRevD.98.044021}{{\em Phys.
  Rev.} {\bfseries D98} no.~4, (2018) 044021},
\href{http://arxiv.org/abs/1712.06517}{{\ttfamily arXiv:1712.06517 [gr-qc]}}.

\bibitem{Oshita:2018fqu}
N.~Oshita and N.~Afshordi, ``{Probing microstructure of black hole spacetimes
  with gravitational wave echoes},''
  \href{http://dx.doi.org/10.1103/PhysRevD.99.044002}{{\em Phys. Rev.}
  {\bfseries D99} no.~4, (2019) 044002},
\href{http://arxiv.org/abs/1807.10287}{{\ttfamily arXiv:1807.10287 [gr-qc]}}.

\bibitem{Burgess:2018pmm}
C.~P. Burgess, R.~Plestid, and M.~Rummel, ``{Effective Field Theory of Black
  Hole Echoes},'' \href{http://dx.doi.org/10.1007/JHEP09(2018)113}{{\em JHEP}
  {\bfseries 09} (2018) 113}, \href{http://arxiv.org/abs/1808.00847}{{\ttfamily
  arXiv:1808.00847 [gr-qc]}}.

\bibitem{Wang:2019rcf}
Q.~Wang, N.~Oshita, and N.~Afshordi, ``{Echoes from Quantum Black Holes},''
  \href{http://dx.doi.org/10.1103/PhysRevD.101.024031}{{\em Phys. Rev. D}
  {\bfseries 101} no.~2, (2020) 024031},
  \href{http://arxiv.org/abs/1905.00446}{{\ttfamily arXiv:1905.00446 [gr-qc]}}.

\bibitem{Cardoso:2019apo}
V.~Cardoso, V.~F. Foit, and M.~Kleban, ``{Gravitational wave echoes from black
  hole area quantization},''
  \href{http://dx.doi.org/10.1088/1475-7516/2019/08/006}{{\em JCAP} {\bfseries
  08} (2019) 006}, \href{http://arxiv.org/abs/1902.10164}{{\ttfamily
  arXiv:1902.10164 [hep-th]}}.

\bibitem{Coates:2019bun}
A.~Coates, S.~H. Völkel, and K.~D. Kokkotas, ``{Spectral Lines of Quantized,
  Spinning Black Holes and their Astrophysical Relevance},''
  \href{http://dx.doi.org/10.1103/PhysRevLett.123.171104}{{\em Phys. Rev.
  Lett.} {\bfseries 123} no.~17, (2019) 171104},
  \href{http://arxiv.org/abs/1909.01254}{{\ttfamily arXiv:1909.01254 [gr-qc]}}.

\bibitem{Buoninfante:2019teo}
L.~Buoninfante, A.~Mazumdar, and J.~Peng, ``{Nonlocality amplifies echoes},''
  \href{http://dx.doi.org/10.1103/PhysRevD.100.104059}{{\em Phys. Rev. D}
  {\bfseries 100} no.~10, (2019) 104059},
  \href{http://arxiv.org/abs/1906.03624}{{\ttfamily arXiv:1906.03624 [gr-qc]}}.

\bibitem{Delhom:2019btt}
A.~Delhom, C.~F. Macedo, G.~J. Olmo, and L.~C. Crispino, ``{Absorption by black
  hole remnants in metric-affine gravity},''
  \href{http://dx.doi.org/10.1103/PhysRevD.100.024016}{{\em Phys. Rev. D}
  {\bfseries 100} no.~2, (2019) 024016},
  \href{http://arxiv.org/abs/1906.06411}{{\ttfamily arXiv:1906.06411 [gr-qc]}}.

\bibitem{Dey:2020lhq}
R.~Dey, S.~Chakraborty, and N.~Afshordi, ``{Echoes from braneworld black
  holes},'' \href{http://dx.doi.org/10.1103/PhysRevD.101.104014}{{\em Phys.
  Rev. D} {\bfseries 101} no.~10, (2020) 104014},
  \href{http://arxiv.org/abs/2001.01301}{{\ttfamily arXiv:2001.01301 [gr-qc]}}.

\bibitem{Buoninfante:2020tfb}
L.~Buoninfante, ``{Echoes from corpuscular black holes},''
  \href{http://dx.doi.org/10.1088/1475-7516/2020/12/041}{{\em JCAP} {\bfseries
  12} (2020) 041}, \href{http://arxiv.org/abs/2005.08426}{{\ttfamily
  arXiv:2005.08426 [gr-qc]}}.

\bibitem{Liu:2021aqh}
H.~Liu, W.-L. Qian, Y.~Liu, J.-P. Wu, B.~Wang, and R.-H. Yue, ``{On an
  alternative mechanism for the black hole echoes},''
  \href{http://arxiv.org/abs/2104.11912}{{\ttfamily arXiv:2104.11912 [gr-qc]}}.

\bibitem{Abedi:2016hgu}
J.~Abedi, H.~Dykaar, and N.~Afshordi, ``{Echoes from the Abyss: Tentative
  evidence for Planck-scale structure at black hole horizons},''
  \href{http://dx.doi.org/10.1103/PhysRevD.96.082004}{{\em Phys. Rev. D}
  {\bfseries 96} no.~8, (2017) 082004},
  \href{http://arxiv.org/abs/1612.00266}{{\ttfamily arXiv:1612.00266 [gr-qc]}}.

\bibitem{Ashton:2016xff}
G.~Ashton, O.~Birnholtz, M.~Cabero, C.~Capano, T.~Dent, B.~Krishnan, G.~D.
  Meadors, A.~B. Nielsen, A.~Nitz, and J.~Westerweck, ``{Comments on: "Echoes
  from the abyss: Evidence for Planck-scale structure at black hole
  horizons"},''
\href{http://arxiv.org/abs/1612.05625}{{\ttfamily arXiv:1612.05625 [gr-qc]}}.

\bibitem{Westerweck:2017hus}
J.~Westerweck, A.~Nielsen, O.~Fischer-Birnholtz, M.~Cabero, C.~Capano, T.~Dent,
  B.~Krishnan, G.~Meadors, and A.~H. Nitz, ``{Low significance of evidence for
  black hole echoes in gravitational wave data},''
  \href{http://dx.doi.org/10.1103/PhysRevD.97.124037}{{\em Phys. Rev.}
  {\bfseries D97} no.~12, (2018) 124037},
\href{http://arxiv.org/abs/1712.09966}{{\ttfamily arXiv:1712.09966 [gr-qc]}}.

\bibitem{Abedi:2018pst}
J.~Abedi, H.~Dykaar, and N.~Afshordi, ``{Comment on: "Low significance of
  evidence for black hole echoes in gravitational wave data"},''
\href{http://arxiv.org/abs/1803.08565}{{\ttfamily arXiv:1803.08565 [gr-qc]}}.

\bibitem{Conklin:2019fcs}
R.~S. Conklin and B.~Holdom, ``{Gravitational wave echo spectra},''
  \href{http://dx.doi.org/10.1103/PhysRevD.100.124030}{{\em Phys. Rev. D}
  {\bfseries 100} no.~12, (2019) 124030},
  \href{http://arxiv.org/abs/1905.09370}{{\ttfamily arXiv:1905.09370 [gr-qc]}}.

\bibitem{Tsang:2019zra}
K.~W. Tsang, A.~Ghosh, A.~Samajdar, K.~Chatziioannou, S.~Mastrogiovanni,
  M.~Agathos, and C.~Van Den~Broeck, ``{A morphology-independent search for
  gravitational wave echoes in data from the first and second observing runs of
  Advanced LIGO and Advanced Virgo},''
  \href{http://dx.doi.org/10.1103/PhysRevD.101.064012}{{\em Phys. Rev. D}
  {\bfseries 101} no.~6, (2020) 064012},
  \href{http://arxiv.org/abs/1906.11168}{{\ttfamily arXiv:1906.11168 [gr-qc]}}.

\bibitem{Uchikata:2019frs}
N.~Uchikata, H.~Nakano, T.~Narikawa, N.~Sago, H.~Tagoshi, and T.~Tanaka,
  ``{Searching for black hole echoes from the LIGO-Virgo Catalog GWTC-1},''
  \href{http://dx.doi.org/10.1103/PhysRevD.100.062006}{{\em Phys. Rev. D}
  {\bfseries 100} no.~6, (2019) 062006},
  \href{http://arxiv.org/abs/1906.00838}{{\ttfamily arXiv:1906.00838 [gr-qc]}}.

\bibitem{Abedi:2020ujo}
J.~Abedi, N.~Afshordi, N.~Oshita, and Q.~Wang, ``{Quantum Black Holes in the
  Sky},'' \href{http://dx.doi.org/10.3390/universe6030043}{{\em Universe}
  {\bfseries 6} no.~3, (2020) 43},
  \href{http://arxiv.org/abs/2001.09553}{{\ttfamily arXiv:2001.09553 [gr-qc]}}.

\bibitem{Bena:2007kg}
I.~Bena and N.~P. Warner, ``{Black holes, black rings and their microstates},''
  \href{http://dx.doi.org/10.1007/978-3-540-79523-0_1}{{\em Lect. Notes Phys.}
  {\bfseries 755} (2008) 1--92},
  \href{http://arxiv.org/abs/hep-th/0701216}{{\ttfamily arXiv:hep-th/0701216}}.

\bibitem{Gibbons:2013tqa}
G.~W. Gibbons and N.~P. Warner, ``{Global structure of five-dimensional
  fuzzballs},'' \href{http://dx.doi.org/10.1088/0264-9381/31/2/025016}{{\em
  Class. Quant. Grav.} {\bfseries 31} (2014) 025016},
  \href{http://arxiv.org/abs/1305.0957}{{\ttfamily arXiv:1305.0957 [hep-th]}}.

\bibitem{Bates:2003vx}
B.~Bates and F.~Denef, ``{Exact solutions for supersymmetric stationary black
  hole composites},'' \href{http://dx.doi.org/10.1007/JHEP11(2011)127}{{\em
  JHEP} {\bfseries 11} (2011) 127},
  \href{http://arxiv.org/abs/hep-th/0304094}{{\ttfamily arXiv:hep-th/0304094}}.

\bibitem{Bianchi:2020bxa}
M.~Bianchi, D.~Consoli, A.~Grillo, J.~F. Morales, P.~Pani, and G.~Raposo,
  ``{Distinguishing fuzzballs from black holes through their multipolar
  structure},'' \href{http://dx.doi.org/10.1103/PhysRevLett.125.221601}{{\em
  Phys. Rev. Lett.} {\bfseries 125} no.~22, (2020) 221601},
  \href{http://arxiv.org/abs/2007.01743}{{\ttfamily arXiv:2007.01743
  [hep-th]}}.

\bibitem{Cvetic:1995uj}
M.~Cvetic and D.~Youm, ``{Dyonic BPS saturated black holes of heterotic string
  on a six torus},'' \href{http://dx.doi.org/10.1103/PhysRevD.53.R584}{{\em
  Phys. Rev. D} {\bfseries 53} (1996) 584--588},
  \href{http://arxiv.org/abs/hep-th/9507090}{{\ttfamily arXiv:hep-th/9507090}}.

\bibitem{Mark:2017dnq}
Z.~Mark, A.~Zimmerman, S.~M. Du, and Y.~Chen, ``{A recipe for echoes from
  exotic compact objects},''
  \href{http://dx.doi.org/10.1103/PhysRevD.96.084002}{{\em Phys. Rev. D}
  {\bfseries 96} no.~8, (2017) 084002},
  \href{http://arxiv.org/abs/1706.06155}{{\ttfamily arXiv:1706.06155 [gr-qc]}}.

\bibitem{Correia:2018apm}
M.~R. Correia and V.~Cardoso, ``{Characterization of echoes: A Dyson-series
  representation of individual pulses},''
  \href{http://dx.doi.org/10.1103/PhysRevD.97.084030}{{\em Phys. Rev.}
  {\bfseries D97} no.~8, (2018) 084030},
\href{http://arxiv.org/abs/1802.07735}{{\ttfamily arXiv:1802.07735 [gr-qc]}}.

\bibitem{Ferrari:1984zz}
V.~Ferrari and B.~Mashhoon, ``{New approach to the quasinormal modes of a black
  hole},''
\href{http://dx.doi.org/10.1103/PhysRevD.30.295}{{\em Phys. Rev.} {\bfseries
  D30} (1984) 295--304}.

\bibitem{Cardoso:2008bp}
V.~Cardoso, A.~S. Miranda, E.~Berti, H.~Witek, and V.~T. Zanchin, ``{Geodesic
  stability, Lyapunov exponents and quasinormal modes},''
  \href{http://dx.doi.org/10.1103/PhysRevD.79.064016}{{\em Phys. Rev.}
  {\bfseries D79} (2009) 064016},
\href{http://arxiv.org/abs/0812.1806}{{\ttfamily arXiv:0812.1806 [hep-th]}}.

\bibitem{Yang:2012he}
H.~Yang, D.~A. Nichols, F.~Zhang, A.~Zimmerman, Z.~Zhang, and Y.~Chen,
  ``{Quasinormal-mode spectrum of Kerr black holes and its geometric
  interpretation},'' \href{http://dx.doi.org/10.1103/PhysRevD.86.104006}{{\em
  Phys. Rev. D} {\bfseries 86} (2012) 104006},
  \href{http://arxiv.org/abs/1207.4253}{{\ttfamily arXiv:1207.4253 [gr-qc]}}.

\bibitem{Cardoso:2016olt}
V.~Cardoso, C.~F.~B. Macedo, P.~Pani, and V.~Ferrari, ``{Black holes and
  gravitational waves in models of minicharged dark matter},''
  \href{http://dx.doi.org/10.1088/1475-7516/2016/05/054}{{\em JCAP} {\bfseries
  05} (2016) 054}, \href{http://arxiv.org/abs/1604.07845}{{\ttfamily
  arXiv:1604.07845 [hep-ph]}}. [Erratum: JCAP 04, E01 (2020)].

\bibitem{Pani:2018flj}
P.~Pani and V.~Ferrari, ``{On gravitational-wave echoes from neutron-star
  binary coalescences},''
  \href{http://dx.doi.org/10.1088/1361-6382/aacb8f}{{\em Class. Quant. Grav.}
  {\bfseries 35} no.~15, (2018) 15LT01},
  \href{http://arxiv.org/abs/1804.01444}{{\ttfamily arXiv:1804.01444 [gr-qc]}}.

\bibitem{webpage}
\url{https://web.uniroma1.it/gmunu}.

\bibitem{Loffler:2011ay}
F.~L{\"{o}}ffler, J.~Faber, E.~Bentivegna, T.~Bode, P.~Diener, R.~Haas,
  I.~Hinder, B.~C. Mundim, C.~D. Ott, E.~Schnetter, G.~Allen, M.~Campanelli,
  and P.~Laguna, ``{{T}he {E}instein {T}oolkit: {A} {C}ommunity {C}omputational
  {I}nfrastructure for {R}elativistic {A}strophysics},''
  \href{http://dx.doi.org/doi:10.1088/0264-9381/29/11/115001}{{\em Class.
  Quantum Grav.} {\bfseries 29} no.~11, (2012) 115001},
  \href{http://arxiv.org/abs/arXiv:1111.3344 [gr-qc]}{{\ttfamily
  arXiv:1111.3344 [gr-qc]}}.

\bibitem{Zilhao:2013hia}
M.~Zilh{\~a}o and F.~L{\"o}ffler, ``{An Introduction to the Einstein
  Toolkit},'' \href{http://dx.doi.org/10.1142/S0217751X13400149}{{\em Int. J.
  Mod. Phys.} {\bfseries A28} (2013) 1340014},
\href{http://arxiv.org/abs/1305.5299}{{\ttfamily arXiv:1305.5299 [gr-qc]}}.

\bibitem{EinsteinToolkit:2019_10}
M.~Babiuc-Hamilton {\em et~al.}, ``The {E}instein {T}oolkit,'' Oct., 2019.
\newblock \url{https://doi.org/10.5281/zenodo.3522086}. To find out more, visit
  http://einsteintoolkit.org.

\bibitem{Schnetter:2003rb}
E.~Schnetter, S.~H. Hawley, and I.~Hawke, ``{Evolutions in 3-D numerical
  relativity using fixed mesh refinement},''
  \href{http://dx.doi.org/10.1088/0264-9381/21/6/014}{{\em Class. Quant. Grav.}
  {\bfseries 21} (2004) 1465--1488},
\href{http://arxiv.org/abs/gr-qc/0310042}{{\ttfamily arXiv:gr-qc/0310042
  [gr-qc]}}.

\bibitem{CarpetCode:web}
\url{http://www.carpetcode.org/}. {Carpet}: Adaptive Mesh Refinement for the
  {Cactus} Framework.

\bibitem{Pollney_2011}
D.~Pollney, C.~Reisswig, E.~Schnetter, N.~Dorband, and P.~Diener, ``High
  accuracy binary black hole simulations with an extended wave zone,''
  \href{http://dx.doi.org/10.1103/physrevd.83.044045}{{\em Physical Review D}
  {\bfseries 83} no.~4, (Feb, 2011) }.
  \url{http://dx.doi.org/10.1103/PhysRevD.83.044045}.

\bibitem{Cunha:2017wao}
P.~V.~P. Cunha, J.~A. Font, C.~Herdeiro, E.~Radu, N.~Sanchis-Gual, and
  M.~Zilh{\~a}o, ``{Lensing and dynamics of ultracompact bosonic stars},''
  \href{http://dx.doi.org/10.1103/PhysRevD.96.104040}{{\em Phys. Rev.}
  {\bfseries D96} no.~10, (2017) 104040},
\href{http://arxiv.org/abs/1709.06118}{{\ttfamily arXiv:1709.06118 [gr-qc]}}.

\bibitem{Ikeda:2020xvt}
T.~Ikeda, L.~Bernard, V.~Cardoso, and M.~Zilh\~ao, ``{Black hole binaries and
  light fields: Gravitational molecules},''
  \href{http://dx.doi.org/10.1103/PhysRevD.103.024020}{{\em Phys. Rev. D}
  {\bfseries 103} no.~2, (2021) 024020},
  \href{http://arxiv.org/abs/2010.00008}{{\ttfamily arXiv:2010.00008 [gr-qc]}}.

\bibitem{Bernard:2019nkv}
L.~Bernard, V.~Cardoso, T.~Ikeda, and M.~Zilh{\~a}o, ``{Physics of black hole
  binaries: Geodesics, relaxation modes, and energy extraction},''
  \href{http://dx.doi.org/10.1103/PhysRevD.100.044002}{{\em Phys. Rev. D}
  {\bfseries 100} no.~4, (2019) 044002},
  \href{http://arxiv.org/abs/1905.05204}{{\ttfamily arXiv:1905.05204 [gr-qc]}}.

\bibitem{Arnowitt:1962hi}
R.~L. Arnowitt, S.~Deser, and C.~W. Misner, ``{The Dynamics of general
  relativity},'' \href{http://dx.doi.org/10.1007/s10714-008-0661-1}{{\em Gen.
  Rel. Grav.} {\bfseries 40} (2008) 1997--2027},
  \href{http://arxiv.org/abs/gr-qc/0405109}{{\ttfamily arXiv:gr-qc/0405109}}.

\bibitem{Price:1971fb}
R.~H. Price, ``{Nonspherical perturbations of relativistic gravitational
  collapse. 1. Scalar and gravitational perturbations},''
  \href{http://dx.doi.org/10.1103/PhysRevD.5.2419}{{\em Phys. Rev. D}
  {\bfseries 5} (1972) 2419--2438}.

\bibitem{Testa:2018bzd}
A.~Testa and P.~Pani, ``{Analytical template for gravitational-wave echoes:
  signal characterization and prospects of detection with current and future
  interferometers},'' \href{http://dx.doi.org/10.1103/PhysRevD.98.044018}{{\em
  Phys. Rev. D} {\bfseries 98} no.~4, (2018) 044018},
  \href{http://arxiv.org/abs/1806.04253}{{\ttfamily arXiv:1806.04253 [gr-qc]}}.

\bibitem{1978CMaPh..63..243F}
J.~L. {Friedman}, ``{Ergosphere instability},''
  \href{http://dx.doi.org/10.1007/BF01196933}{{\em Commun. Math. Phys.}
  {\bfseries 63} (Oct., 1978) 243--255}.

\bibitem{Cardoso:2007az}
V.~Cardoso, P.~Pani, M.~Cadoni, and M.~Cavaglia, ``{Ergoregion instability of
  ultracompact astrophysical objects},''
  \href{http://dx.doi.org/10.1103/PhysRevD.77.124044}{{\em Phys.Rev.}
  {\bfseries D77} (2008) 124044},
\href{http://arxiv.org/abs/0709.0532}{{\ttfamily arXiv:0709.0532 [gr-qc]}}.

\bibitem{Chirenti:2008pf}
C.~B. Chirenti and L.~Rezzolla, ``{On the ergoregion instability in rotating
  gravastars},'' \href{http://dx.doi.org/10.1103/PhysRevD.78.084011}{{\em
  Phys.Rev.} {\bfseries D78} (2008) 084011},
\href{http://arxiv.org/abs/0808.4080}{{\ttfamily arXiv:0808.4080 [gr-qc]}}.

\bibitem{Pani:2010jz}
P.~Pani, E.~Barausse, E.~Berti, and V.~Cardoso, ``{Gravitational instabilities
  of superspinars},'' \href{http://dx.doi.org/10.1103/PhysRevD.82.044009}{{\em
  Phys.Rev.} {\bfseries D82} (2010) 044009},
\href{http://arxiv.org/abs/1006.1863}{{\ttfamily arXiv:1006.1863 [gr-qc]}}.

\bibitem{Cardoso:2008kj}
V.~Cardoso, P.~Pani, M.~Cadoni, and M.~Cavaglia, ``{Instability of
  hyper-compact Kerr-like objects},''
  \href{http://dx.doi.org/10.1088/0264-9381/25/19/195010}{{\em
  Class.Quant.Grav.} {\bfseries 25} (2008) 195010},
\href{http://arxiv.org/abs/0808.1615}{{\ttfamily arXiv:0808.1615 [gr-qc]}}.

\bibitem{Maggio:2017ivp}
E.~Maggio, P.~Pani, and V.~Ferrari, ``{Exotic Compact Objects and How to Quench
  their Ergoregion Instability},''
  \href{http://dx.doi.org/10.1103/PhysRevD.96.104047}{{\em Phys. Rev.}
  {\bfseries D96} no.~10, (2017) 104047},
\href{http://arxiv.org/abs/1703.03696}{{\ttfamily arXiv:1703.03696 [gr-qc]}}.

\bibitem{Maggio:2018ivz}
E.~Maggio, V.~Cardoso, S.~R. Dolan, and P.~Pani, ``{Ergoregion instability of
  exotic compact objects: electromagnetic and gravitational perturbations and
  the role of absorption},''
  \href{http://dx.doi.org/10.1103/PhysRevD.99.064007}{{\em Phys. Rev.}
  {\bfseries D99} no.~6, (2019) 064007},
\href{http://arxiv.org/abs/1807.08840}{{\ttfamily arXiv:1807.08840 [gr-qc]}}.

\bibitem{Eperon:2016cdd}
F.~C. Eperon, H.~S. Reall, and J.~E. Santos, ``{Instability of supersymmetric
  microstate geometries},''
  \href{http://dx.doi.org/10.1007/JHEP10(2016)031}{{\em JHEP} {\bfseries 10}
  (2016) 031}, \href{http://arxiv.org/abs/1607.06828}{{\ttfamily
  arXiv:1607.06828 [hep-th]}}.

\bibitem{Keir:2014oka}
J.~Keir, ``{Slowly decaying waves on spherically symmetric spacetimes and
  ultracompact neutron stars},''
  \href{http://dx.doi.org/10.1088/0264-9381/33/13/135009}{{\em Class. Quant.
  Grav.} {\bfseries 33} no.~13, (2016) 135009},
  \href{http://arxiv.org/abs/1404.7036}{{\ttfamily arXiv:1404.7036 [gr-qc]}}.

\bibitem{Cardoso:2014sna}
V.~Cardoso, L.~C.~B. Crispino, C.~F.~B. Macedo, H.~Okawa, and P.~Pani, ``{Light
  rings as observational evidence for event horizons: long-lived modes,
  ergoregions and nonlinear instabilities of ultracompact objects},''
  \href{http://dx.doi.org/10.1103/PhysRevD.90.044069}{{\em Phys. Rev. D}
  {\bfseries 90} no.~4, (2014) 044069},
  \href{http://arxiv.org/abs/1406.5510}{{\ttfamily arXiv:1406.5510 [gr-qc]}}.

\bibitem{Dimitrov:2020txx}
V.~Dimitrov, T.~Lemmens, D.~R. Mayerson, V.~S. Min, and B.~Vercnocke,
  ``{Gravitational Waves, Holography, and Black Hole Microstates},''
  \href{http://arxiv.org/abs/2007.01879}{{\ttfamily arXiv:2007.01879
  [hep-th]}}.

\bibitem{Bacchini:2021fig}
F.~Bacchini, D.~R. Mayerson, B.~Ripperda, J.~Davelaar, H.~Olivares, T.~Hertog,
  and B.~Vercnocke, ``{Fuzzball Shadows: Emergent Horizons from
  Microstructure},'' \href{http://arxiv.org/abs/2103.12075}{{\ttfamily
  arXiv:2103.12075 [hep-th]}}.

\bibitem{Addazi:2020obs}
A.~Addazi, M.~Bianchi, M.~Firrotta, and A.~Marciano, ``{String Memories ...
  Lost and Regained},''
\href{http://arxiv.org/abs/2008.02206}{{\ttfamily arXiv:2008.02206 [hep-th]}}.

\bibitem{Aldi:2020qfu}
A.~Aldi, M.~Bianchi, and M.~Firrotta, ``{String memories... openly retold},''
  \href{http://dx.doi.org/10.1016/j.physletb.2020.136037}{{\em Phys. Lett.}
  {\bfseries B813} (2021) 136037},
\href{http://arxiv.org/abs/2010.04082}{{\ttfamily arXiv:2010.04082 [hep-th]}}.

\bibitem{Aldi:2021zhh}
A.~Aldi, M.~Bianchi, and M.~Firrotta, ``{Spinning-off stringy electro-magnetic
  memories},''
\href{http://arxiv.org/abs/2101.07054}{{\ttfamily arXiv:2101.07054 [hep-th]}}.

\bibitem{Strominger:2014pwa}
A.~Strominger and A.~Zhiboedov, ``{Gravitational Memory, BMS Supertranslations
  and Soft Theorems},'' \href{http://dx.doi.org/10.1007/JHEP01(2016)086}{{\em
  JHEP} {\bfseries 01} (2016) 086},
\href{http://arxiv.org/abs/1411.5745}{{\ttfamily arXiv:1411.5745 [hep-th]}}.

\bibitem{JMaRT}
V.~Jejjala, O.~Madden, S.~F. Ross, and G.~Titchener, ``Nonsupersymmetric smooth
  geometries and d1-d5-p bound states,''
  \href{http://dx.doi.org/10.1103/physrevd.71.124030}{{\em Physical Review D}
  {\bfseries 71} no.~12, (Jun, 2005) }.
  \url{http://dx.doi.org/10.1103/PhysRevD.71.124030}.

\bibitem{Cardoso:2005gj}
V.~Cardoso, O.~J.~C. Dias, J.~L. Hovdebo, and R.~C. Myers, ``{Instability of
  non-supersymmetric smooth geometries},''
  \href{http://dx.doi.org/10.1103/PhysRevD.73.064031}{{\em Phys. Rev. D}
  {\bfseries 73} (2006) 064031},
  \href{http://arxiv.org/abs/hep-th/0512277}{{\ttfamily arXiv:hep-th/0512277}}.

\bibitem{Bianchi:2019lmi}
M.~Bianchi, M.~Casolino, and G.~Rizzo, ``{Accelerating strangelets via Penrose
  process in non-BPS fuzzballs},''
  \href{http://dx.doi.org/10.1016/j.nuclphysb.2020.115010}{{\em Nucl. Phys. B}
  {\bfseries 954} (2020) 115010},
  \href{http://arxiv.org/abs/1904.01097}{{\ttfamily arXiv:1904.01097
  [hep-th]}}.

\bibitem{Chandrasekhar:1975zza}
S.~Chandrasekhar and S.~L. Detweiler, ``{The quasi-normal modes of the
  Schwarzschild black hole},''
  \href{http://dx.doi.org/10.1098/rspa.1975.0112}{{\em Proc. Roy. Soc. Lond. A}
  {\bfseries 344} (1975) 441--452}.

\bibitem{Pani:2013pma}
P.~Pani, ``{Advanced Methods in Black-Hole Perturbation Theory},''
  \href{http://dx.doi.org/10.1142/S0217751X13400186}{{\em Int. J. Mod. Phys. A}
  {\bfseries 28} (2013) 1340018},
  \href{http://arxiv.org/abs/1305.6759}{{\ttfamily arXiv:1305.6759 [gr-qc]}}.

\bibitem{Aminov:2020yma}
G.~Aminov, A.~Grassi, and Y.~Hatsuda, ``{Black Hole Quasinormal Modes and
  Seiberg-Witten Theory},''
\href{http://arxiv.org/abs/2006.06111}{{\ttfamily arXiv:2006.06111 [hep-th]}}.

\bibitem{QNMvsSWwip}
M.~Bianchi, D.~Consoli, A.~Grillo, and F.~Morales, ``{Black Hole and Fuzzball
  Quasinormal Modes from quantum Seiberg-Witten curves},''.
to appear (2021).

\end{thebibliography}\endgroup

\end{document}